\documentclass[aps,pre,groupedaddress]{revtex4-2}
\usepackage{fullpage}
\usepackage{todonotes,color}
\usepackage[centertags]{amsmath}
\usepackage{amscd,amsfonts,amsmath,amssymb,latexsym,mathtools}
\usepackage{graphicx}
\usepackage{theorem}
\usepackage{enumerate}
\usepackage[toc,page]{appendix}
\usepackage{float}
\usepackage{subcaption,overpic}
\usepackage{todonotes,color}
\graphicspath{{.}{./final_figures/}}
\usepackage{overpic}
\usepackage{verbatim}

\DeclareMathOperator{\sech}{sech}

\newcommand{\mat}[4]{\left[\begin{array}{cc} #1 & #2 \\ #3 & #4  \end{array}
  \right]}
\newcommand{\B}{b}
\newcommand{\Aa}{a}
\newcommand{\RHS}{\rho}
\newcommand{\eq}[2]{\sdd{#1}{x} + V_0^4(2 #2 + 3 #1) - #1}
\newcommand{\eqr}[1]{\sdd{#1}{x} + 5 V_0^4 #1 - #1}

\newcommand{\Vn}{V}
\newcommand{\Vexp}{V_{\mathrm{exp}}}
\newcommand{\Uexp}{U_{\mathrm{exp}}}
\newcommand{\Wexp}{W_{\mathrm{exp}}}
\newcommand{\tst}{\mbox{T.S.T.}}
\newcommand{\vtwo}[2]{\left[\begin{array}{c} #1 \\ #2  \end{array}
  \right]}

\newcommand{\rd}{\mbox{d}}
\newcommand{\dd}{\partial}

\newcommand{\la}{\lambda}
\newcommand{\non}{\nonumber}
\newcommand{\eps}{\epsilon}
\newcommand{\beqa}{\begin{eqnarray}}
\newcommand{\eeqa}{\end{eqnarray}}
\newcommand{\beqas}{\begin{eqnarray*}}
\newcommand{\eeqas}{\end{eqnarray*}}
\newcommand{\ba}{\begin{align}}
\newcommand{\ea}{\end{align}}
\newcommand{\bas}{\begin{align*}}
\newcommand{\eas}{\end{align*}}
\newcommand{\beq}{\begin{equation}}
\newcommand{\eeq}{\end{equation}}
\newcommand{\re}{\mathrm{Re}}
\newcommand{\im}{\mathrm{Im}}
\newcommand{\ra}{\rightarrow}
\newcommand{\al}{\alpha}

\newcommand{\ee}{\mathrm{e}}
\newcommand{\ii}{\mathrm{i}}
\newcommand{\pdhfrac}[2]{\mathchoice{\frac{#1}{#2}}{#1/#2}{#1/#2}{#1/#2}}
\newcommand{\fdd}[2]{\pdhfrac{\mathrm{d}#1}{\mathrm{d}#2}}
\newcommand{\sdd}[2]{\pdhfrac{\mathrm{d}^2#1}{\mathrm{d}#2^2}}
\newcommand{\pd}[2]{\pdhfrac{{\partial}#1}{{\partial}#2}}
\newcommand{\spd}[2]{\pdhfrac{\partial^2#1}{{\partial}#2^2}}

\renewcommand{\d}[1]{\mathrm{d}#1}

\newcommand{\Ve}{V_G}

\newcommand{\Vs}{V_{s}}

\newcommand{\V}{V}

\newcommand{\A}{A}

\graphicspath{{./}{./Figures/}}
\makeatletter
\def\input@path{{./}{./Figures/}}
\makeatother

\begin{document}

\title{Multi-hump Collapsing Solutions in the Nonlinear Schr{\"o}dinger Problem: Existence, Stability and Dynamics}

\author{S. Jon Chapman}
\affiliation{Mathematical Institute, University of Oxford, AWB, ROQ, Woodstock Road, Oxford OX2 6GG}

\author{M. Kavousanakis}
\affiliation{School of Chemical Engineering, National Technical University of Athens, 15780, Athens, Greece}

\author{E.G. Charalampidis}
\affiliation{Department of Mathematics and Statistics, and Computational Science Research Center, San Diego State University, San Diego, CA 92182-7720, USA}

 \author{I.G. Kevrekidis}
\affiliation{Department of Chemical and Biomolecular Engineering \& \\
Department of Applied Mathematics and Statistics, Johns Hopkins University, Baltimore, MD 21218, USA}

\author{P.G. Kevrekidis}
\affiliation{Department of Mathematics and Statistics, University of
  Massachusetts, Amherst MA 01003-4515, USA}

\affiliation{Department of Physics, University of
  Massachusetts, Amherst MA 01003, USA}

\date{\today}

\begin{abstract}
In the present work we examine multi-hump solutions of
the nonlinear Schr{\"o}dinger equation in the blowup
regime of the one-dimensional model with power law
nonlinearity, bearing a suitable exponent of $\sigma>2$.
We find that families of
such solutions exist for arbitrary pulse numbers,
with all of them bifurcating from the critical
case of $\sigma=2$. Remarkably, all of them involve
``bifurcations from infinity'', i.e., the pulses come
inward from an infinite distance as the exponent $\sigma$ increases past the critical point. The position
of the pulses is quantified and 
the stability of the
waveforms is also systematically examined in the so-called
``co-exploding frame''. Both the 
equilibrium distance between the pulse peaks and the point spectrum eigenvalues associated with the multi-hump configurations
are obtained as a function of the blowup rate $G$ theoretically, and these
findings are supported by detailed numerical computations.~
Finally, some prototypical dynamical scenarios are
explored, and an outlook towards such multi-hump solutions
in higher dimensions is provided.
\end{abstract}

\maketitle

\section{Introduction}
The nonlinear Schr{\"o}dinger (NLS) model
is undoubtedly one of the quintessential ones within
the realm of dispersive nonlinear partial differential
equations~\cite{ablowitz1981solitons,ablowitz2,sulem,AblowitzPrinariTrubatch}. It arises as a canonical or as an 
envelope model description in a diverse range of
physical settings ranging from the evolution of 
optical beams in fibers and lasers~\cite{hasegawa:sio95,Kivshar2003} to the description
of atomic wavefunctions in ultracold Bose-Einstein 
condensates~\cite{Pitaevskii2003,Pethick2008,siambook}
and from plasmas~\cite{kono} to water waves~\cite{ablowitz2}.

While the focus of a large number of studies on
the NLS model concerns the dynamics of its solitary
wave solutions (either bright~\cite{Kivshar2003} or
dark~\cite{siambook,Frantzeskakis_2010}), the presence 
of {\em collapse} is another key feature of the equation when 
nonlinear effects  overcome dispersive ones.~In this case,
the relevant ground state solution to the NLS becomes 
(orbitally) unstable leading to the formation of singular
solutions~\cite{sulem,fibich2015}.~From a physical perspective,
the focusing nature of the problem induces self-similar collapse
when the nonlinearity becomes ``too strong''
(e.g., considering a power nonlinearity $|u|^{2 \sigma} u$
for a sufficiently large $\sigma$)~\cite{siettos,jon1,jon2}, or when 
the dimensionality of the problem increases for fixed nonlinearity
(e.g., considering
a radially symmetric problem where the 
Laplacian reads $\Delta u= u_{rr} + (d-1) u_r/r$ for
sufficiently large $d$~\cite{sulem,lemesurier,fibich2015,budd:1999,eva}).

There has been a series of classic experiments 
that address relevant collapse features in NLS-type
models, such as ones in nonlinear optics~\cite{moll} 
observing the famous Townes soliton and its collapse,
or more recent ones examining the collapse of structures
with topological charge~\cite{gaeta2}. 
Intriguingly, after numerous years, experiments
in other fields such as ultracold atomic systems
have also been catching up, and have enabled alternative
Bose-Einstein condensate (BEC) realizations of
Townes solitons in two separate recent experiments~\cite{dalibard,chenlung}. Indeed,
additional directions related to collapse have
also been recently explored. These include the collapse
of copropagating beams with different wavelengths
in so-called two-color systems~\cite{twocolor}, 
as well as the consideration of vortical beams quenched
from repulsive to attractive interactions in BECs
and thus accordingly led to collapse~\cite{banerjee2024collapse}.

It is some aspects of this very recent experimental
endeavor of~\cite{banerjee2024collapse} that
have motivated the present work. Indeed, in that
work it was found that the original ring of atomic
mass, prevented by the vortical structure from 
collapsing in the center of the condensate system,
broke into a necklace structure consisting of
individual (Townes-like) blobs which subsequently
collapsed in the periphery of the ring pattern.
This is also reminiscent of a scenario proposed
earlier in the context of azimuthal modulational
instability of a vortex solitary wave, e.g., 
in~\cite{CAPLAN20091399}. This scenario produces
multiple concurrently  collapsing structures, a
setting that has, indeed,  been theoretically
considered earlier in the work of~\cite{Nawa1998}.
Arguably, the simplest setup enabling the study of such
multi-bump states potentially leading to multiple
concurrent blowups arises in one spatial dimension, i.e.,
$d=1$.~Indeed, the earlier work of~\cite{budd:1999}  has
partially examined such states in the realm of
variable dimension $d$ (i.e., used as a bifurcation parameter), 
which is rather unphysical (albeit interesting from a bifurcation perspective).

Here, motivated from our earlier studies which have
considered the mathematically equivalent, yet more
tractable setting of a variable nonlinear
exponent $\sigma$, we revisit the examination of multi-pulse
self-similarly exploding solutions. Firstly, we provide
the mathematical setup (in Section~\ref{sec:theor_numer_setup})
that 
facilitates the 
numerical computation of these solutions, as a
systematic bifurcation problem from the critical
point of $\sigma d=2$.
Varying the nonlinear exponent $\sigma$ with fixed $d=1$, we construct
the full bifurcation diagrams of such multi-pulse
collapsing states, observing all of them emerging
from {\it infinity} at the respective bifurcation point. As
part of our analytical considerations, we derive the
asymptotic distance of the pulses as a function of the deviation
from the critical point. Moreover, following up on the analysis
of the stability of the single-bump collapsing branch that appeared
in our earlier works~\cite{jon1,jon2}, here
we provide a systematic spectral analysis of the
multi-bump branches, examining how the number
of unstable modes of the solutions grows as a function
of the blowup rate $G$. 
The summary of our theoretical and numerical results is provided in 
section III.~Section \ref{sec:steady} focuses on the details 
of the theoretical analysis for the steady problem;  the corresponding
stability analysis is considered in Section~\ref{sec:stab}.
Indeed, we have identified some
unexpected dependencies of the relevant eigenvalues 
and explain the systematics of how the relevant
power laws on $G$ emerge.~Both the eigenvalues and
the eigenvectors of the resulting instabilities 
are provided together with a number of direct numerical computations
performed on the half domain.~The latter showcase the 
destabilization of the relevant multi-humped solutions 
towards the single-humped one.~Finally, we summarize 
our findings and conclusions in Section~\ref{sec:conclusions},
and discuss exciting avenues of future research emanating from 
the present work. A number of technical details
concerning the numerical computations (involving, e.g.,
the comparison with alternative numerical approaches,
as well as the dynamics in the full domain), and the
analysis (such as the consideration of higher order
terms in the bifurcation diagram) are presented in suitable
Appendices.

\section{Theoretical and Numerical Setup}
\label{sec:theor_numer_setup}
Our starting point for the examination of the
multi-pulse collapse solutions will be the 
one-dimensional, nonlinear Schr\"odinger (NLS)
equation with a power-law nonlinearity~\cite{jon1,jon2} given by:
\begin{align}
\ii \pd{\psi}{z} + \spd{\psi}{x} + |\psi|^{2 \sigma} \psi = 0,
\label{eq:NLS}
\end{align}
where $\psi\coloneqq\psi(x,z)$ is the complex-valued wavefunction,
and $\sigma$, the strength of the (power-law) nonlinearity.~The 
Hamiltonian associated with the NLS 
reads:
\begin{align}
H = \int_{-\infty}^\infty \left(
\left| \pd{\psi}{x}\right|^2 -\frac{1}{\sigma+1} |\psi|^{2\sigma+2}
\right) \, \d x,
\label{eq:NLSHam}
\end{align}
and satisfies:
\[ \ii \pd{\psi}{z} = \frac{\delta H}{\delta \psi^\ast}, \qquad \ii 
 \pd{\psi^\ast}{z} = -\frac{\delta H}{\delta \psi},\]
where $\ast$ stands for complex conjugation.~It should be noted 
that $H$ is finite for the Cauchy problem of the NLS.

The study of self-similar blowup solutions to the NLS necessitates
the introduction of the so-called stretched variables:
\[  \xi \coloneqq \frac{x}{L}, \quad \tau \coloneqq \int_0^z
  \frac{\d z'}{L^2(z')}, \quad \psi(x,z) \coloneqq L^{-1/\sigma}
  v(\xi,\tau) \]
whose substitution to the NLS and Hamiltonian [cf.~Eqs.~\eqref{eq:NLS} and~\eqref{eq:NLSHam}] gives respectively
\begin{align}
  \ii \pd{v}{\tau} + \spd{v}{\xi} +  |v|^{2 \sigma} v
  - \ii \xi L L_z \pd{v}{\xi}- \frac{\ii L L_z}{\sigma} v= 0,
\label{eq:NLScoexplod}
\end{align}  
and
  \[ H = L^{-2/\sigma-2}\int_{-\infty}^\infty \left(
\left| \pd{v}{\xi}\right|^2 -\frac{1}{\sigma+1} |v|^{2\sigma+2}
\right) \, \d x.
\]
We will refer to Eq.~\eqref{eq:NLScoexplod} as the NLS in the co-exploding frame hereafter, and define $G\coloneqq -L\,L_{z}$ 
corresponding to the blowup rate.~This way, Eq.~\eqref{eq:NLScoexplod}
is conveniently written as
\begin{equation}
\label{eq:selfsimilarNLS}
i \frac{\partial v}{\partial \tau} + \frac{\partial^2 v}{\partial \xi^2} +
    |v|^{2 \sigma} v -v + i G \left(\xi \frac{\partial v}{ \partial \xi} + \frac{1}{\sigma} v \right) = 0,
\end{equation}
where we have factored out phase rotations via the transformation of 
the (complex-valued) field $v\mapsto v^{\ii \tau}$.~Self-similar blowup solutions to 
Eq.~\eqref{eq:selfsimilarNLS} are identified as \textit{stationary} 
solutions therein, that is, $v(\xi,\tau)\mapsto v(\xi)$ when
$\sigma>2$ corresponding to the supercritical case for the one-dimensional NLS.

\noindent Indeed, we are able to numerically identify self-similar solutions
by posing Eq.~\eqref{eq:selfsimilarNLS} on a computational
yet finite spatial domain $[-K,K]\ni \xi$ supplemented with homogeneous
Neumann boundary conditions, i.e., $v_{\xi}|_{\xi=\pm K}=0,\,\,\forall \tau$.~We
consider a uniform spatial discretization with resolution $d \xi$, and replace
the spatial derivatives with respect to $\xi$ in Eq.~\eqref{eq:selfsimilarNLS}
with a centered, fourth-order accurate finite difference scheme.
We further validated our numerical findings in this work by employing 
the finite element method as implemented in the open-source
software \texttt{FreeFem\,$++$}~\cite{freefem} (see also Sec.~\ref{app:num_comp} in the Appendix).

Then, the identification of stationary solutions, i.e., self-similar ones
to Eq.~\eqref{eq:selfsimilarNLS} involves the following twofold process.~At
first, we perform time-stepping using the BDF1 (i.e., backward Euler) scheme
in order the self-similar dynamics to approach a stationary solution,
i.e.,  $v(\xi,\tau)\mapsto v(\xi)$.~We note that we corroborated our
simulations here by using \textsc{MATLAB}'s \texttt{ode23t} ODE solver
(employing the trapezoidal method).~Since the blowup rate $G$ is an extra unknown to the problem, we close
the system by imposing a pointwise, i.e., pinning condition of the form: 
$\mathrm{Im}\left[v(\xi=0,\tau)\right]=C\equiv\mathrm{const.}$ stemming from:
\begin{align}
\label{eq:pointwise_cond}
 \int_{-K}^{K} \textrm{Im} \left[ v (\xi,\tau) \right] T (\xi)\textrm{d}\xi = C,
\end{align}
where $T(\xi)$ is the so-called template function~\cite{rowley_2003}.~For the pointwise condition given by Eq.~\eqref{eq:pointwise_cond}, we choose 
$T(\xi)=\delta(\xi)$, and in most of our numerical computations, we set $C=0$, thus seeking for a self-similar profile $v$ satisfying:
$\mathrm{Im}[v(0, \tau)]=0$.~We would like to note at this point that 
our numerical experimentation
has suggested that pointwise conditions may be a bit restrictive (especially during parametric continuations), and thus as an alternative, we have also employed alternative phase conditions
obtaining the same results in terms of the amplitude $|v|$ (and density $|v|^{2}$) of the solution although the real and imaginary parts of $v$ may be different.~This is not surprising as each such approach  respectively ``pins down" a solution out of the group orbit of self-similar transformations.

Once the self-similar dynamics approaches  a stationary profile $v(\xi)$ (and fixed blowup rate $G$), the second step ``corrects" the solution by seeking to identify it as a numerically exact solution using Newton's method.~The waveform obtained from the self-similar dynamics is fed as an initial guess to the Newton solver, and it converges typically, within 2-3 iterations with very high accuracy.~The resulting
numerically exact solution (up to a prescribed accuracy) enables us subsequently to perform its spectral stability analysis in the co-exploding frame.~For our stability computations and the results shown next, we use \textsc{MATLAB}'s \texttt{eigs} (sparse matrix) eigenvalue solver.%
~Our stability results are corroborated using also the FreeFEM's eigenvalue solver \cite{freefem} (see also Sec.~\ref{app:num_comp} in the Appendix).%
~The spectral stability results we obtained using both eigensolvers match precisely with each other.
The above described process works very well
for the fundamental (single hump) self-similar
solution of Eq.~(\ref{eq:selfsimilarNLS}).

Our primary focus, however, in this work is on multi-humped, self-similar solutions of the NLS for $\sigma>2$.
Since these solutions are dynamically unstable, an alternative approach is required, as self-similar dynamics will not 
(typically) converge to a stable stationary profile.~To identify suitable profiles that can serve as initial guesses for Newton's method in solving the stationary two-point boundary value problem, we adopt the approach (known as zero-Hamiltonian solutions method) introduced in~\cite{budd:1999} (see, also~\cite{fibich2015}).~In brief, we consider the radially symmetric case (for $d=1$ in~\cite{budd:1999}) that renders Eq.~\eqref{eq:selfsimilarNLS} now defined on the half-line, i.e., $[0,\infty)\ni\xi$ with $v_{+}$ being the symmetric solution.~For computational purposes, we truncate the semi-infinite domain to a finite one $[0,K]$ with $K$ typically being $200$ to $1000$.~The stationary problem that we aim to solve is:
\begin{equation}\label{eq:bvp}
    v''_{+} - v_{+} +i G \left(\frac{1}{\sigma}v_+ + \xi v_{+}' \right)+|v_{+}|^{2\sigma}v_{+} = 0,
\end{equation}
supplemented with boundary conditions: 
\begin{equation}\label{eq:bvp_bc}
 v'_{+}(0)=0, \quad \mathrm{Im}[v_{+}(0)]=0, \quad \Big|K v'_{+}(K)+\left(1 + i/G\right) v_{+}(K)\Big| = 0,
\end{equation}
where primes stand for (total) differentiation with respect to $\xi$.

\begin{figure}[!pt]
\centering
\includegraphics[width=0.65\textwidth]{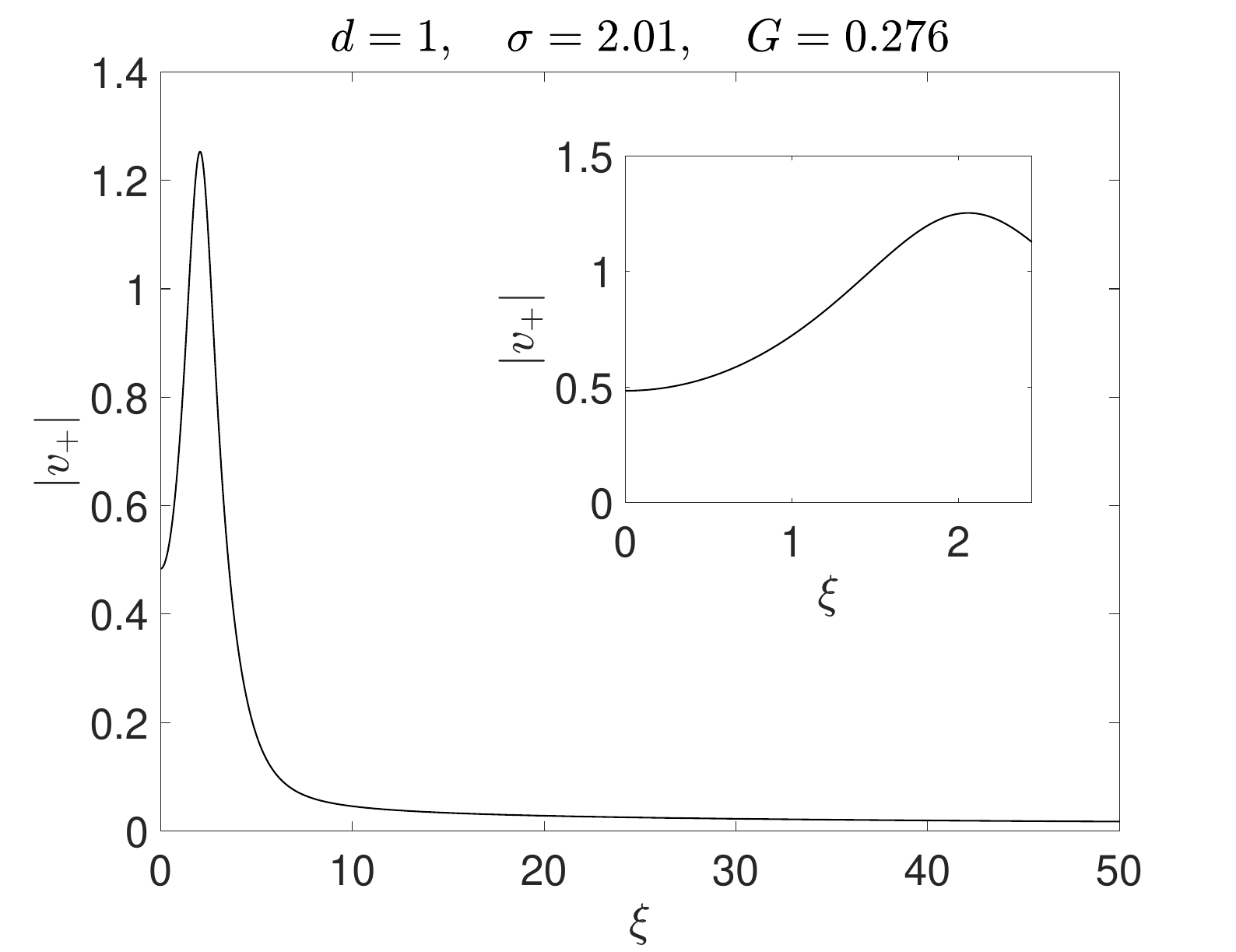}
\caption{The amplitude $|v_+|$ of a double-humped solution to the boundary value problem of Eqs.~\eqref{eq:bvp}-\eqref{eq:bvp_bc} for $\sigma=2.01$ and $K=200$.~The blowup rate for this solution is $G \approx 0.276$, and the bump, i.e., the highest in-amplitude peak appears at $\xi \approx 2.05$.~To ease visualization, the amplitude $|v_+|$ is shown on $[0,50]$.%
The inset displays a zoom of $|v_+|$ near the origin confirming the vanishing derivative of the amplitude.
\label{fig:initial}
}
\end{figure}

The above boundary value problem is solved by a combination of a shooting method and a (gradient-free) minimization technique that ultimately determines the unknown parameters $(\mathrm{Re}[v_{+}(0)],G)$, and thus the profile of interest.~Indeed, starting from an initial guess $(\mathrm{Re}[v_{+}^{(0)}(0)],G^{(0)})$, the latter 
is treated as an initial condition to Eq.~\eqref{eq:bvp} which itself is viewed as an initial-value problem.~Upon performing a forward integration in $\xi$ up to $\xi=K$, we treat the boundary condition at $\xi=K$ in Eq.~\eqref{eq:bvp_bc} as our objective function we want to minimize.~For our purposes, we used \textsc{MATLAB}'s \texttt{fminsearch} function (combined with the \texttt{ode23t} integrator), and corroborated our results using \textsc{NAG}'s \texttt{e04jyf} minimization routine (and \texttt{dop853} integrator).~Upon convergence, we determine $(\mathrm{Re}[v_{+}(0)],G)$ and the profile $v_{+}(\xi)$ of interest.~Then, the solution $v(\xi)$ on $[-K,K]$ is constructed via symmetrization:
\begin{equation} 
    v(\xi) = \Bigg\{\begin{array}{lr}
        v_{+}(\xi),  & \text{for } 0\leq \xi\leq K\\
        v_{+}(-\xi), & \,\,\,\,\,\text{for\!\!}\,\,\, -\!\!K\leq \xi\le 0
        \end{array}
\end{equation}
and fed to our Newton solver in order to obtain the numerically exact solution.~We 
performed a systematic scan of different initial guesses $(\mathrm{Re}[v_{+}^{(0)}(0)],G^{(0)})$
taken from the $(\mathrm{Re}[v_{+}(0)],G)$-plane, that gave us a plethora of multi-bump solutions to the NLS.~Then, those were continued over $\sigma$ parametrically by using pseudo-arclength continuation~\cite{kuznetsov_book},
thus obtaining branches of self-similar, multi-humped solutions, and constructing the respective bifurcation diagrams, i.e., $G$ vs $\sigma$.%
~For illustration purposes, Figure~\ref{fig:initial} displays an example of a double-humped solution to the boundary-value problem of Eqs.~\eqref{eq:bvp}-\eqref{eq:bvp_bc} obtained on $[0,200]$, where 
its amplitude $|v_+|$ is depicted as a function of $\xi$.~The localized bump is located around $\xi \approx 2.05$.
For this particular solution, the determined parameters are: $\textrm{Re}[v_+(0)] = 0.4836 $ and $G =  0.275549$.

\section{Multihumped 1d self-similar solutions: Principal Existence and Stability Results}
\label{sec:numerics}

\subsection{Existence Results}

\begin{figure}[!pt]  
\centering
\includegraphics[width=0.8\textwidth]{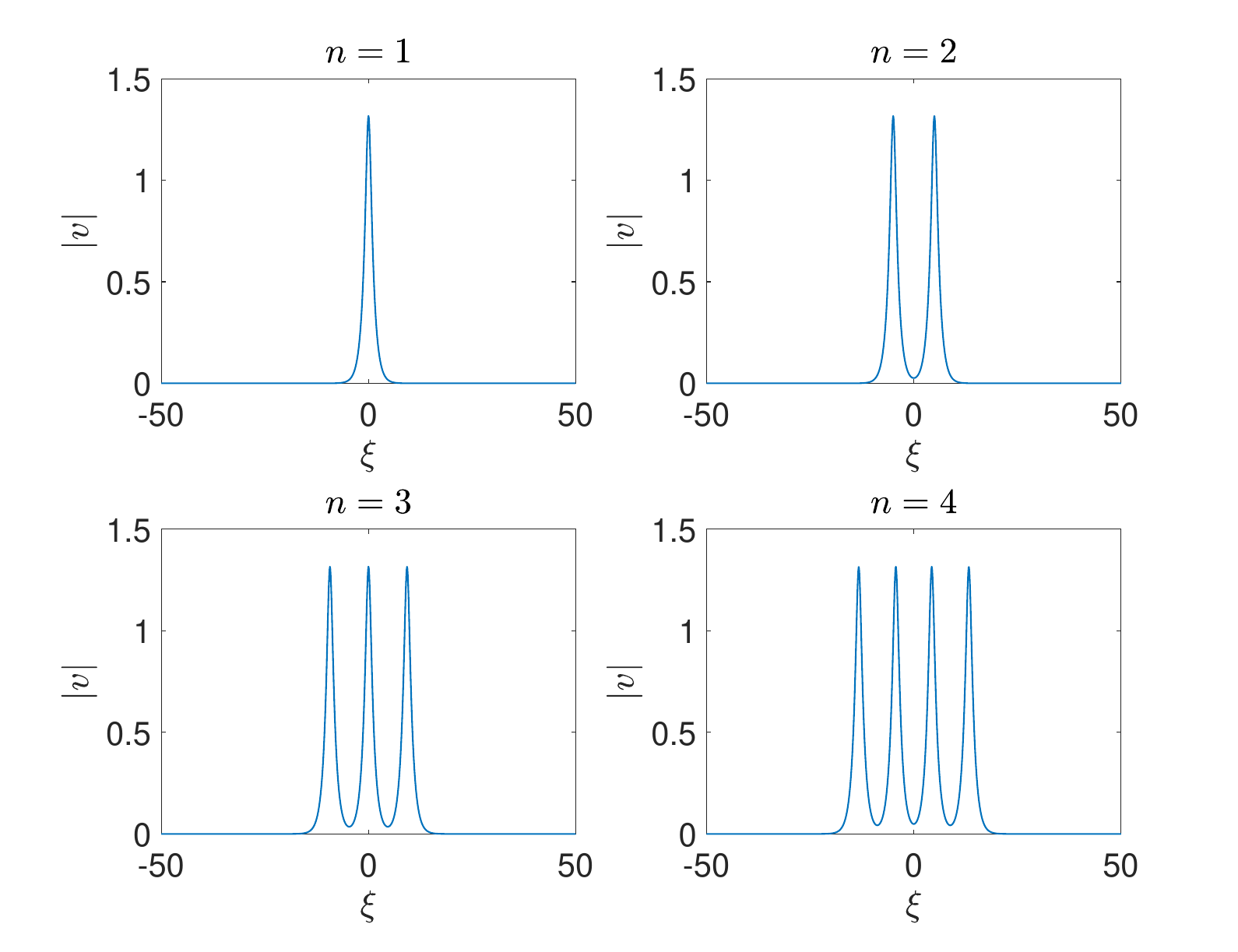}
\caption{Comparison of self-similar blowup solutions with $n=1,2,3$ and 4 humps for $G=0.01$ and $K=50$.
}
\label{fig:profiles}
\end{figure}

\begin{figure}[!pt]
\centering
\begin{tabular}{cc}
\includegraphics[width=0.49\textwidth]{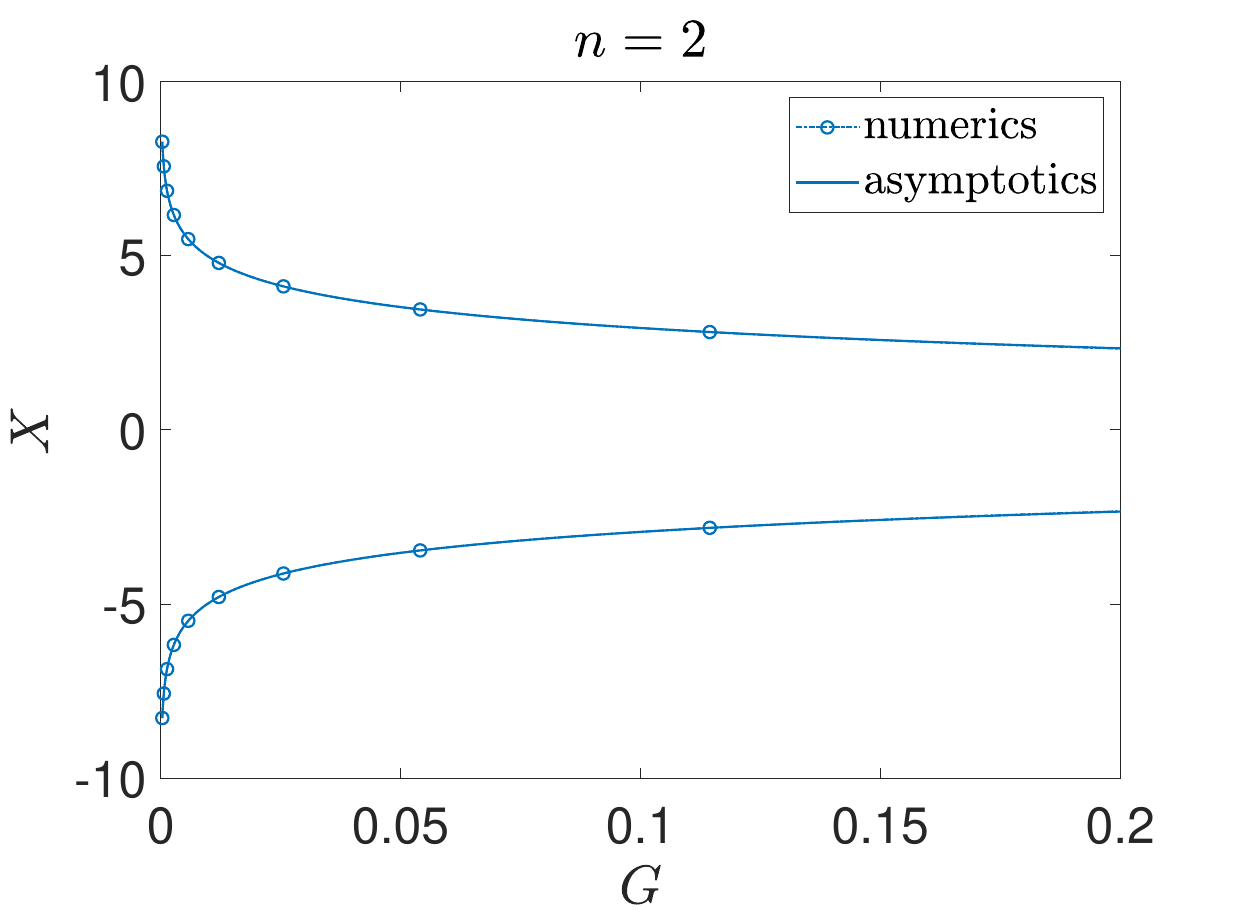}&\includegraphics[width=0.49\textwidth]{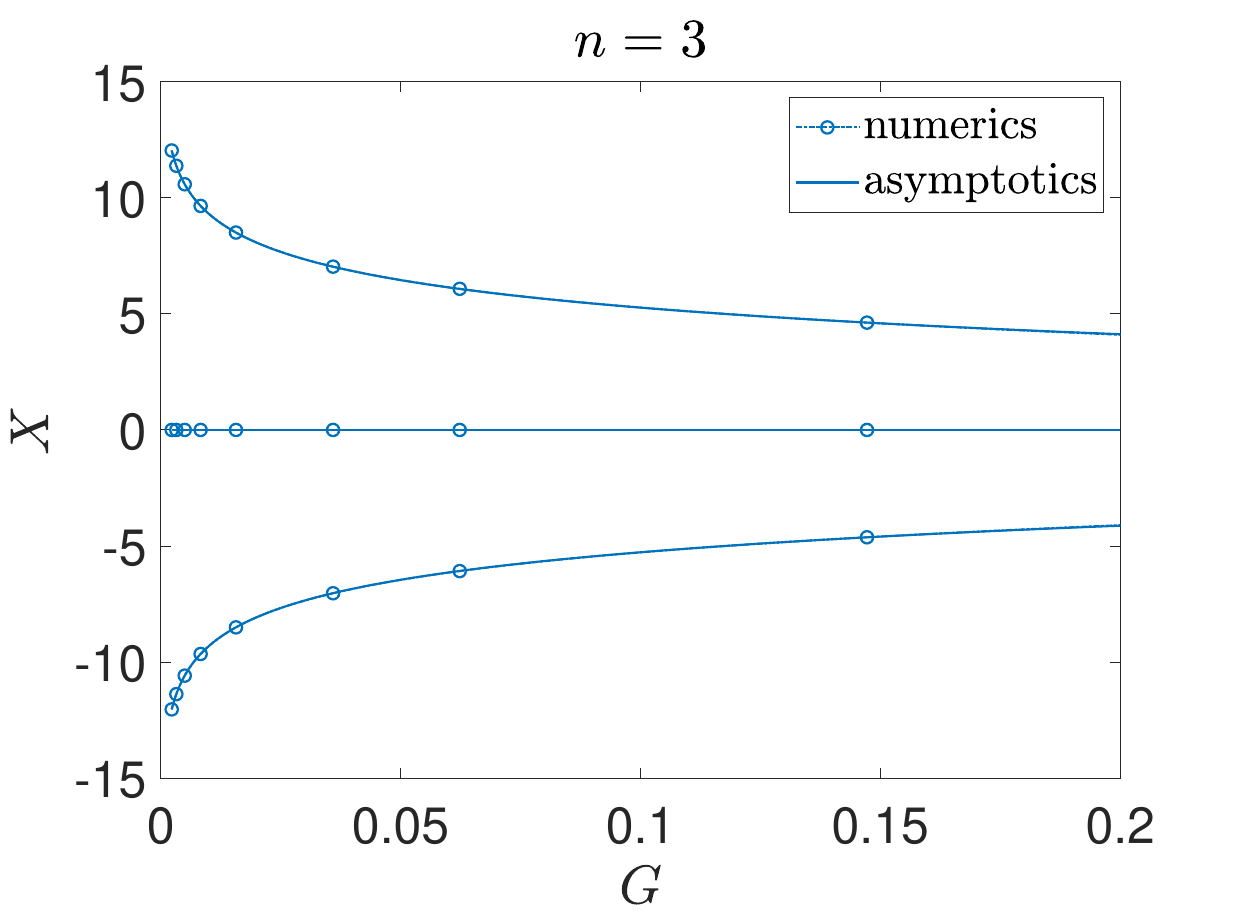}
  \\
(a)&(b)\\
\end{tabular}
\begin{tabular}{c}     
\includegraphics[width=0.49\textwidth]{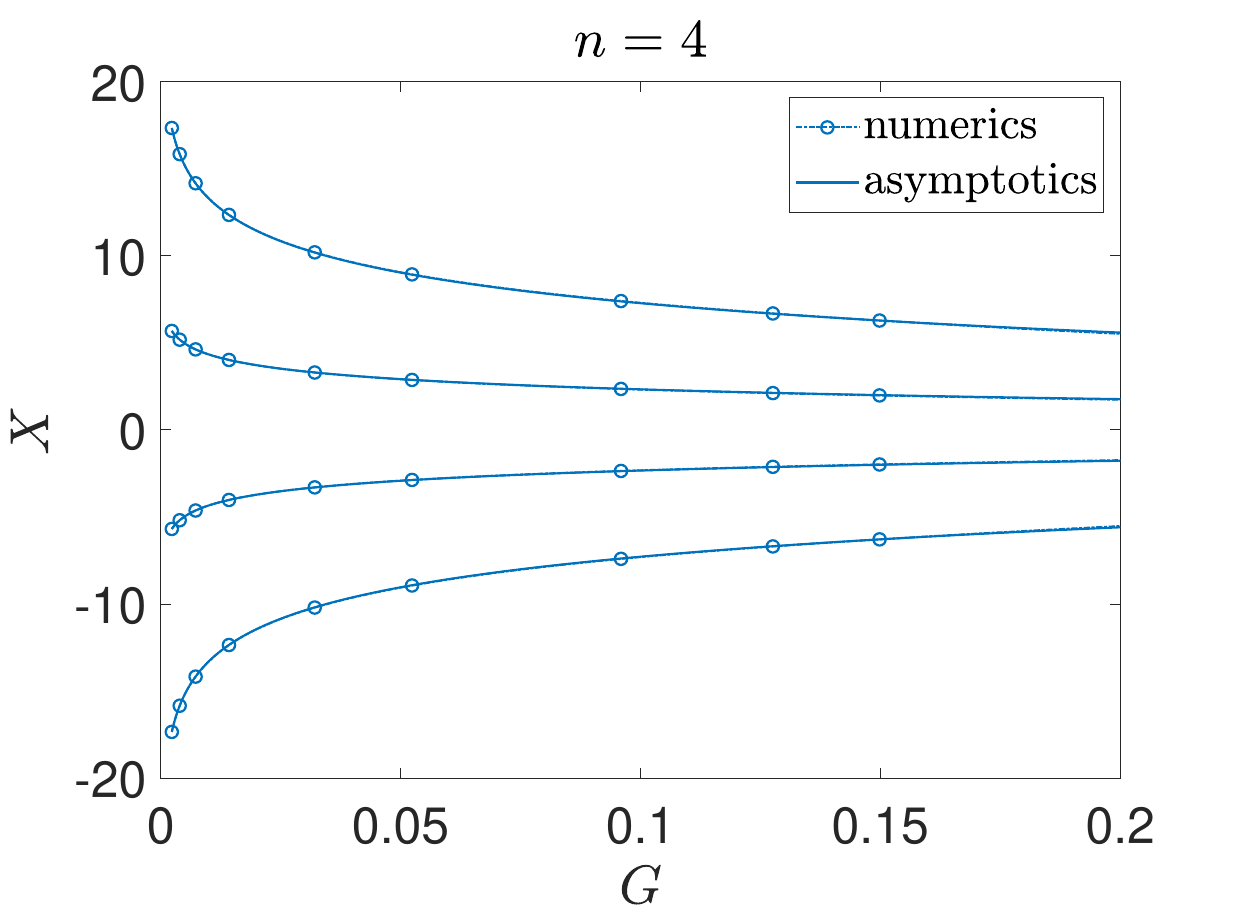}
  \\
(c)
\end{tabular}
\caption{Variation of hump positions, $X$, with blowup rate, $G$ for two, three and four-humped self-similar solutions. 
The  dash-dotted lines with open circles lines correspond to numerical solutions with $K=50$. 
The full lines represent the asymptotic prediction of hump positions.
} 
\label{fig:humpposition}
\end{figure}

Our prototypical results in terms of the existence
of multi-pulse collapse states are presented 
in Figures~\ref{fig:profiles} and~\ref{fig:humpposition}.
More specifically,
Figure~\ref{fig:profiles} presents the spatial profiles 
of self-similar solutions with $n=1$, $2$, $3$ and $4$ humps, and blowup rate $G=0.01$ ($\sigma>2$).
Indeed, we have identified even higher order
states (not shown here), always coming in two sets
of families (see also the relevant discussion of~\cite{budd:1999}), namely ones with odd numbers
of peaks (featuring a peak at the center of symmetry
of the configuration), as well as ones with an
even number of peaks, {\em which are symmetric around
a local minimum of the profile density}.

The variation of hump positions with blowup rate, $G$ is displayed in Fig.~\ref{fig:humpposition}. This is a 
point of contact and indeed of excellent
agreement with our theoretical analysis that we now
summarize (the analysis is presented in full detail
in Section \ref{sec:steady}). 
%
For the case of $n=2$ equations, the theoretical
prediction for the equilibrium positions of the two-pulse
configurations is given (symmetrically around $0$) by
the $G$-dependent expressions:
\[
X_1  =  -\frac{32}{\pi G^2}\ee^{-X_{2}+X_1},\qquad
X_2  =  \frac{32}{\pi G^2}  \ee^{-X_2+X_{1}},
\]
so that $X_2=-X_1=X$, say, with
\[ X = \frac{32}{\pi G^2}  \ee^{-2X}.\]
The case of 
$n=3$  equations leads our general expression detailed
in section \ref{sec:steady} to result in the predictions of:
\[
X_1  =  -\frac{32}{\pi G^2}\ee^{-X_{2}+X_1},\qquad
X_2  =  \frac{32}{\pi G^2}\left(  \ee^{-X_2+X_{1}}-  \ee^{-X_3+X_{2}}\right),\qquad
X_3  =  \frac{32}{\pi G^2}  \ee^{-X_3+X_{2}},
\]
so that  $X_3=-X_1=X$, say, $X_2=0$, with
\[ X = \frac{32}{\pi G^2}  \ee^{-X}.\]

Finally, among the cases shown in Fig.~\ref{fig:humpposition}, for $n=4$, our theoretical predictions yield:
\begin{align*}
X_1  &=  -\frac{32}{\pi G^2}\ee^{-X_{2}+X_1},&
X_2  &=  \frac{32}{\pi G^2}\left(  \ee^{-X_2+X_{1}}-  \ee^{-X_3+X_{2}}\right),\\
X_3  &=  \frac{32}{\pi G^2} \left(  \ee^{-X_3+X_{2}}-  \ee^{-X_4+X_{3}}\right),&
X_4  &=  \frac{32}{\pi G^2}  \ee^{-X_4+X_{3}},
\end{align*}
so that, with  $X_3=-X_2=Y_1$, $X_4=-X_1=Y_2$, say, with
\[
Y_1  =  \frac{32}{\pi G^2} \left(  \ee^{-2Y_1}-  \ee^{-Y_2+Y_1}\right),\qquad
Y_2 =  \frac{32}{\pi G^2}  \ee^{-Y_2+Y_1}.
\]

We can observe that Fig.~\ref{fig:humpposition} 
presents {\it excellent agreement} with the observed numerical
results throughout the interval of consideration for the
values of $G$. Similar predictions can be made
for arbitrary values of $n$, as will be seen in 
Section~\ref{sec:steady}.

\begin{figure}[h!]
\centering
\includegraphics[width=0.75\textwidth]{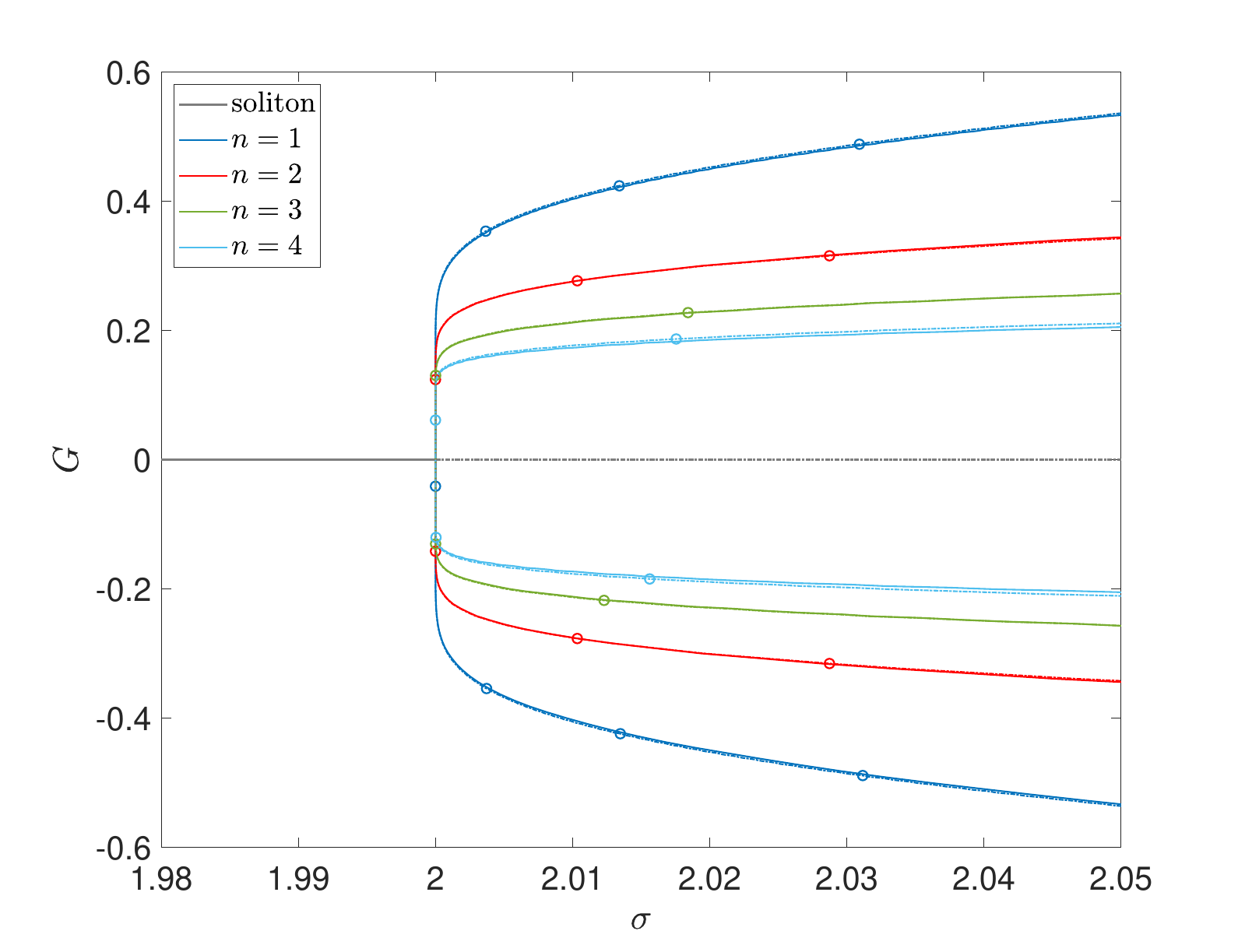}
\caption{Variation of the blowup rate $G$ as a function of $\sigma$ for domain size, $K=50$.
The gray lines represent the blowup rate of soliton solutions, and the blue line displays the blowup rate of $n=1$-humped self-similar solutions.
The blowup rates of self-similar solutions with $n=2,3,4$ humps are depicted with red, green, and cyan lines, respectively.
The dot-dashed lines with open circles represent the numerical computations, and the solid lines show the asymptotic prediction for the blowup rate $G$.
} \label{fig:bifdiagram}

\end{figure}

Moreover, having the positions of the pulses, we 
also derive in section \ref{sec:steady} a systematic expression that
provides the corresponding branch's bifurcation diagram,
i.e., connects the blowup rate $G$ with the nonlinearity
power $\sigma$ in line with our earlier calculation in~\cite{jon1}. More specifically:
\beq
 \sigma-2= \frac{8\sigma}{n\pi}
(\ee^{2X_n}+\ee^{-2X_1}) \frac{\ee^{-\pi/G}}{G}. \label{sigeqn1}
\eeq
 Note that with $n=1$ and $X_1=X_n=0$ this reduces to the formula in \cite{jon1}, and also that we anticipate the solutions of (\ref{sigeqn1}) to satisfy $X_1 = -X_n$ so that 
\beq
 \sigma-2= \frac{16\sigma}{n\pi}
\frac{\ee^{2X_n-\pi/G}}{G}. \label{sigeqn2}
\eeq
Figure~\ref{fig:bifdiagram} illustrates the dependence of $G$ with the parameter $\sigma$ for self-similar solutions of different numbers of humps, as obtained through our numerical
computations.
Among all self-similar profiles, only the single-humped (n=1) solution
with $G>0$ is found to be stable (as will be 
discussed below). However, importantly
for all the branches considered, we find
again excellent agreement between the theoretical
bifurcation curve prediction and the numerically obtained
one. Only in the case of $n=4$ can we detect a nearly
imperceptible discrepancy at the level of the figure (at least with the
eyeball metric)  between theory and computation.
While this discrepancy is modestly amplified as $n$ becomes
higher, it is clear that all the relevant branches
are accurately captured by our asymptotics. Moreover,
and quite importantly, this bifurcation analysis definitively
shows that all of the relevant solutions emerge {\it concurrently} at the blowup threshold of $\sigma d=2$.
Indeed, as our position diagrams clearly indicate,
the relevant waveforms constitute an example of
``bifurcation from infinity'', with the relevant pulses
being ``summoned back'' (from an infinite)  to a finite distance from each other 
as the blowup rate $G$ increases.

\subsection{Spectral Properties}

We now turn to the examination of the stability features
of each one of the branches that are identified herein.
Before we do so, it is instructive to remind ourselves
of the findings for the single-hump $n=1$ branch
as obtained earlier in~\cite{jon2}. In the relevant
case, as the destabilization of the relevant solitonic
branch proceeds (for the $G=0$ solution), we have identified
$6$ point spectrum eigenvalues, i.e., $3$-pairs near the
origin. When we move to the co-exploding frame
to obtain the self-similarly collapsing solution as
a stationary state, it was argued in~\cite{jon2}
that two exact eigenvalues can be computed
($\lambda=2 G$ and $\lambda=G$) associated with the
symmetries of rescaling and of translation, respectively, in the
case of the infinite domain.
One more eigenvalue was found to stay at $\lambda=0$, while
3 negative real 
eigenvalues slightly deviating from the opposites of
the above values (as the system is no longer Hamiltonian)
were also identified. 

To consider the stability problem in further mathematical 
detail, we
first perform a transformation:
$v(\xi,\tau) = \V(\xi,\tau) \ee^{\ii \tau-\ii G(\tau) \xi^2/4}$
to turn our PDE problem to:
\[ 
  \ii \pd{\V}{\tau} + \frac{G' \xi^2}{4} \V+ \spd{\V}{\xi}+ |\V|^{2 \sigma}\V - \V 
- \frac{\ii (\sigma-2) G}{2 \sigma} \V
+ \frac{G^2 \xi^2 }{4}\V= 0,\]
where $G' = \fdd{G}{\tau}$.

Then, the steady state solutions satisfy:
\beq
 \sdd{\Vs}{\xi}+ |\Vs|^{2 \sigma}\Vs - \Vs 
- \frac{\ii (\sigma-2) G}{2 \sigma} \Vs
+ \frac{G^2 \xi^2 }{4}\Vs= 0,\label{sseqn}
\eeq
with $G$ constant equal to $G(\sigma)$.
We linearize about the steady state by writing
\begin{eqnarray}
\V(\xi,\tau) = \Vs(\xi) + \eps\left( f(\xi) \ee^{\la \tau} + g^*(\xi) \ee^{\la^* \tau} \right)
\label{linear14}
\end{eqnarray} 
and accordingly obtain the eigenvalue problem
\beqa
\ii\la f+\sdd{ f}{\xi} +
\sigma |\Vs|^{2\sigma-2}\Vs^{2}g + ( \sigma+1)|\Vs|^{2 \sigma}  f 
- f
- \frac{\ii (\sigma-2) G}{2 \sigma} f
+ \frac{G^2 \xi^2 }{4} f &=& 0,\label{geqn}\\
-\ii  \la g
+ \sdd{g}{\xi} 
+  \sigma |\Vs|^{2\sigma-2}(\Vs^*)^{2} f 
+ ( \sigma+1)|\Vs|^{2 \sigma} g 
-  g 
+ \frac{\ii (\sigma-2) G}{2 \sigma}  g 
+ \frac{G^2 \xi^2 }{4} g  &=& 0.\label{feqn}
\eeqa

Then, the two exact eigenvalues and their corresponding
eigenvectors ---for the infinite domain case--- 
can be explicitly written as:
$\la = 2G$,
\beq
f  =  \ii \Vs + G \left(\frac{\Vs}{\sigma} + \xi \fdd{\Vs}{\xi} - \frac{\ii G \xi^2 \Vs}{2}\right),\quad
g  =  -\ii \Vs^*+ G \left(\frac{\Vs^*}{\sigma} + \xi \fdd{\Vs^*}{\xi} + \frac{\ii G \xi^2 \Vs^*}{2}\right).\label{FG2G}
\eeq
and $\la = G$,
\beq
f  =  \fdd{\Vs}{\xi} - \frac{\ii G \xi \Vs}{2},\qquad
g  =  \fdd{\Vs^*}{\xi} + \frac{\ii G \xi \Vs^*}{2}.\label{FGG}
\eeq

\begin{figure}[h!]
\centering
\begin{tabular}{cc}
\includegraphics[width=0.49\textwidth]{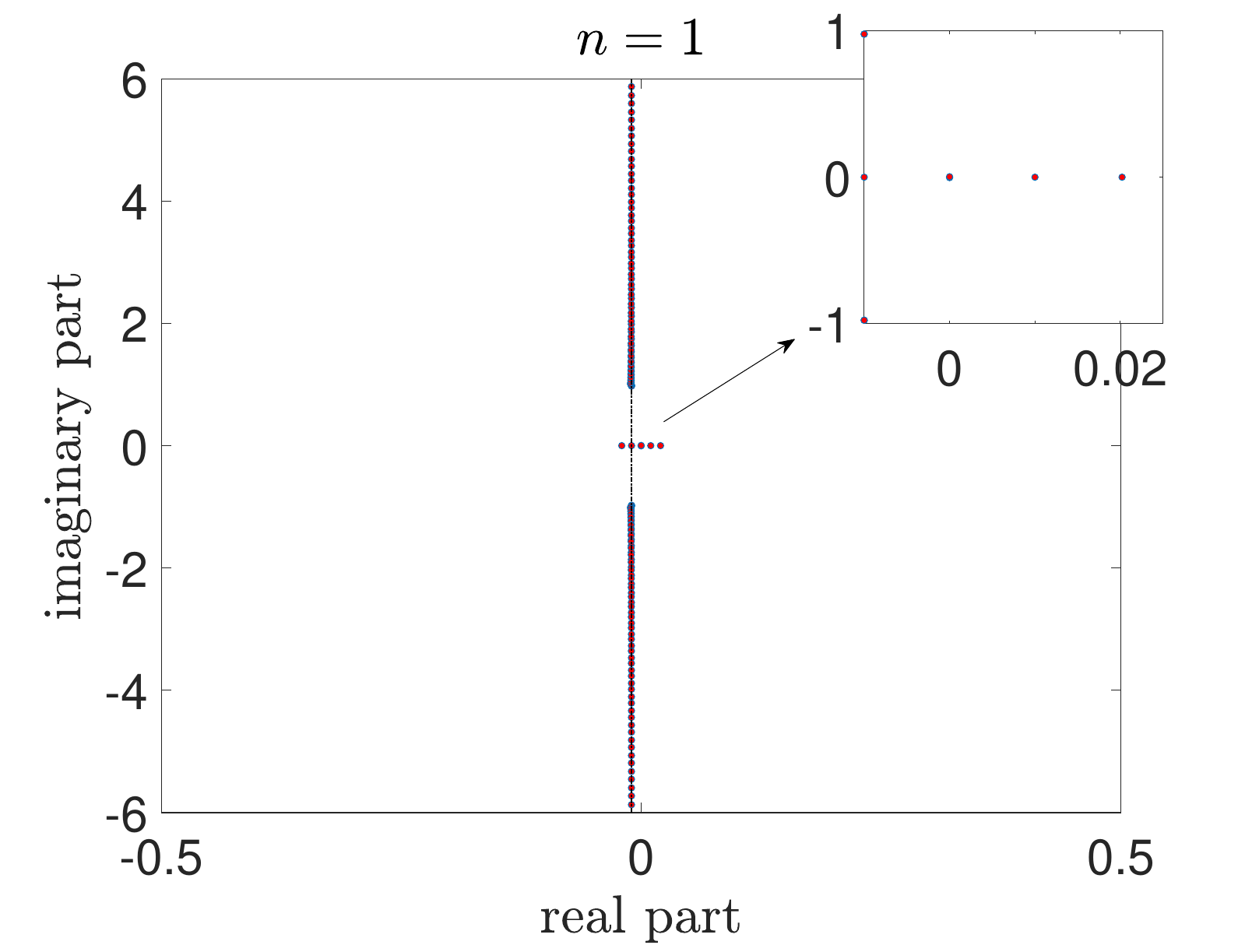}&\includegraphics[width=0.49\textwidth]{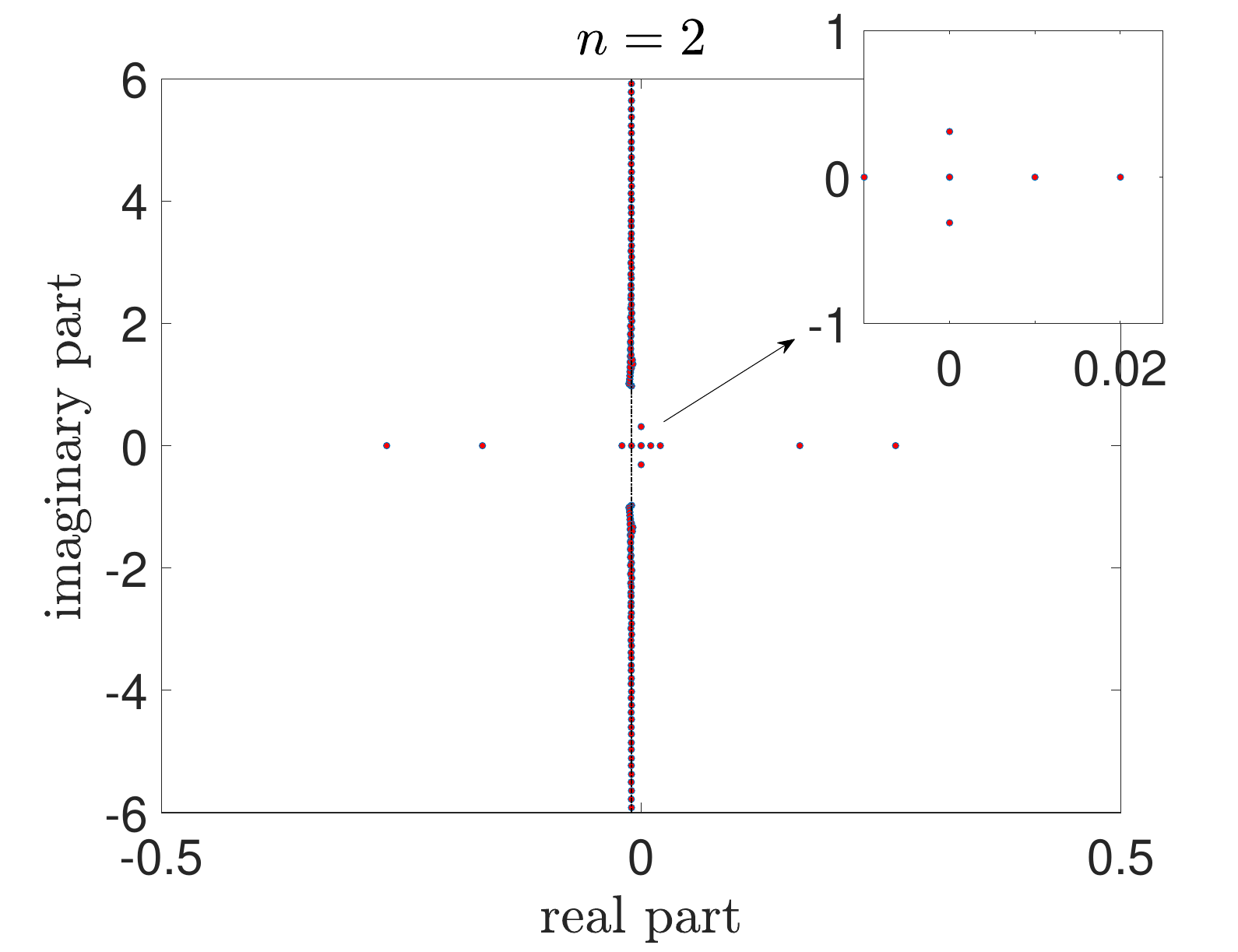}
  \\
(a)&(b)\\
\includegraphics[width=0.49\textwidth]{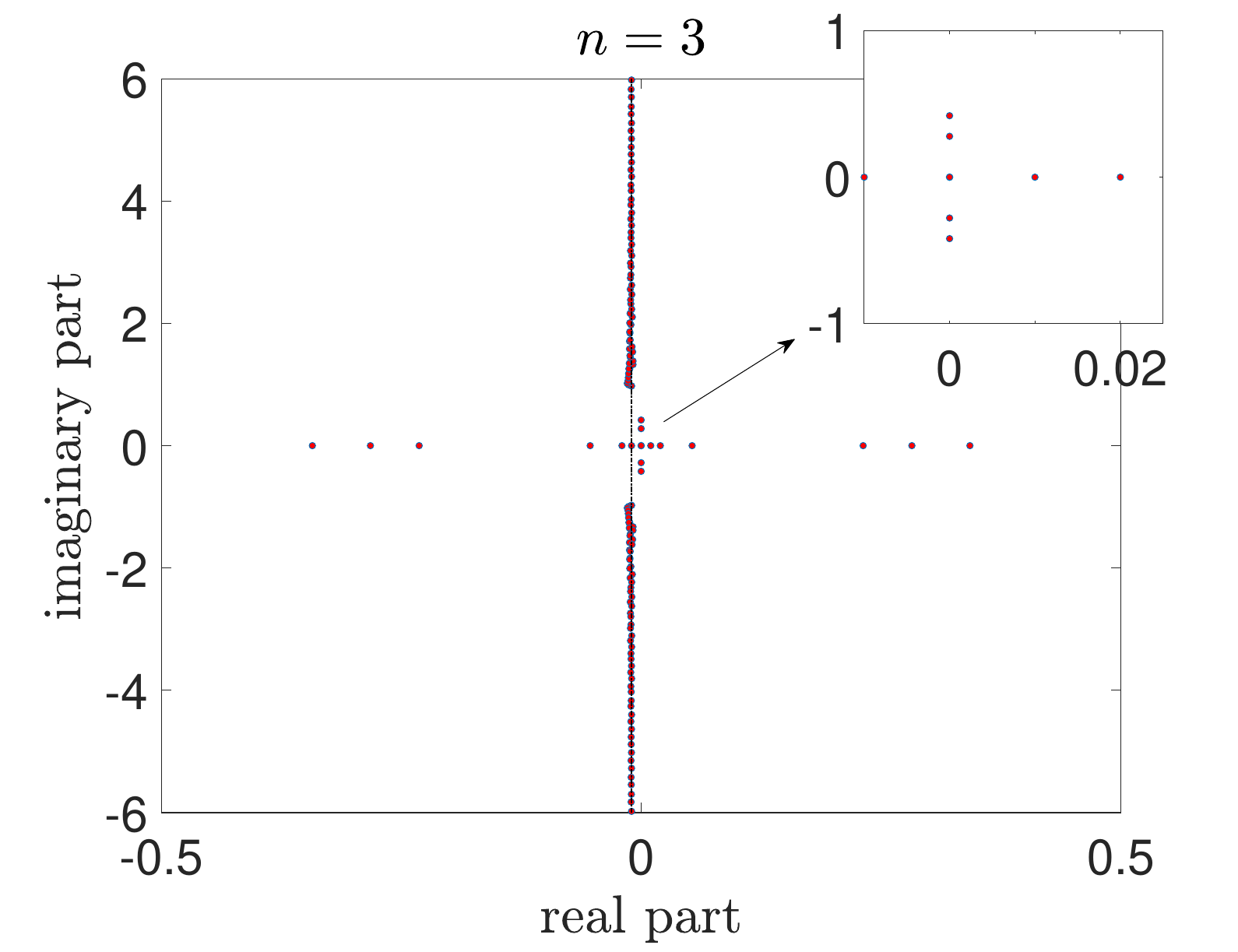}&\includegraphics[width=0.49\textwidth]{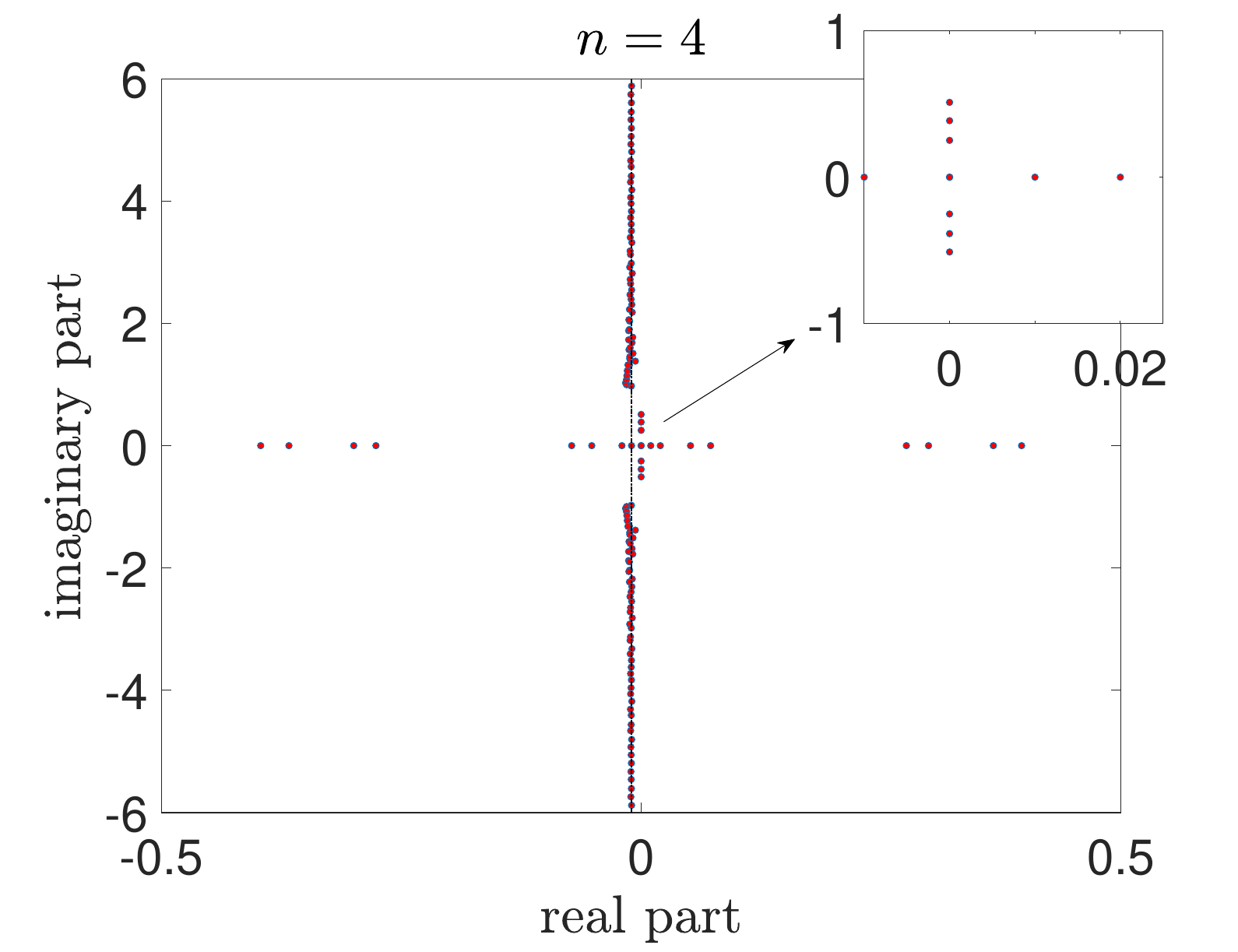}
  \\
(c)&(d)
\end{tabular}
\caption{Single ($n=1$) and multi-humped ($n=2,3,4$) self-similar solution spectra for $G=0.01$. 
The vertical part of the spectra align at $-G=-0.01$ for all cases (black dash-dotted lines).
The insets illustrate real eigenvalues $\lambda \approx G$, $\lambda \approx 2G$, and eigenvalues with zero real part. 
One can observe the existence of a single conjugate pair of eigenvalues with zero real part for self-similar solutions with $n=2$ humps, two pairs for $n=3$, and three pairs for $n=4$. 
For single-humped self-similar solutions there is only an eigenvalue at the origin. 
} \label{fig:spectra_multi}

\end{figure}

What we find in the case of the multi-bump collapse solutions
is that the number of eigenvalues is effectively 
doubled when we consider two-pulse solutions instead of one.
I.e., in the case of $n=2$ we need to account for 12 point
spectrum eigenvalues instead of $6$. In the case of
$3$-pulse solutions this number rises to $18$ and
for $n=4$ to $24$. Remarkably (for us!) we have been able
to fully categorize the eigenvalues of each one of
these scenarios. Indeed, what we find in each
case of $n$-pulse states is the following:
the configuration ends up possessing {\it exactly}
$n-1$ real eigenvalues that, to leading order, depend on the 
blowup rate in a predominantly square root form, i.e.,
$\lambda \propto G^{1/2}$. There are similarly $n+1$ real
eigenvalues that end up with a dependence $\lambda \propto G$.
Then, there exist $n-1$ pairs that end up being purely
imaginary for an $n$-pulse state. Finally, one pair
ends up near the origin. Given that, for each of the real
eigenvalues, nearly the opposite $\lambda$ is also
(very close) to an eigenvalue, this accounts for
$(n-1)+(n+1)+(n-1)+1$ i.e., $3 n$ (approximate) pairs, i.e.,
$6n$ eigenvalues in total, which is what we observe
for solutions consisting of $n$-copies of the
original pulse.~The spectra of the multi-humped self-similar
solutions are illustrated in Fig.~\ref{fig:spectra_multi} for $G=0.01$.
In all cases, one can observe that a vertical part of each spectrum aligns at $-G$.
In all cases, one can observe the existence of real positive eigenvalues, the number of which changes with the number of humps, $n$.
In particular, single-humped solutions have two dominant real
eigenvalues, $n=2$ solutions have four, $n=3$ have six, and $n=4$ have
eight dominant real eigenvalues. 
All  multi-humped self-similar solution have two real eigenvalues $\lambda_1 \approx G$, and $\lambda_2 \approx 2G$ [rather than equal
to $G$ or $2G$ as our computations are affected by the finiteness
of the computational domain].
The relevant count is fully in line with our
qualitative discussion above.

\begin{figure}[h!]
\centering
\includegraphics[width=0.99\textwidth]{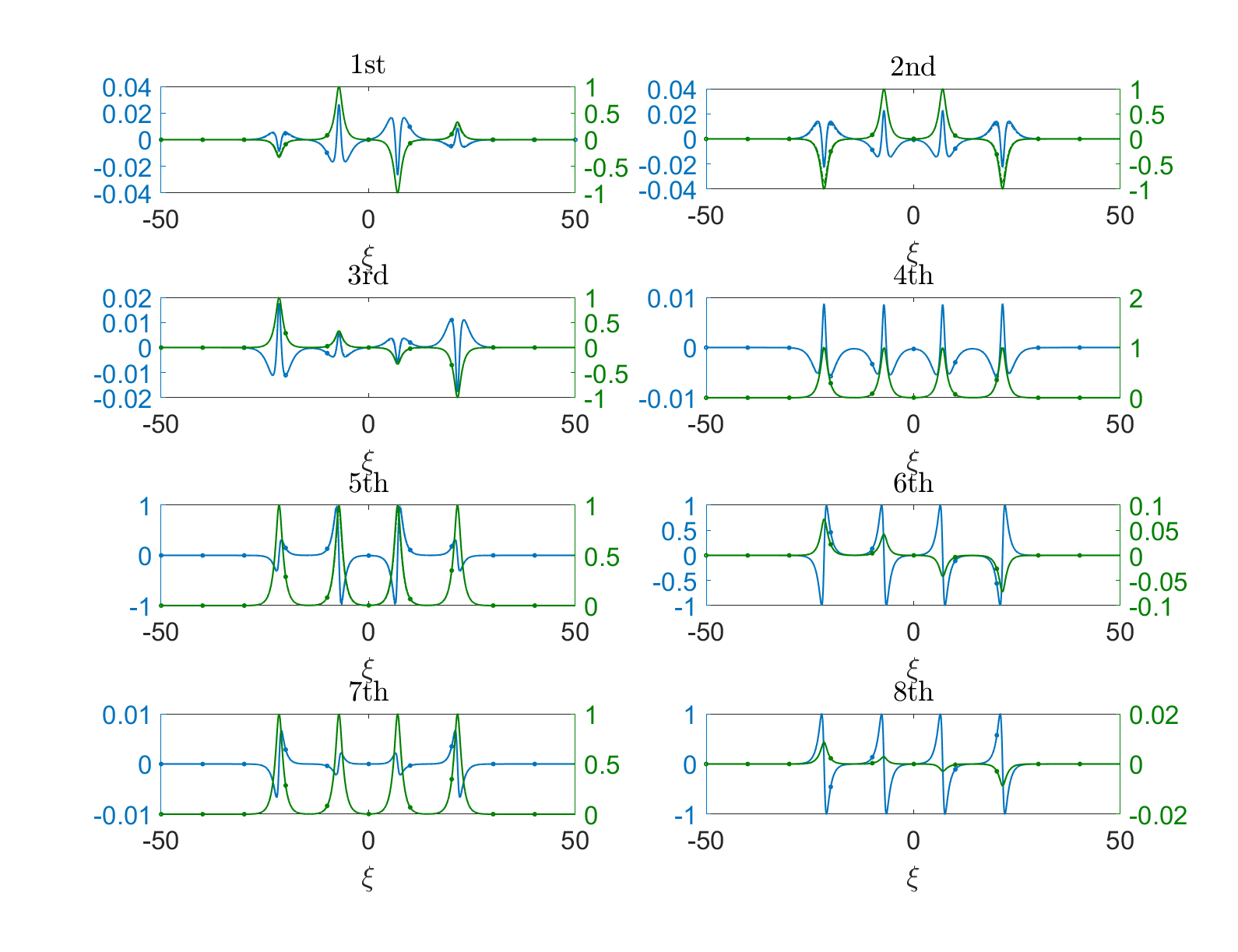}
\caption{Eigenfunctions for 4-humped solutions and $G=5\times10^{-4}$.
Blue depicts the real part and green corresponds to the imaginary parts. 
The numerical eigenfunctions are displayed with dash-dotted lines with
open circles and asymptotic predictions (from \S\ref{sec:stab},
truncated at $O(G)$) are displayed with solid
lines.
The two are practically coincident with each other.
  } \label{fig:eigenfunctions}

\end{figure}

\begin{figure}[h!]
\centering
\begin{tabular}{cc}
\includegraphics[width=0.49\textwidth]{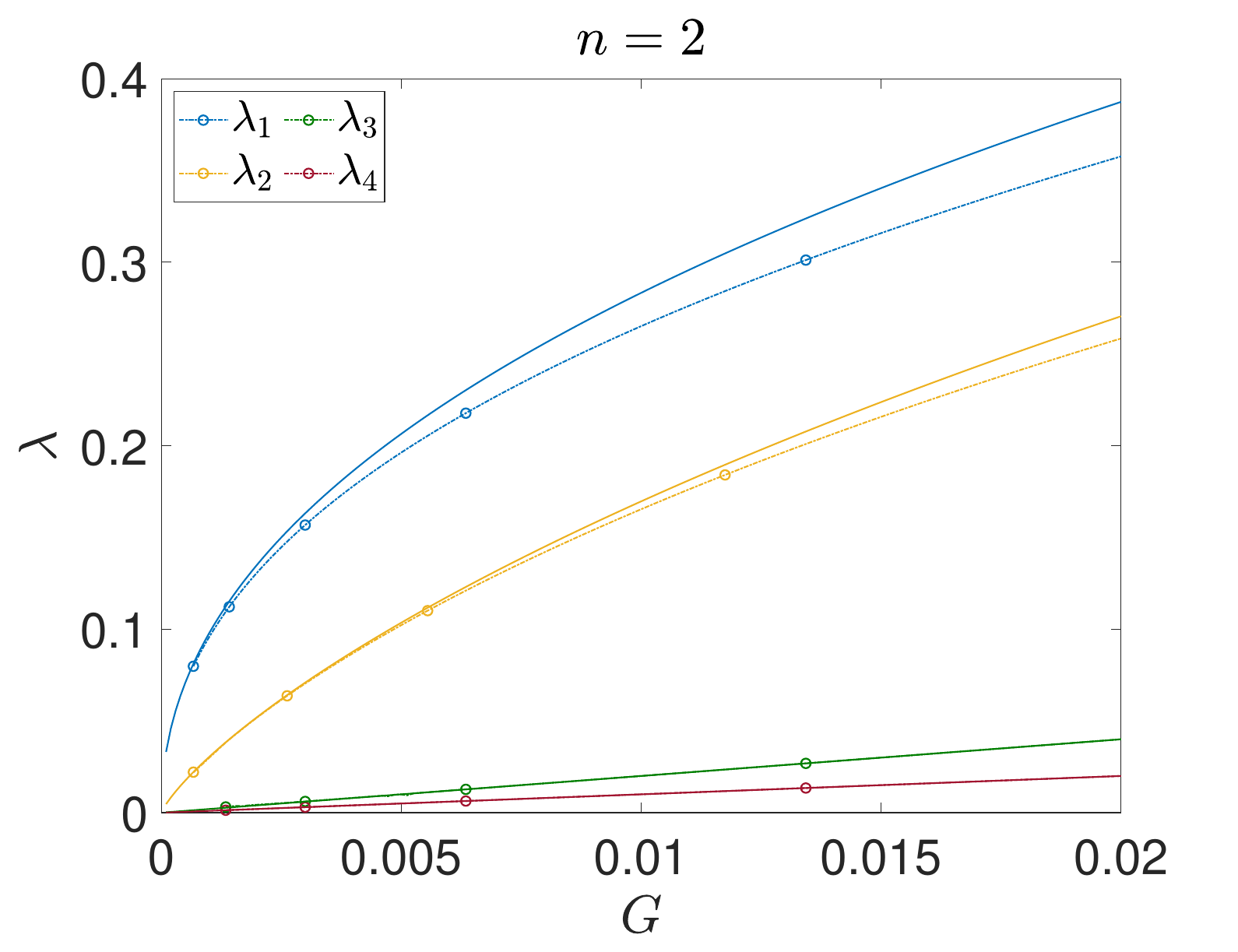}&\includegraphics[width=0.49\textwidth]{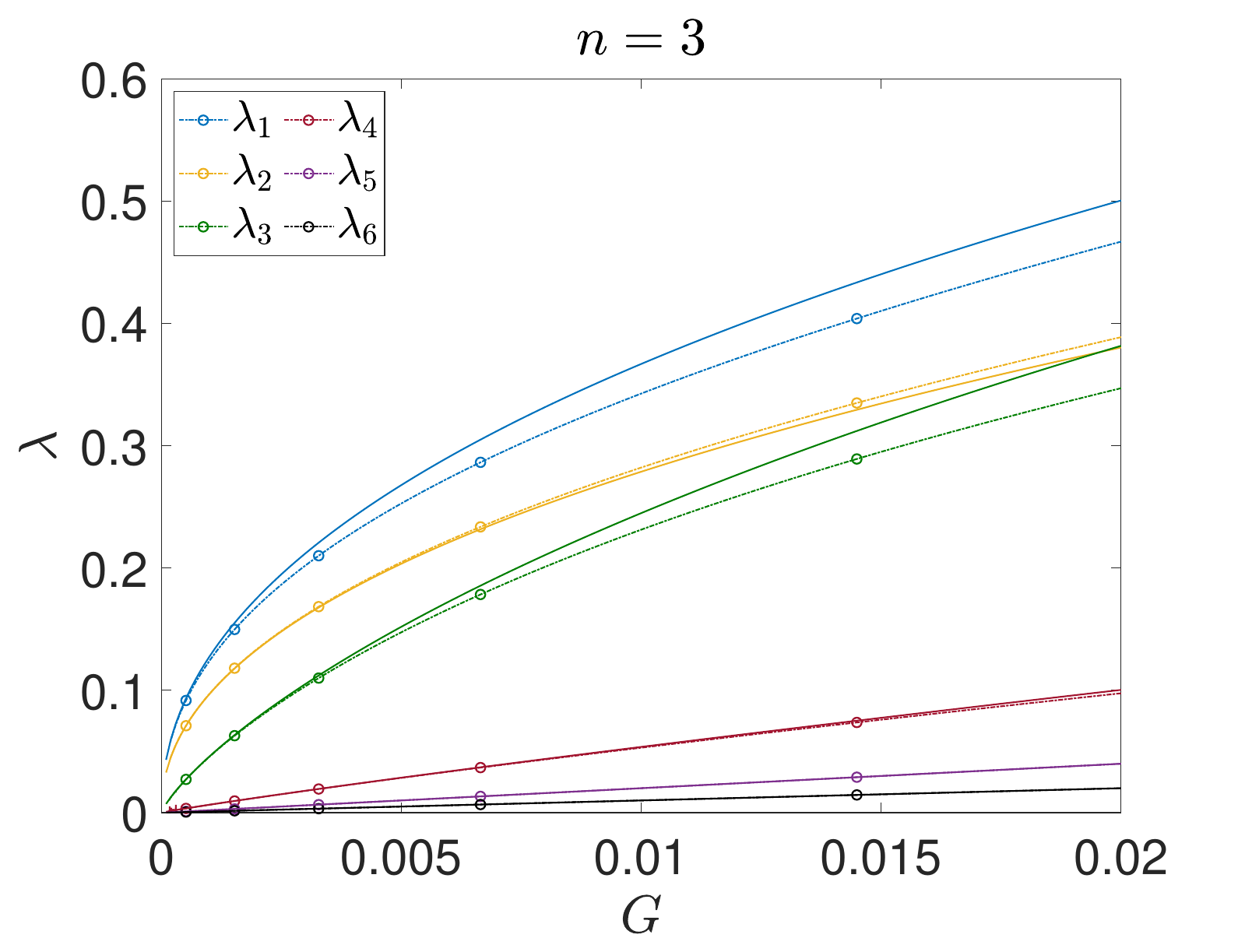}
  \\
(a)&(b)\\
\end{tabular}
\begin{tabular}{c}
\includegraphics[width=0.49\textwidth]{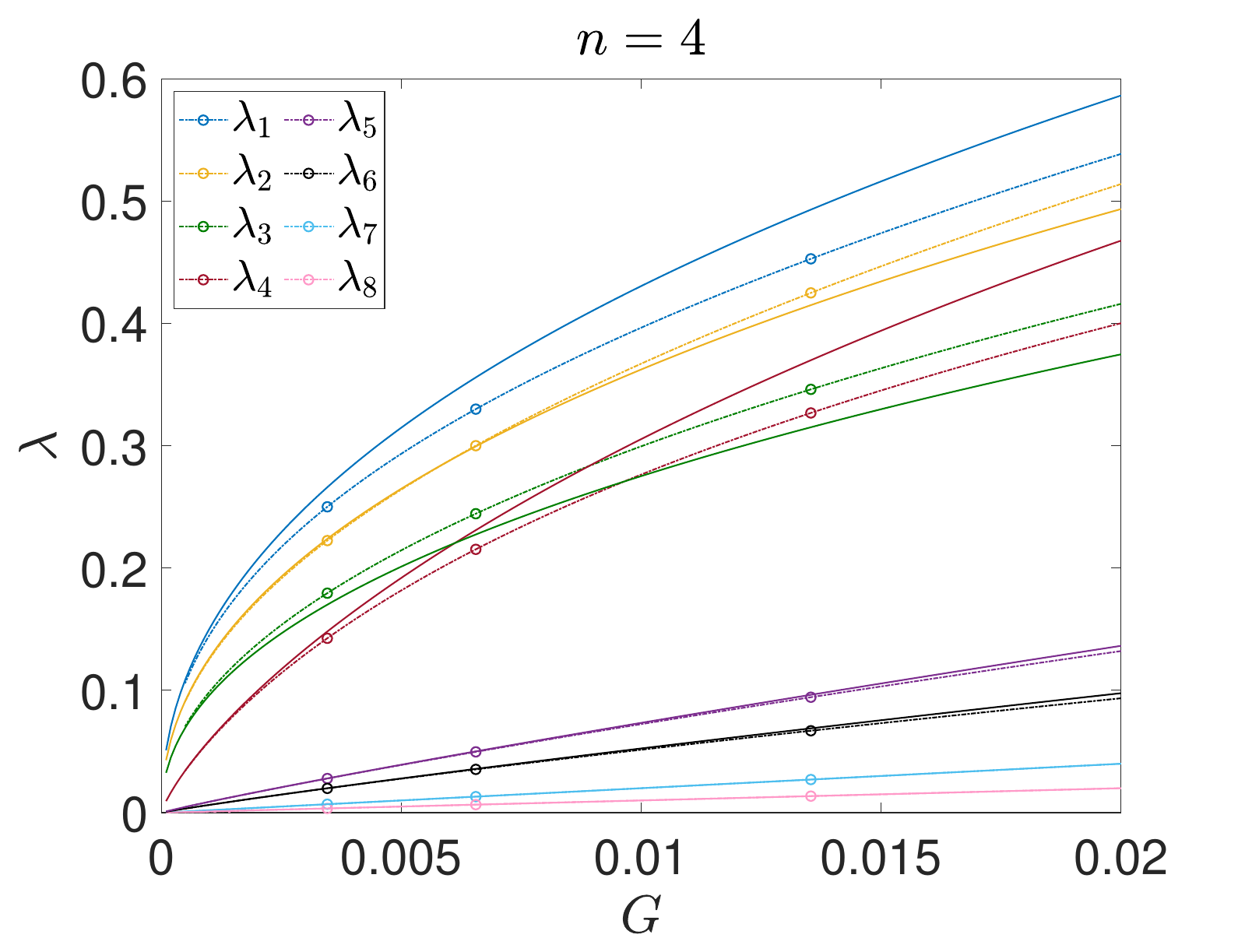}
  \\
(c)\\
\end{tabular}
\caption{  
  Variation of dominant real eigenvalues with blowup rate, $G$ for (a) two, (b) three, and (c) four humped self-similar solutions. 
All multi-humped solutions share the $\lambda \approx G$ and $\lambda \approx 2G$ eigenvalues ($\lambda_4$ and $\lambda_3$ for $n=2$, $\lambda_6$ and $\lambda_5$ for $n=3$, and $\lambda_8$ and $\lambda_7$ for $n=4$, respectively) . 
Numerical computations for each eigenvalue are displayed with dash-dotted lines with open circles and asymptotic predictions are depicted with solid lines of the same color for each eigenvalue. 
} \label{fig:dominanteig}

\end{figure}

Importantly, our methodology does not only provide
the qualitative theoretical backdrop for the
eigenvalue dependence on $G$. It provides, as will
be evident in section \ref{sec:stab}, detailed quantitative
estimates for the relevant eigenvalues.
Moreover, it provides explicit predictions for
the corresponding eigenfunctions.
These can be seen in comparison to 
the numerically obtained
eigenvectors in Fig.~\ref{fig:eigenfunctions} and 
to the corresponding numerical eigenvalues
in Fig.~\ref{fig:dominanteig}. 
The former are in excellent agreement with the numerical computations.
For the latter, the
$G^{1/2}$ dependence of $n-1$ real eigenvalues,
and the $G$ dependence of another $n+1$ ones
is evident, and we see that for most
of them the agreement is also quantitatively very accurate.
For some of the eigenvalues, we observe a slight disparity (e.g., for
$\lambda_3$ (green) one for the $n=2$ case, $\lambda_4$
and $\lambda_6$
(black and red)  for $n=3$ and perhaps most notably
$\lambda_5$ and $\lambda_8$ (purple and pink ones)
for $n=4$). This is, however, chiefly true when the
eigenvalues are sufficiently large and takes place
due to higher order corrections not accounted for
in our theory. Interestingly, for some other
eigenvalues, these higher order corrections appear to
be negligible and the prediction remains highly accurate
throughout the interval of considered values of
$G$. At the moment, we do not have an explanation about
why this takes place for some of the eigenvalues but
not for others, other than to allude to reasons of symmetry.
However, this is a potential interesting topic for further
exploration.

\subsection{Dynamics in the Self-similar Frame}

Our methodology captures the associated dynamics in the rescaled $\xi-\tau$ spatio-temporal domain. 
In particular, we aim to validate the  instability of multi-humped self-similar NLS solutions --discussed in the previous sections-- in contrast to the stable single-humped self-similar solutions. 
For this purpose, we solve Eq.~(\ref{eq:selfsimilarNLS}) in $\xi \in [0,K]$ with zero flux boundary conditions at both ends of the domain. 
The blowup rate, $G$, is computed by imposing a pinning condition, which ``freezes'' the imaginary part of $v$ at $\xi=0$ to a constant value, $C$, i.e.: $\textrm{Im}\left[ v(0,\tau) \right] = C$, in line
with the above discussion.
Time-stepping is performed by applying the backward Euler scheme,
as also discussed above.

\begin{figure}
 \centering
\begin{tabular}{c c}
\includegraphics[width=0.5\linewidth]{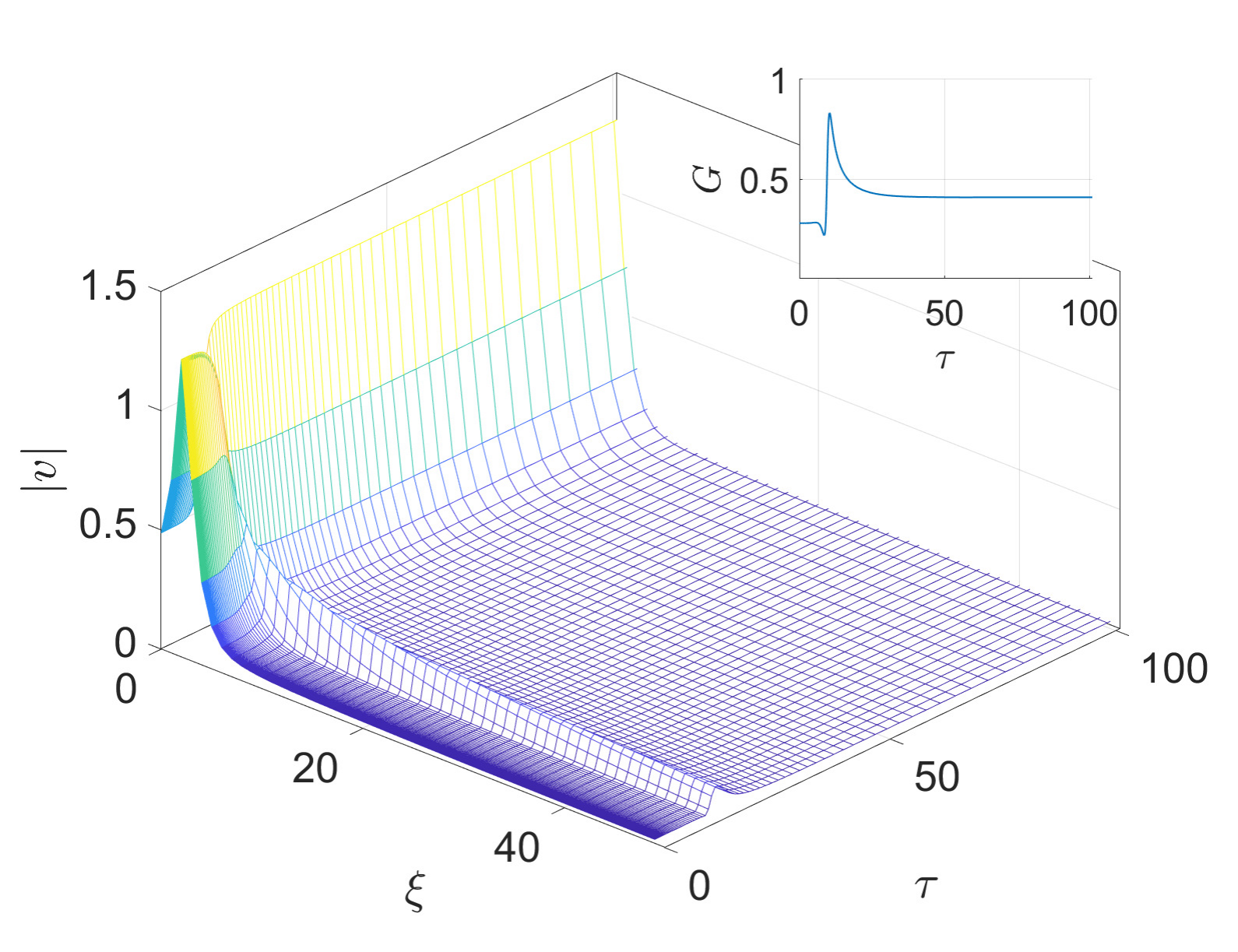}     & \includegraphics[width=0.5\linewidth]{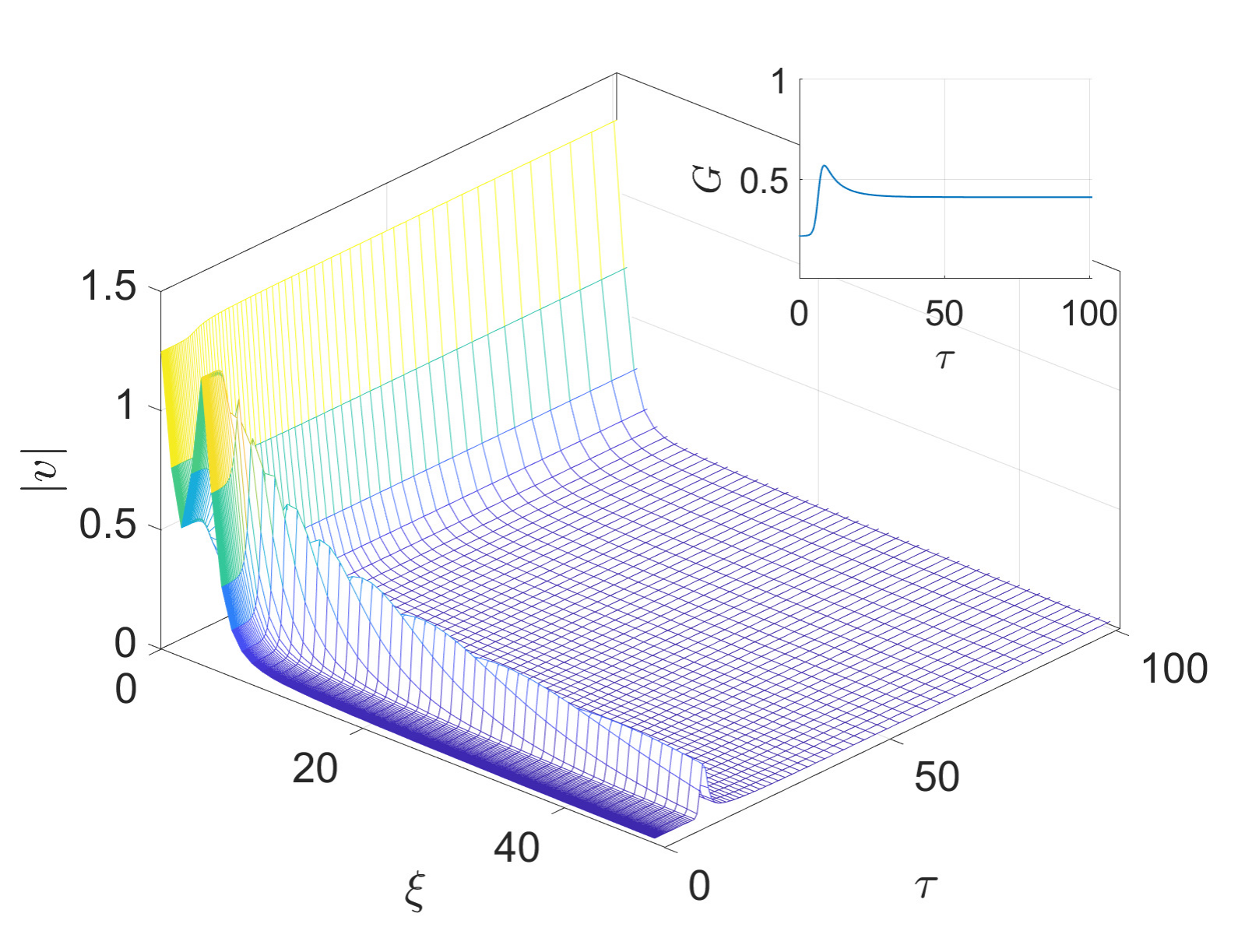}  \\
(a)     & (b) 
\end{tabular}      
 \caption{Long time stability of single-humped self-similar solutions in dynamical simulations.
 (a) A two-humped self-similar solution, initially computed at $\sigma=2.011$, evolves toward a stable single-humped self-similar solution after a slight decrease to $\sigma=2.0109$. 
 (b) A three-humped self-similar solution ($\sigma=2.011$) transitions to a stable single-humped self-similar state when $\sigma$ in increased to $\sigma=2.0111$.
The insets display the corresponding evolution of the blowup rate $G$. 
Simulations are performed in the domain $\xi \in [0, 50]$, with a nodal distance of $\Delta \xi=0.002$.
    }
    \label{fig:mndynamicssuccess}
\end{figure}

Figure~\ref{fig:mndynamicssuccess}(a)-(b) illustrates the evolution of unstable two-humped and three-humped self-similar solutions toward a stable single-humped self-similar state.
The insets display the evolution of the blowup rate, $G$, with both cases achieving convergence after $\tau \approx 40$.
For the two-humped case (Fig.~\ref{fig:mndynamicssuccess}(a)), the initial condition is the self-similar solution at $\sigma=2.011$ which is perturbed by decreasing $\sigma$ by $10^{-4}$, i.e., setting $\sigma=2.0109$. 
For the three-humped case (Fig.~\ref{fig:mndynamicssuccess}(b)), we start with the self-similar solution computed at $\sigma=2.011$ and perturb it by increasing $\sigma$ by  $10^{-4}$, i.e., using 
$\sigma=2.0111$ in the dynamical integration.
In both cases, it is clear that asymptotically once some of
the system's ``mass'' is ejected outward, the profiles can
converge to the single-hump state.

\begin{figure}
 \centering
\begin{tabular}{c c}
\includegraphics[width=0.5\linewidth]{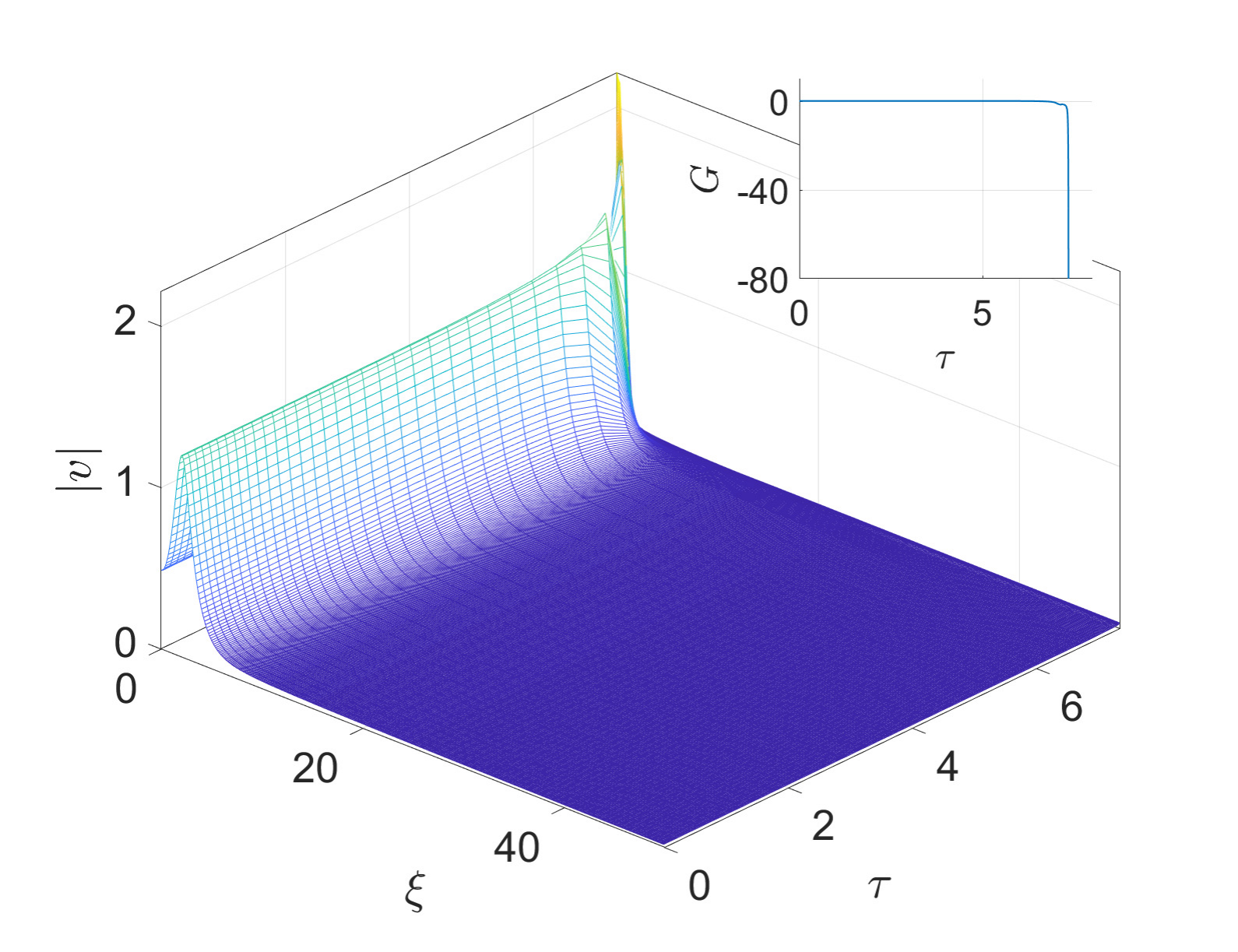}     & \includegraphics[width=0.5\linewidth]{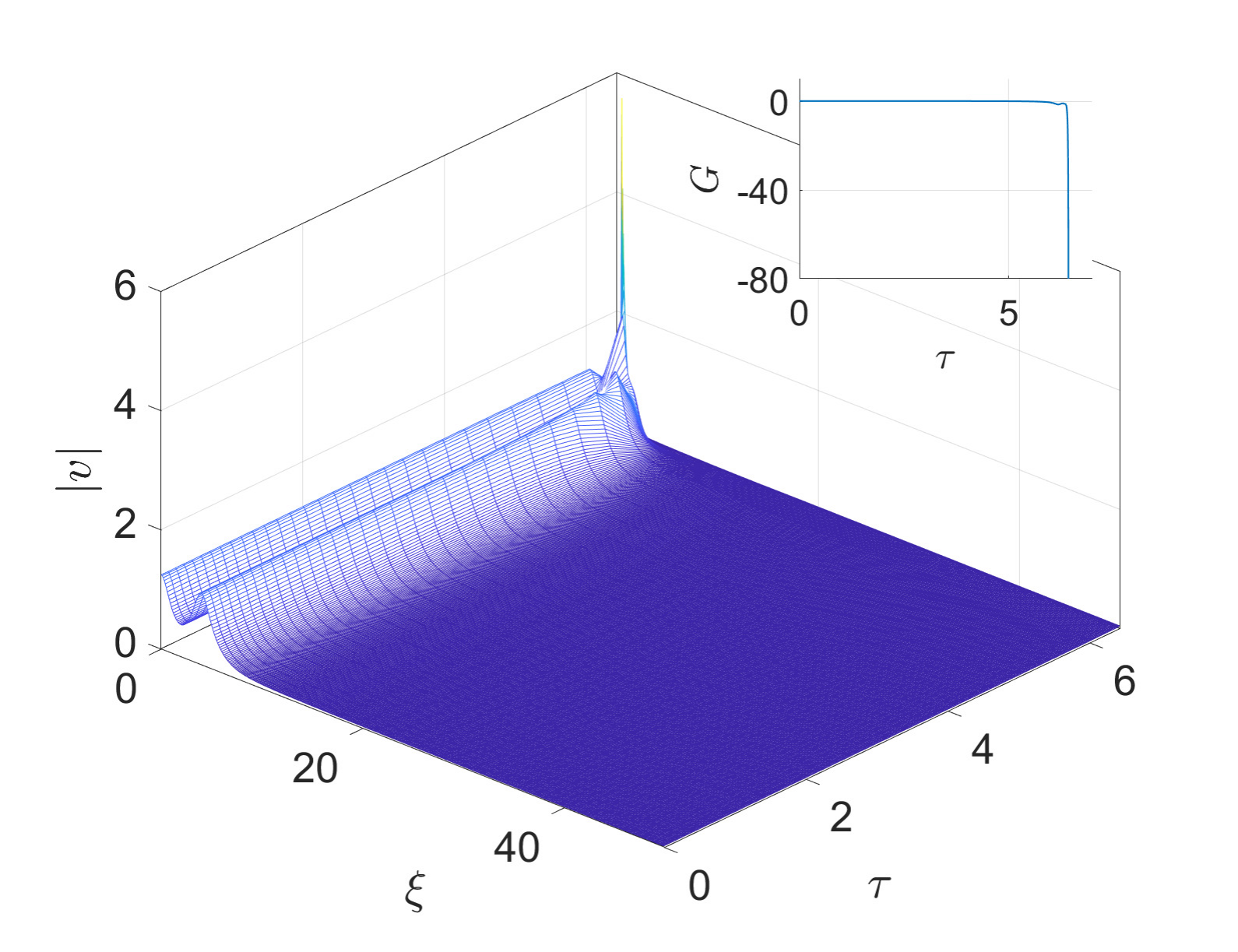}  \\
(a)     & (b) 
\end{tabular}      
 \caption{ %
 (a) Dynamical simulation starting from a two-humped self-similar solution at $\sigma=2.011$;
 the $\sigma$ value is now slightly increased to $\sigma=2.0111$, and dynamics evolve away from the stable single-humped self-similar solution.
 (b) An initial three-humped self-similar solution computed at $\sigma=2.011$ is now perturbed setting $\sigma=2.0109$.
 Again, one observes that rescaled dynamics fail to converge to a stable self-similar state.
The insets display the corresponding evolution of the blowup rate $G$. 
Simulations are performed in the domain $\xi \in [0, 50]$, with a nodal distance of $\Delta \xi=0.002$.
    }
    \label{fig:mndynamicsfailure}
\end{figure}

However, it is important to note that convergence to the stable single-humped self-similar solution is not always guaranteed, and depends on the nature of the initial perturbation.
For instance, reversing the sign of the $\sigma$ perturbation in the two previous simulations prevents the dynamics from converging to a stable single-humped solution, as shown in  Figs.~\ref{fig:mndynamicsfailure}(a)-(b).
Specifically, Fig.~\ref{fig:mndynamicsfailure}(a) illustrates the evolution of the two-humped solution computed at $\sigma=2.011$ and perturbed by increasing $\sigma$ by $10^{-4}$ (i.e., $\sigma=2.011$).
Similarly, Fig.~\ref{fig:mndynamicsfailure}(b) depicts the evolution of the three-humped solution, also computed at $\sigma=2.011$, this time perturbed by decreasing $\sigma$ to $2.0109$.
In both cases, the rescaled dynamics fail to reach a stable single-humped self-similar state. 

\begin{figure}
 \centering
\begin{tabular}{c c}
\includegraphics[width=0.5\linewidth]{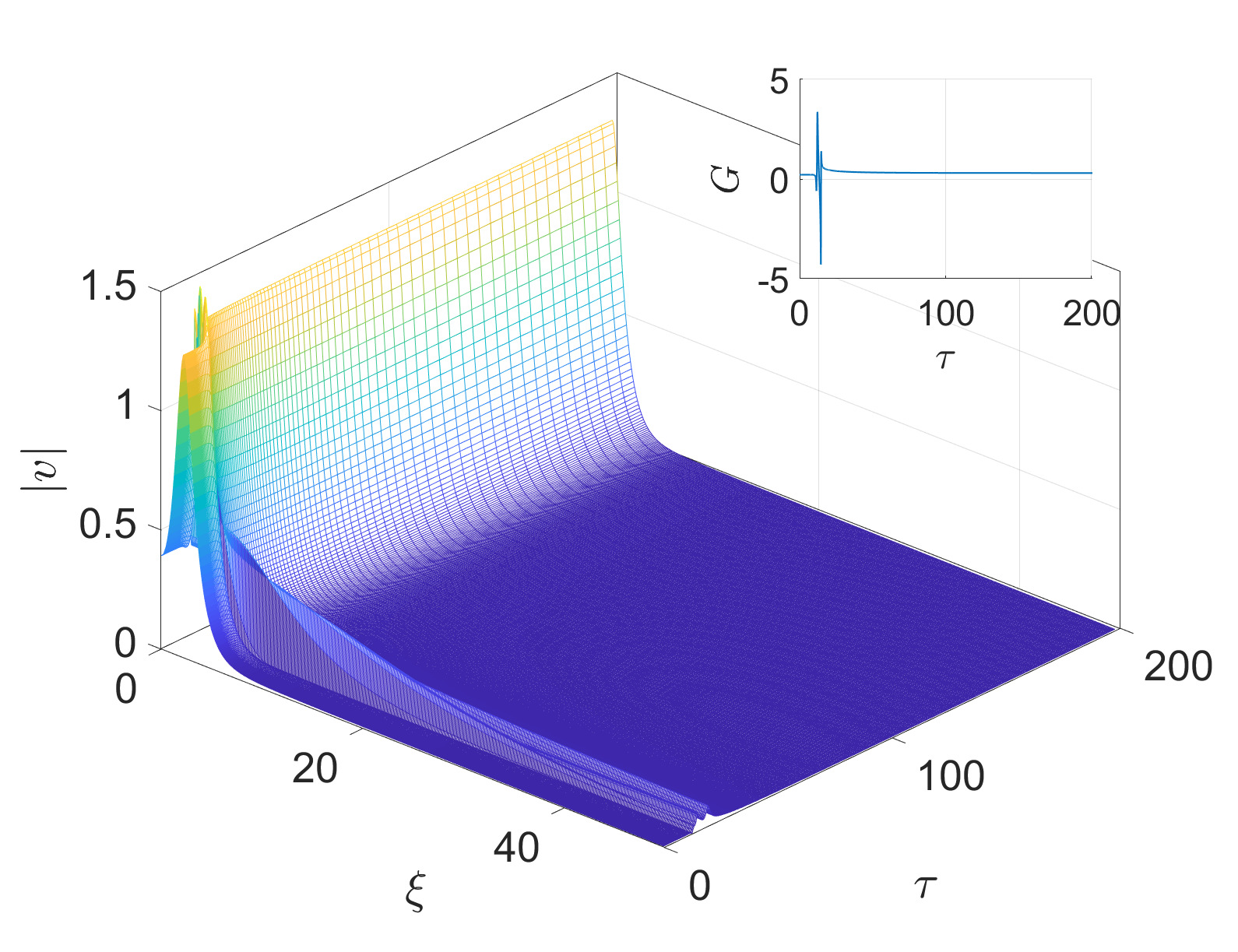}     & \includegraphics[width=0.5\linewidth]{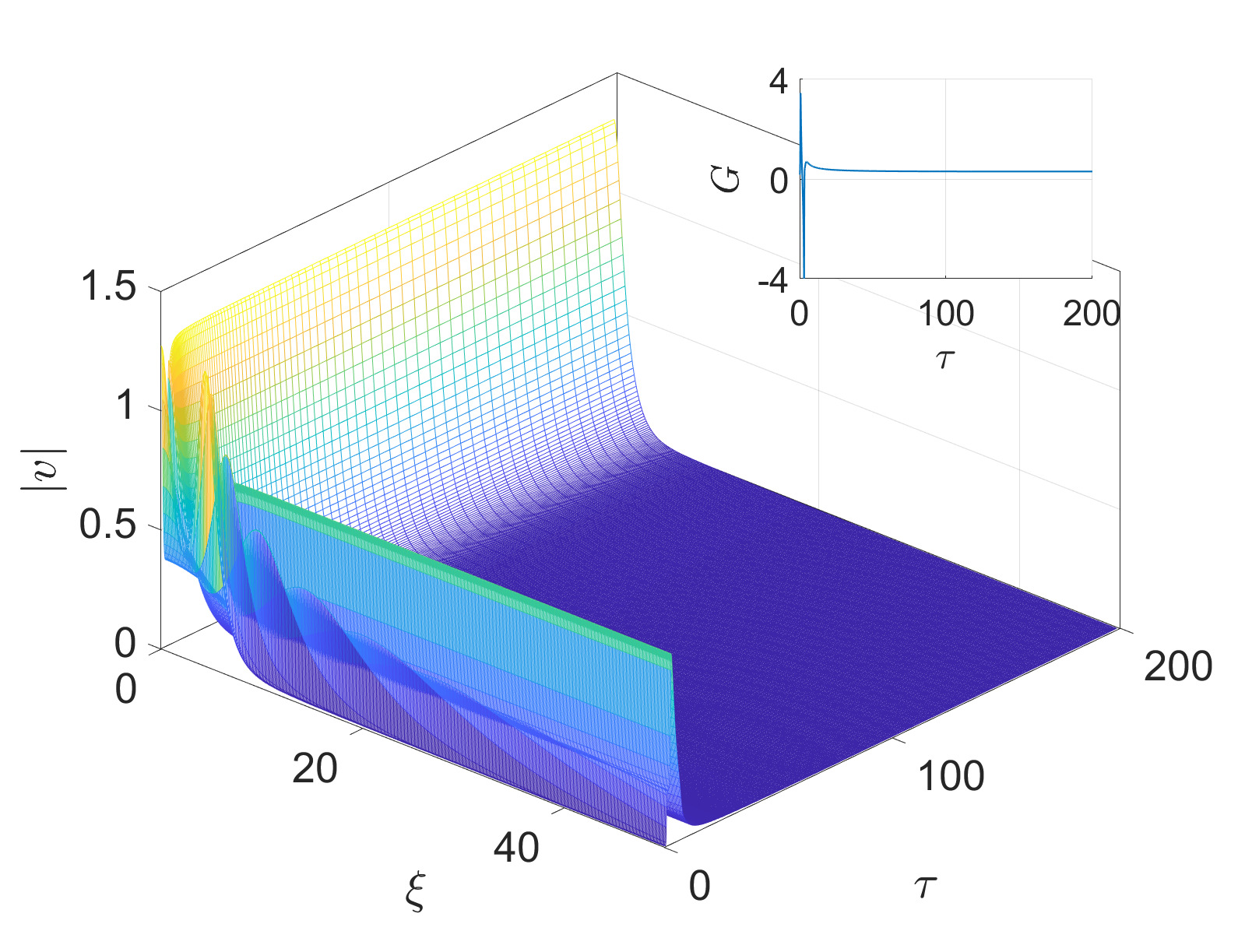}  \\
(a)     & (b) \\
\end{tabular}
\begin{tabular}{c}
\includegraphics[width=0.5\linewidth]{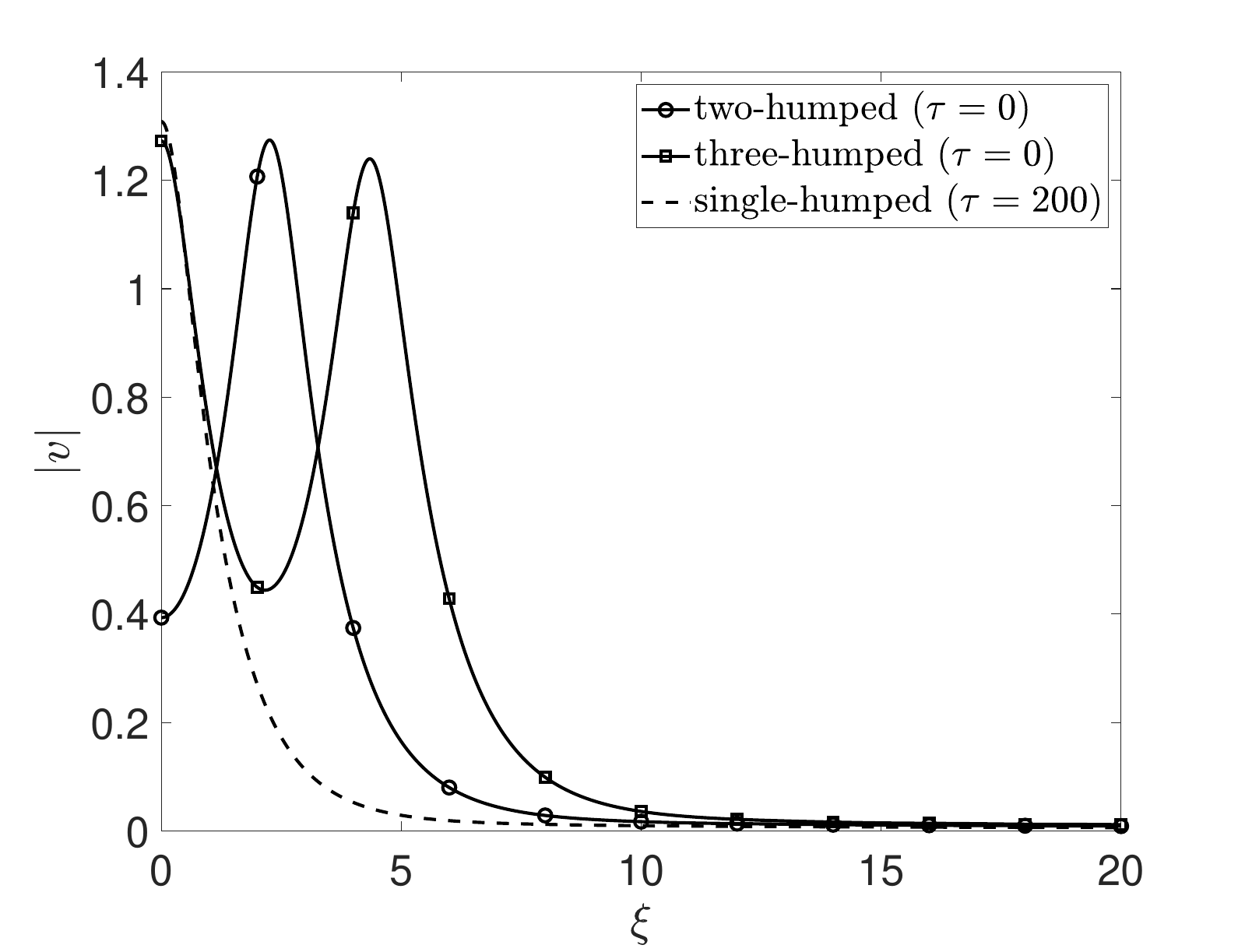} \\
(c) 
\end{tabular}

    \caption{Dynamical simulations starting from (a) two-humped, and (b) three-humped self-similar solutions at $\sigma=2.001$.
    By perturbing the initial condition along an unstable eigendirection (the one associated with the second largest real eigenvalue), the self-similar dynamics converged to the corresponding stable single-humped solution at large $\tau$.
    The inset of each figure shows the corresponding evolution of the blow-up rate $G$. 
    (d) The two different multi-humped states (2-humped and 3-humped) ultimately converge to the same stable single-humped self-similar profile (dashed line).
    }
    \label{fig:mndynamicsmultipletostable}
\end{figure}

To further reinforce our statement we perform additional simulations, where the initial multi-humped state is perturbed along an unstable eigenvector associated with one of the dominant real eigenmodes. 
Here, we apply perturbations along the eigenmode corresponding to the largest real eigenvalue.
In Figs.~\ref{fig:mndynamicsmultipletostable}(a)-(b), we present the spatio-temporal evolution of $v$ starting from perturbed (a) two-humped, and (b) three-humped self-similar solutions computed at $\sigma=2.001$.
All different multi-modal self-similar states {\em 
are expected to evolve toward the same stable single-humped self-similar solution}.
This is illustrated in Fig.~\ref{fig:mndynamicsmultipletostable}(c), where we show the initial state for each simulation (2-humped and 3-humped at $\tau=0$) alongside the final, stable single-humped self-similar state at $\tau=200$ for $\sigma=2.001$. 
The converged blow-up rate ($G(\tau=200) \approx 0.302$ is identical across all three simulations.

\section{Detailed Asymptotic Analysis of the multi-hump steady state}
\label{sec:steady}
We suppose that $\sigma$ is exponentially close to $2$ as $G \ra 0$
so that, in steady state,
\[
   \sdd{\Vs}{\xi}+ |\Vs|^{4}\Vs - \Vs 
   + \frac{G^2 \xi^2 }{4}\Vs + \tst= 0,\]
 where $\tst$ represents transcendently small terms.
We look for a general multi-hump solution, with humps located at $\xi = 
X_i(G)$ with $|X_i|\gg 1$, $i = 1,\ldots,n$, where we label such that $X_i<X_{i+1}$.
We aim to determine the $X_i$ along with the
dependence of $G$ on $\sigma$. We will find that each $X_i = O(\log
G)$.
By symmetry we expect to find $X_i = - X_{n+1-i}$, though we will not assume it.
We use the pinning condition $\im(\Vs(0))=0$, for which  we will find
$\Vs$ is real to all algebraic orders in $G$.

For the single hump we found in \cite{jon1} that the
solution comprised a near field in the vicinity of the hump and  a
far field away from the hump. The far field contained a turning point,
so there was also an inner region in the vicinity of the turning point
which gave an exponentially-small reflection back towards the near
field. This exponentially small component ultimately led to the
relationship between $\sigma$ and $G$ describing the
bifurcation.
In the multi-hump case we will find that there is again a
near field in the vicinity of each hump, along with new regions
between adjacent humps. The far field, including the turning point,
remains largely the same as in the single-hump case. We will detail
the asymptotic solution in each region in turn, matching them to
obtain the positions of the humps and the relationship between $\sigma$ and $G$.

\subsection{Near field of the $i$th hump}
\label{sec:nearhump}
We write $\xi = X_i + x$ 
to give
\beq
   \sdd{\Vs}{x}+ |\Vs|^{4}\Vs - \Vs 
   + \frac{G^2 (x+X_i)^2 }{4}\Vs= 0.\label{inner}
\eeq
Now expand
\beq
 \Vs =\sum_{n=0}^\infty G^{2n} \Vn_n\label{expansion}
\eeq
to give the leading-order equation
\beq
 \sdd{\Vn_0}{x}+ |\Vn_0|^4\Vn_0 - \Vn_0 = 0,\label{inner0}
\eeq
where we have assumed (and will verify a posteriori) that $G^2X_i^2 \ll 1$.
The relevant solution is the soliton,
\beq
V_0 = 3^{1/4} \sech^{1/2}\left(2 x\right) %
= 3^{1/4} \sech^{1/2}\left[2(\xi-X_i)\right].\label{solitonI}
\eeq

\subsection{Solution in between the humps}
\label{sec:gaps}
Consider the gap between $X_i$ and $X_{i+1}$. 
In this region $\Vs$ is (algebraically) small, so we can neglect the nonlinear terms at leading order. Then, expanding again as in (\ref{expansion}) gives
\beq
 \sdd{V_0}{\xi} - V_0 = 0.\label{outer0}
\eeq
The inner solution (\ref{solitonI}) near $X_{i+1}$ written in terms of $\xi$ is  
\[ V_0 = 3^{1/4} \sech^{1/2}\left[2 (\xi-X_{i+1})\right].\]
For large $X_{i+1}$ with $\xi\ll X_{i+1}$ this is
\beq
V_0 \sim 12^{1/4} \ee^{\xi-X_{i+1}}.\label{refl}
\eeq
Similarly the inner solution near $X_{i}$ written in terms of $\xi$ is  
\[ V_0 = 3^{1/4} \sech^{1/2} \left[2 (\xi-X_{i})\right].\]
For large $X_{i}$ with $X_{i}\ll\xi$ this is
\beq
V_0 \sim 12^{1/4} \ee^{-\xi+X_{i}}.\label{refl2}
\eeq
Imposing these two matching conditions the solution of (\ref{outer0}) for $X_i\ll \xi \ll X_{i+1}$ yields
\beq
V_0 = 12^{1/4} \ee^{-\xi+X_{i}} + 12^{1/4} \ee^{\xi-X_{i+1}}.\label{farorigin}
\eeq
There is an exponentially small (in $X_i-X_{i+1}$) interaction term from each hump on the other. We will see that this term will match with the next term in the inner expansion, which is $O(G^2)$, so that $X_i-X_{i+1}$ will be logarithmically large in $G$.
Writing \eqref{farorigin} in terms of the inner variable $x = \xi -
X_{i+1}$ gives
\beq
V_0 = 12^{1/4} \ee^{x} + 12^{1/4} \ee^{-x+X_i-X_{i+1}} \label{grow1};
\eeq
thus the extra interaction term corresponds to an exponentially growing term
leaving the inner region near $X_{i+1}$.
Similarly, in terms of the inner variable $x = \xi -
X_{i}$ we have
\beq
V_0 = 12^{1/4} \ee^{-x} +12^{1/4} \ee^{x+X_i-X_{i+1}}, \label{grow2}
\eeq
again
corresponding to an exponentially growing term leaving the inner
region near $X_i$.

In the regions $\xi>X_n$ and $\xi<X_1$ there is no additional interaction term---the solution to (\ref{outer0}) is a single exponential which decays at  $\pm \infty$ respectively. We may include these regions in the general framework by adopting the convention that $X_0=-\infty$, $X_{n+1} = \infty$, in which case (\ref{farorigin}) holds in these regions also.

\subsection{Next order in the near field}
\label{sec:nearhump1}
Equating coefficients of $G^2$ in (\ref{inner}) gives
\beq
 \sdd{V_1}{x}+ 5 V_0^4V_1 - V_1 
   = -\frac{(x+X_i)^2 }{4}V_0.\label{inner1}
\eeq
The homogeneous version of (\ref{inner1}) is satisfied by $V_0'$, so that there is a solvability condition.
Multiplying by $V_0'$ and integrating gives, on the LHS, 
\[
  \int_{-R}^R \fdd{V_0}{x}\left(\sdd{V_1}{x}+ 5 V_0^4V_1 - V_1
  \right)\, \d x = \left[ \fdd{V_0}{x} \fdd{V_1}{x} - \sdd{V_0}{x} V_1
  \right]^R_{-R} +  \int_{-R}^R\left( \frac{\dd^3 V_0}{\dd x^3}+ 5 V_0^4  \fdd{V_0}{x}- 
  \fdd{V_0}{x}\right)V_1 \, \d x
\]
so that the solvability condition is (when accounting for the
vanishing of the integral term upon differentiation of the
expression for $V_0$ in Eq.~(\ref{inner0}))
\beq
- \int_{-R}^R  \fdd{V_0}{x}\frac{(x+X_i)^2 }{4}V_0\, \d x = \left[ \fdd{V_0}{x} \fdd{V_1}{x} - \sdd{V_0}{x} V_1
  \right]^R_{-R} .\label{sol}
\eeq
Since $V_0$ is even the LHS may be simplified to
\[ - X_i \int_{-R}^R  \fdd{V_0}{x}\frac{x}{2}V_0\, \d x = X_i \int_{-R}^R
  \frac{V_0^2}{4}\, \d x \ra  \frac{\sqrt{3}\,\pi X_i }{8} \mbox{ as } R \ra \infty.\]
For the RHS we evaluate the boundary terms by matching with (\ref{grow1}) and (\ref{grow2}).
As $x \ra \infty$,
\beq G^2 V_1 \sim 12^{1/4} \ee^{-X_{i+1}+X_i} \ee^{x}, \qquad V_0 \sim 12^{1/4} \ee^{-x},\label{V1inf}
\eeq
so that
\[ \fdd{V_0}{x} \fdd{V_1}{x} - \sdd{V_0}{x} V_1\sim - 4 \sqrt{3} \, 
  \frac{\ee^{-X_{i+1}+X_i}}{G^2}.
\]
Similarly, as $x \ra -\infty$,
\beq
 G^2 V_1 \sim 12^{1/4} \ee^{-X_i+X_{i-1}} \ee^{-x}, \qquad V_0 \sim 12^{1/4} \ee^{x},\label{V1minf}
\eeq
so that
\[ \fdd{V_0}{x} \fdd{V_1}{x} -  \sdd{V_0}{x} V_1\sim -4\sqrt{3}\,  
  \frac{\ee^{-X_i+X_{i-1}}}{G^2}.
\]
Thus the solvability condition (\ref{sol}) is, as $R \ra \infty$, 
\beq X_i  =  \frac{32}{\pi G^2} \left(  \ee^{-X_i+X_{i-1}}
    -\ee^{-X_{i+1}+X_i}
  \right).\label{Xeqns}
\eeq
Equations (\ref{Xeqns}) form a closed system of $n$ equations for the $n$ unknowns $X_1,\ldots,X_n$. 
It is relevant to note here that this set of equations 
(that is sometimes referred to as the ---steady state 
form of--- Toda lattice with nonzero masses)
has appeared in other settings recently, such as, e.g.,
the case of $N$ dark solitary waves in the presence
of a parabolic confinement applicable to atomic
Bose-Einstein condensates~\cite{Coles_2010}.
In the latter case, the exponential tail-tail interaction
of the solitary waves is balanced by the presence
of the parabolic trap (onsite) effect on each of the
solitary waves. On the other hand, here the linear
effective potential stems from the role of the phase
of the complex wavefunction past the critical point of
collapse.
For the present case, 
we still need to identify $\sigma$, and to do so we need to consider the regions $\xi>X_n$ and $\xi<X_1$ more carefully.

\subsection{Far field}
\label{sec:farfield}
Consider first the region $\xi>X_n$. 
When $\xi$ is large the term $G^2 \xi^2 \Vs/4$ can no longer be neglected. 
Thus there is a different balance in the eqution in the far field. 
The solution is almost exactly the same as for the single-hump  soliton reported in \cite{jon1}, so we merely sketch the analysis here.

We rescale $\xi = \rho/G$ to give
\[
 G^2  \spd{\Vs}{\rho}+ |\Vs|^{4}\Vs - \Vs 
 + \frac{\rho^2}{4} \Vs= 0.\]
Since $\Vs$ is  exponentially small in the far field we
can neglect the nonlinear term.
We look for a WKB solution in the form
\begin{equation}
  \Vs \sim  \ee^{\phi(\rho)/G} \sum_{n=0}^\infty \A_n(\rho) G^{n}.\label{farfieldexp}
\end{equation}
At leading order this gives
\[ (\phi')^2 =1-\frac{\rho^2}{4}.\]
Note the turning point at $\rho=2$. For $\rho<2$ the relevant solution is
\[ \phi' = -\left(1-\frac{\rho^2}{4}\right)^{1/2}\]
so that $\phi$  is decreasing (and $V$ is exponentially decaying) as $\rho$ increases.
Let us fix the constant by writing
\[ \phi = -\int_0^\rho \left(1-\frac{\bar{\rho}^2}{4}\right)^{1/2}\, \d \bar{\rho}.\]
The leading-order amplitude equation is
\[ 2 \phi' \A_0' + \phi'' \A_0 = 0 \]
so that
\[ \A_0 = \frac{a_0}{(- \phi')^{1/2}} = \frac{2^{1/2}\,
    a_0}{(4-\rho^2)^{1/4}}.\]
The constant $a_0$ is determined by matching with the near field
solution in the vicinity of $X_i$.
As $\rho \ra 0$
\beq
\ee^{\phi(\rho)/G}A_0 \sim a_0 \ee^{ -\rho/G } .\label{in0}
\eeq
Since the right-most soliton is not positioned at the origin, the matching condition is slightly different to that in \cite{jon1}.
Writing (\ref{solitonI}) (or indeed (\ref{farorigin})) in terms of $\rho$  and expanding for small $G$ gives
\[ V_0 \sim 12^{1/4} \ee^{- \rho/G + X_n}.\]
Matching with \eqref{in0} gives
\[ a_0 = 12^{1/4} \ee^{X_n}.\]
Beyond the turning point, as $\rho \ra \infty$ only the solution in which
\[ \phi' = \ii\left(\frac{\rho^2}{4}-1\right)^{1/2}\]
has a finite Hamiltonian.
Thus for $\rho>2$, the WKB solution is
\[ \Vs = \al \ee^{\ii \phi_2(\rho)/G} \sum_{n=0}^{\infty} A_n(\rho) (\ii G)^n,
\]
for some constant $\alpha$, where
\[ \phi_2 = \int_2^\rho \left(\frac{\bar{\rho}^2}{4}-1\right)^{1/2}\, \d \bar{\rho}, \qquad A_0(\rho) =\frac{2^{1/2}a_0}{(\rho^2-4)^{1/4}}.
\]
The turning point causes an exponentially small reflection back into the near field.
The solution in the turning point region is exactly the same as in \cite{jon1}, with the result that $\alpha = \ee^{\ii \pi/4}$ and the WKB solution in $\rho<2$ is modified to 
\beqas
\Vs &=&  \ee^{\phi(\rho)/G} \sum_{n=0}^{\infty} A_n(\rho) G^n + \gamma\ee^{- \phi(\rho)/G} \sum_{n=0}^{\infty} A_n(\rho) (-G)^n \qquad \rho<2,
\eeqas
where, for an infinite domain, 
\[ \gamma = \frac{\ii}{2}\ee^{2\phi(2)/G}=\frac{\ii}{2}\ee^{-\pi/G}.\]

\subsection{Matching back into the near field}
\label{sec:sol}
We have found that the exponentially small correction to the far field in $0<\rho<2$ is
\[ \Vexp = \gamma\ee^{- \phi(\rho)/G} \sum_{n=0}^{\infty} A_n(\rho) (-G)^n
\]
As $\rho \ra 0$
\beq
\Vexp \sim a_0 \gamma  \ee^{\rho/G}. \label{match}
\eeq
Writing $\Vs = \Ve + \Vexp$ 
where
$\Ve$
is the original (multi-hump) algebraic expansion, and neglecting quadratic terms in $\Vexp$ but keeping now the term involving $\sigma-2$ gives, noting that $\Ve$ is real, we obtain
\[
\sdd{\Vexp}{\xi}+ \Ve^4 (2 \Vexp^*+3 \Vexp)  
  -\Vexp
  + \frac{G^2 \xi^2 }{4}\Vexp
  =  -2(\sigma-2) \Ve^5 \log \Ve+\frac{\ii (\sigma-2) G}{2 \sigma} \Ve.
\]
We  separate into real and imagingary parts. Writing $\Vexp = \Uexp + \ii \Wexp$ gives
\beqa
   \sdd{\Uexp}{\xi}+
 5 \Ve^4 \Uexp   
 -\Uexp 
  + \frac{G^2  \xi^2 }{4}\Uexp&=& - 2(\sigma-2) \Ve^5 \log \Ve,\label{Ueq}\\ 
  \sdd{\Wexp}{\xi}+
 \Ve^4 \Wexp   
 -\Wexp 
  + \frac{G^2  \xi^2 }{4}\Wexp&=& 
  \frac{ (\sigma-2) G}{2 \sigma} \Ve. \label{Weq}
\eeqa
Note that $\Ve$ satisfies the homogeneous version of (\ref{Weq}).
The resulting solvability condition will determine $\sigma$.
Unlike \S\ref{sec:nearhump}-\S\ref{sec:nearhump1}, it is easiest now to consider this equation throughout the whole near-field region, rather than considering each hump separately.
Multiplying (\ref{Weq}) by $\Ve$ and integrating from $-R$ to $R$,
while also using the equation satisfied by $\Ve$, gives
\beqa
\lefteqn{\int_{-R}^R \left(\Ve  \sdd{\Wexp}{\xi}+
 \Ve^5 \Wexp   
 -\Ve\Wexp 
 + \frac{G^2 \xi^2 }{4}\Ve\Wexp\right)\, \d \xi\non }\qquad&& \\
&=&\left[\Ve \fdd{\Wexp}{\xi} - \Wexp \fdd{\Ve}{\xi}\right]^R_{-R}+
\int_{-R}^R \left(\Wexp  \sdd{\Ve}{\xi}+
 \Ve^5 \Wexp   
 -\Ve\Wexp 
 + \frac{G^2 \xi^2 }{4}\Ve\Wexp\right)\, \d \xi \non \\
 & = &\left[\Ve \fdd{\Wexp}{\xi} - \Wexp \fdd{\Ve}{\xi}\right]^R_{-R} 
 =  
   \frac{ (\sigma-2) G}{2 \sigma} \int_{-R}^R\Ve^2\, \d x\label{sigsol}
\eeqa
As $R \ra \infty$ we evaluate the boundary terms by matching using (\ref{match}). We find
\beqas
\Ve(R) \fdd{\Wexp}{x}(R) - \Wexp(R) \fdd{\Ve}{x}(R)
& = & 2 \sqrt{3}\, \ee^{2X_n} \ee^{-\pi/G} \qquad \mbox{ as }R \ra \infty.
\eeqas
A similar calculation in $\xi<X_1$ shows that
\beqas
\Ve(-R) \fdd{\Wexp}{x}(-R) - \Wexp(-R) \fdd{\Ve}{x}(-R)
& = & -2 \sqrt{3}\, \ee^{-2X_1} \ee^{-\pi/G} \qquad \mbox{ as }R \ra \infty.
\eeqas
The dominant contribution to the RHS of (\ref{sigsol}) comes from each of the humps, since $\Ve$ is $O(G^2)$ in between the humps. 
Since
\[ \int_{-\infty}^\infty V_0(x)^2\, \d x = \frac{\sqrt{3}\,\pi}{2},\]
we find, finally
\beq
 \sigma-2= \frac{8\sigma}{n\pi}
(\ee^{2X_n}+\ee^{-2X_1}) \frac{\ee^{-\pi/G}}{G}. \label{sigeqn1secIV}
\eeq
 Note that with $n=1$ and $X_1=X_n=0$ this reduces to the formula in \cite{jon1}, and also that we anticipate the solutions of (\ref{sigeqn1secIV}) to satisfy $X_1 = -X_n$ so that 
\beq
 \sigma-2= \frac{16\sigma}{n\pi}
\frac{\ee^{2X_n-\pi/G}}{G}. \label{sigeqn2secIV}
\eeq

\section{Detailed Asymptotic Analysis of the multi-hump eigenvalues}

\label{sec:stab}

Since $\Vs$ is real to all orders, and $\sigma$ is
exponentially close to 2, we may write the
linearization problem of Eqs.~(\ref{geqn})-(\ref{feqn}) as 
\beqa
\ii\la f+\sdd{ f}{\xi} +
2 \Vs^{4}g +3 \Vs^{4} f 
- f
+ \frac{G^2 \xi^2 }{4} f &=& \tst,\label{geqn1}\\
-\ii  \la g
+ \sdd{g}{\xi} 
+  2 \Vs^{4} f 
+ 3\Vs^{4} g 
-  g 
+ \frac{G^2 \xi^2 }{4} g  &=& \tst.\label{feqn1}
\eeqa
We aim to find asymptotic approximations to the real eigenvalues found numerically in \S\ref{sec:numerics}. Since $\Vs$ is real to all orders, when $\la$ is real we may write $g = f^* + \tst$, so that 
\beq
\ii\la f+\sdd{ f}{\xi} +
2 \Vs^{4}f^* +3 \Vs^{4} f 
- f
+ \frac{G^2 \xi^2 }{4} f = \tst
\eeq
Anticipating that $\la$ is small when $G$ is small, the leading-order eigenfunctions satisfy
\beq
\sdd{f_0}{\xi} +
2 V_0^{4}f_0^* +3 V_0^{4} f_0 
- f_0  = 0.\label{eqnf0}
\eeq
There are two solutions to this equation, namely
\[ f_0 = \fdd{V_0}{\xi}, \qquad f_0 = \ii V_0.\]
For a multi-hump steady state we may take a different multiple of $\fdd{V_0}{\xi}$ or $\ii V_0$ at each hump.
We begin in \S\ref{sec:iV0} by analysing the eigenfunctions which are proportional to $\ii V_0$ near each hump, which give the dominant eigenvalues. We then, in \S\ref{sec:Vp}, analyse those eigenfunctions which are proportional to $\fdd{V_0}{\xi}$, which are much more difficult to find.

Throughout the analysis we will encounter inhomogeneous versions of (\ref{eqnf0}) in the form
\beq
\sdd{f}{x} +
2 V_0^{4}f^* +3 V_0^{4} f 
- f  = \RHS.\label{eqnf0rhs}
\eeq
Since the imaginary part of the homogeneous equation has the nontrivial solution $V_0$, and the real part of the homogeneous equation has the nontrivial solution $\fdd{V_0}{x}$, there are two solvability conditions on (\ref{eqnf0rhs}), namely, 
\beqa
\im \left[\fdd{f}{x}V_0 - f \fdd{V_0}{x}\right]^R_{-R}  &=& \int_{-R}^R \im(\RHS)V_0\, \d x,\label{solv1}\\
\re \left[\fdd{f}{x}\fdd{V_0}{x} - f \sdd{V_0}{x}\right]^R_{-R}  &=& \int_{-R}^R \re(\RHS)\fdd{V_0}{x}\, \d x,\label{solv2}
\eeqa
for all $R$. We will of course be interested in the limit $R \ra \infty$.

\subsection{Perturbation of $f = \ii V_0$: eigenvalues of $O(G^{1/2})$.}
\label{sec:iV0}
Guided by our numerical results let us anticipate that $\la =
O(G^{1/2})$ and write $\la = G^{1/2}\la_1$. We will perturb in powers
of $G^{1/2}$. The expansion has a similar structure to that of
\S\ref{sec:steady}---there are inner regions in the vicinity of each
hump, and outer regions between humps, which much be matched together
to give $\la$.

\subsubsection{Inner expansion near  the $i$th hump}
With $\xi = X_i + x$, we expand
\[ f = \B_i \sum_{k=0}^\infty G^{k/2}f_k,\]
with 
 $f_0 = \ii  V_0$.
At $O(G^{1/2})$ we have
\[
\eq{f_1}{f_1^*} = \la_1 V_0.
\]
The solvability conditions (\ref{solv1}), (\ref{solv2}) are automatically satisfied, so that $\la_1$ is undetermined, and 
\[ f_1 = \la_1\left(\frac{V_0}{4} + \frac{ x}{2} \fdd{V_0}{x}\right).
\]
At $O(G)$ we find \footnote{Really we should expand $\la$ (and $\B_i$) in powers of $G$, so that there would be additional terms on the RHS due to $\la_2$. However, these terms will simply mirror the terms involving $\la_1$ at the previous order, and we can ignore them if we are only interested in the leading approximation of the eigenvalue.}
\[
\eq{f_2}{f_2^*} 
=-\ii \la_1^2\left( \frac{V_0}{4} + \frac{ x}{2} \fdd{V_0}{x}\right).
\]
Again the solvability conditions (\ref{solv1}), (\ref{solv2}) are automatically satisfied, so that $\la_1$ is still undetermined, and the solution is
\[
f_2  = - \frac{\ii \la_1^2}{8} x^2 V_0.
\]
At $O(G^{3/2})$ we find
\[
\eq{f_3}{f_3^*} 
= - \frac{\la_1^3 x^2}{8} V_0.
\]
As before, the solvability conditions are automatically satisfied, and the solution is
\[
f_3  =  \frac{ \la_1^3}{2}  \hat{V}_1, 
\qquad
g_3  =  \frac{\la_1^3}{2 }\hat{V}_1,
\]
where $\hat{V}_1$ 
satisfies
\[
\eqr{\hat{V}_1} = -\frac{x^2 V_0}{4}, \qquad \hat{V_1} \ra 0 \mbox{ as } x \ra \pm \infty,
\]
that is, $\hat{V_1}$ is the next term in the expansion of the standard
single hump solution \cite{jon1} (i.e. not
accounting for the shift in origin).
At $O(G^{2})$ we find 
\beq
\eq{f_4}{f_4^*} 
= - \frac{\ii  \la_1^4}{2}  \hat{V}_1 - \frac{\ii (x+X_i)^2}{4}V_0 - 4 \ii V_1 V_0^4.\label{iV4}
\eeq
Here at last the solvability condition (\ref{solv1}) is not automatically satisfied, and  will determine $\la_1$. Note that $V_1$ grows exponentially as $x \ra \pm\infty$ (in contrast
to $\hat{V}_1$)---see (\ref{V1inf}), (\ref{V1minf})---so that we must
be careful in evaluating the solvability condition.
Multiplying (\ref{inner1}) by $V_0$ and integrating gives
\beqas
-\int_{-R}^R \frac{(x+X_i)^2}{4}V_0^2\, \d x & = & \int_{-R}^R
 \left( \sdd{V_1}{x} + 5 V_0^4 V_1 - V_1\right)V_0\, \d x \\
& = & \int_{-R}^R
 \left( \sdd{V_0}{x} + 5 V_0^5  - V_0\right)V_1\, \d x +
 \left[\fdd{V_1}{x} V_0 - V_1 \fdd{V_0}{x} \right]^R_{-R}\\
& = & \int_{-R}^R
4 V_0^5V_1\, \d x +
 \left[\fdd{V_1}{x} V_0 - V_1 \fdd{V_0}{x} \right]^R_{-R}.
\eeqas
Thus the solvability conditions on (\ref{iV4}) may be written as
\beqa
 \im \left[\fdd{f_4}{x}V_0 - f_4 \fdd{V_0}{x}\right]^R_{-R}  &=& 
 -   \frac{\la_1^4}{2}  \int_{-R}^R 
  \hat{V}_1 V_0\, \d 
x  
+  \left[\fdd{V_1}{x} V_0 - V_1 \fdd{V_0}{x} \right]^R_{-R}\label{sol1}\\
 \re \left[\fdd{f_4}{x}\fdd{V_0}{x} - f_4 \sdd{V_0}{x}\right]^R_{-R}  &=& 
0.
\label{sol2}
\eeqa
The boundary terms on the RHS can be evaluated using (\ref{V1inf}), (\ref{V1minf}).
To evaluate the boundary terms on the LHS we need to consider $f$ in the region between humps.

\subsubsection{In between the humps}
Consider the gap between $X_i$ and $X_{i+1}$.
In this region $\Vs = O(G^2 \log G)$ so that we can neglect the nonlinear term.
Thus
\beq
\ii G^{1/2}\la_1 f+\sdd{ f}{\xi}
- f
+ \frac{G^2 \xi^2 }{4} f = O(f G^8 \log^4 G),
\eeq
Expanding
\[ f = \sum_{k=0}^\infty G^{k/2} f_k,\] 
gives
\beq
\sdd{f_0}{\xi}
- f_0 = 0,
\eeq
at leading order.
Matching requires
\[
f_0 \sim 12^{1/4}\ii \B_{i+1} \ee^{\xi-X_{i+1}} \quad
\mbox{ as } \xi \ra X_{i+1},\qquad
f_0 \sim 12^{1/4} \ii  \B_{i}\ee^{-\xi+X_i}\quad\mbox{ as } \xi \ra X_i.
\]
Thus
\[ f_0 = 12^{1/4}\ii \B_{i+1} \ee^{\xi-X_{i+1}} +12^{1/4}\ii \B_i \ee^{-\xi+X_i}.\]
In terms of $\xi = X_{i+1} + x$ this is
\[ f_0 = 12^{1/4}\ii \B_{i+1} \ee^{x} +12^{1/4}\ii \B_i \ee^{-x+X_i-X_{i+1}}.\]
Since $\ee^{X_i-X_{i+1}} = O(G^2)$ the additional term must match with $G^2 f_4$ from the $(i+1)$th hump solution, giving
\beq 
\B_{i+1}f_4G^2 \sim 12^{1/4}\ii \B_i \ee^{-x+X_i-X_{i+1}} \label{f4minf}
\eeq
as $x \ra -\infty$.
Similarly, in terms of $\xi = X_{i} + x$,
\[ f_0 = 12^{1/4}\ii \B_{i+1} \ee^{x+X_i-X_{i+1}} +12^{1/4}\ii \B_i \ee^{-x},\]
giving the matching condition on the $i$\,th hump solution
\beq
 \B_{i}f_4G^2 \sim 12^{1/4}\ii \B_{i+1} \ee^{x+X_i-X_{i+1}},\label{f4inf}
\eeq
as $x \ra \infty$.
Under our convention that $X_0 = -\infty$, $X_{n+1} = \infty$
equations (\ref{f4minf}), (\ref{f4inf}) can be extended to include
$i=0$ and $i=n$, respectively.

\subsubsection{Solvability condition}
We now have enough information to evaluate the solvability conditions
(\ref{sol1}), (\ref{sol2}).
Using (\ref{f4minf}), (\ref{f4inf}), (\ref{V1inf}) and (\ref{V1minf}) gives
\beqa
\lim_{R \ra \infty}\re \left[\fdd{f_4}{x}\fdd{V_0}{x} -  f_4 \sdd{V_0}{x}\right]^R_{-R} & = &0,\\
\lim_{R \ra \infty}\im \left[\B_i\fdd{f_4}{x}V_0 - \B_i f_4 \fdd{V_0}{x}\right]^R_{-R} & = &4 \sqrt{3}\,\ii \B_{i+1} \frac{\ee^{X_i-X_{i+1}}}{G^2}+4 \sqrt{3}\,\ii \B_{i-1} \frac{\ee^{X_{i-1}-X_{i}}}{G^2},\\
\lim_{R \ra \infty}\left[\fdd{V_1}{x} V_0 - V_1 \fdd{V_0}{x} \right]^R_{-R}
& = & 4\sqrt{3}\,  \frac{\ee^{X_i - X_{i+1}}}{G^2}+ 4\sqrt{3}\,
\frac{\ee^{X_{i-1} - X_{i}}}{G^2}.\label{V1x}
\eeqa
Thus, letting $R\ra\infty$,  (\ref{sol2}) is satisfied, and,  since
\[ \int_{-\infty}^\infty  \hat{V}_1 V_0\, \d x =\frac{\sqrt{3}\,
    \pi^3}{256},\]
the remaining condition (\ref{sol1})  is
\beq 
\frac{\B_i \la_1^4 \pi^3}{2048} =   
 \left(\B_i-\B_{i+1}\right)\frac{\ee^{X_i - X_{i+1}}}{G^2}+
  \left(\B_i-\B_{i-1}\right)\frac{\ee^{X_{i-1} - X_{i}}}{G^2}, \qquad i = 1,\ldots,n.
\label{Ghalfeigs}
 \eeq
Equation (\ref{Ghalfeigs}) is a set of $n$ homogeneous linear equations for the $n$ coefficients $\B_i$, $i = 1,\ldots,n$. For a nontrivial solution the determinant must vanish, and it is this condition that determines the eigenvalue $\la_1$.
Note that  $\B_i =\B$ for all $i$, $\la_1=0$ is a solution  of
\eqref{Ghalfeigs} for any
$n$. This corresponds to  global phase invariance, with corresponding
eigenvalue $\la=2G$ (we would need to include $\la_2$ and proceed to
higher orders to see that through this expansion, but fortnuately we
have an exact eigenfunction for $\la=2G$ in \eqref{FG2G}). Thus we
expect that there are $n-1$ remaining eigenvalues
which are  $O(G^{1/2})$.
Note also that the eigenvalues appear in pairs---for each positive eigenvalue there is a corresponding negative eigenvalue with the same (leading order) weights $\B_i$.

\subsubsection{Examples}
\paragraph{Two humps:}
With $-X_1=X_2 = X$, (\ref{Ghalfeigs}) is 
\[
 \la_1^4 =   \frac{2048}{\pi^3 G^{2}}
 \left(1-\frac{\B_{2}}{\B_1}\right)\ee^{-2X}
=   
 \frac{2048}{\pi^3 G^{2}}
 \left(1-\frac{\B_{1}}{\B_2}\right)\ee^{-2X},
\]
Thus $\B_2^2=\B_1^2$ so that $\B_2 = \pm \B_1$. The additional solution
$\B_2 = - \B_1$ has positive eigenvalue
\[ \la = \frac{8\ee^{-X/2}}{\pi^{3/4}} = \frac{2^{7/4} X^{1/4}}{\pi^{1/2}}G^{1/2}.\]

\paragraph{Three humps:}
With $X_2=0$, $-X_1=X_3 = X$,  (\ref{Ghalfeigs}) is 
\[
 \frac{\pi^3 G^{2}\la_1^4}{2048} =  
 \left(1-\frac{\B_{2}}{\B_1}\right)\ee^{-X} =   
 \left(1-\frac{\B_{3}}{\B_2}\right)\ee^{- X}+
\left(1-\frac{\B_{1}}{\B_2}\right)\ee^{-X}= 
  \left(1-\frac{\B_{2}}{\B_3}\right)\ee^{- X},
 \]
with solution $(\B_1,\B_2,\B_3) =(1,1,1)$ as expected along with the additional solutions 
\[ (\B_1,\B_2,\B_3) =  (1,-2,1), \qquad  (1,0,-1), 
\]
which have corresponding positive eigenvalues
\[ \la = \frac{2^{3/2}3^{1/4}X^{1/4}}{\pi^{1/2}}G^{1/2}, \qquad   \la = \frac{2^{3/2}X^{1/4}}{\pi^{1/2}}G^{1/2}.\]

\paragraph{Four humps:}
With $-X_1=X_4=Y_2$, $-X_2=X_3=Y_1$, in addition to ${\bf \B}=(\B_1,\B_2,\B_3,\B_4)=(1,1,1,1)$ the solutions of  (\ref{Ghalfeigs}) are
\beqas
\mathbf{\B} & = & (-1,1,1,-1),\\
\mathbf{\B} & = & \left(1+r - (r^2+2r + 2)^{1/2}, 1,-1,-1-r + (r^2+2r + 2)^{1/2}\right),\\
\mathbf{\B} & = & \left(1+r + (r^2+2r + 2)^{1/2}, 1,-1,-1-r- (r^2+2
r + 2)^{1/2}\right),
\eeqas
where $r = Y_1/Y_2$, 
with corresponding positive eigenvalues
\beqas
\la & = & \frac{2^{7/4} Y_2^{1/4}}{\pi^{1/2}}G^{1/2},\\
\la & = & \frac{2^{3/2} }{\pi^{1/2}}\left(2Y_2+Y_1+ \left(Y_1^2+2 Y_1 Y_2
    + 2 Y_2^2\right)^{1/2}\right)^{1/4}G^{1/2} ,\\
\la & = &\frac{2^{3/2} }{\pi^{1/2}}\left(2Y_2+Y_1- \left(Y_1^2+2 Y_1 Y_2
    + 2 Y_2^2\right)^{1/2}\right)^{1/4}G^{1/2}.
\eeqas

\subsection{Perturbation of $f= V_0'$: eigenvalues of $O(G)$.}
\label{sec:Vp}
Let us anticipate that $\la = O(G)$ and write 
$\la = G\la_1$. We will see that these eigenvalues are much more
difficult to approximate, and that there is some subtlety in the
calculation which is not immediately apparent. In order to motivate
the rather detailed calculation which follows later, we first present 
a plausible  analysis along the same lines as
\S\ref{sec:iV0},  which will turn out to be incorrect.

\subsubsection{Inner expansion near  the $i$th hump}
With $\xi = X_i + x$, we try an expansion near the $i$th hump of the form
\[ f = \Aa_i \sum_{k=0}^\infty G^{k}f_k,\]
with 
 $f_0 = \fdd{V_0}{x}$. 
At $O(G)$ we have
\beqas
\eq{f_1}{f_1^*}
 &=& -\ii \la_1 \fdd{V_0}{x}.
\eeqas
The solvability conditions are automatically satisfied, $\la_1$ is
undetermined, and 
\[ f_1 = - \frac{\la_1 \ii x }{2}V_0.
\]
At $O(G^2)$
\beqa
\eq{f_2}{f^*_2}
 & = &- \frac{\la_1^2 x}{2}V_0 - \frac{ (x+X_i)^2}{4}\fdd{V_0}{x} - 20 V_1V_0^3\fdd{V_0}{x}.\label{VpG2}
\eeqa
The solvability condition (\ref{solv2}) is not automatic, and will
determine $\la_1$. We can simplify the right-hand side of
(\ref{solv2}) using (\ref{inner1}).
Differentiating (\ref{inner1}) gives
\[ \frac{\d^3V_1}{\d x^3} + 5 V_0^4 \fdd{V_1}{x} + 20 V_0^3 V_1
  \fdd{V_0}{x} - \fdd{V_1}{x} = -\frac{(x+X_i)}{2}V_0- \frac{(x+X_i)^2}{4}\fdd{V_0}{x}.\]
Thus
\beqas
\lefteqn{ \int^R_{-R}  \left(
    \frac{(x+X_i)^2}{4}\fdd{V_0}{x}+ 20 V_1
    V_0^3\fdd{V_0}{x}\right)\fdd{V_0}{x}\, \d x
  = 
  \int^R_{-R}  \left(-\frac{(x+X_i)}{2}V_0-
    \frac{\d^3 V_1}{\d x^3}- 5 V_0^4
  \fdd{V_1}{x} + \fdd{V_1}{x} 
\right)\fdd{V_0}{x}\, \d x}\qquad && \\
& = &
  \int^R_{-R}  -\frac{(x+X_i)}{2}V_0\fdd{V_0}{x} -\left(
    \frac{\d^3 V_0}{\d x^3}+ 5 V_0^4
  \fdd{V_0}{x} - \fdd{V_0}{x} 
  \right)\fdd{V_1}{x}\, \d x-
 \left[\sdd{V_1}{x}\fdd{V_0}{x} - \fdd{V_1}{x}\sdd{V_0}{x}
\right]^R_{-R}\\
& = &
  \int^R_{-R}  \frac{1}{4}V_0^2 \, \d x -
 \left[\sdd{V_1}{x}\fdd{V_0}{x} - \fdd{V_1}{x}\sdd{V_0}{x}
\right]^R_{-R}
\eeqas
Thus the solvability conditions on (\ref{VpG2}) may be written
\beqa
\im\left[\fdd{f_2}{x}V_0 - f_2 \fdd{V_0}{x}\right]^R_{-R} & =&  0
\label{sol1ajon},\\
\re \left[\fdd{f_2}{x}\fdd{V_0}{x} - f_2 \sdd{V_0}{x}\right]^R_{-R} & =&    \frac{(\la_1^2-1)
  }{4}\int_{-R}^R  V_0^2 \, \d x  +\left[\sdd{V_1}{x}\fdd{V_0}{x} - \fdd{V_1}{x}\sdd{V_0}{x}
\right]^R_{-R}.\qquad\label{sol2ajon}
\eeqa
To evaluate the left-hand we need to consider the region between the humps.

\subsubsection{In between the humps}
As usual consider the gap between $X_i$ and $X_{i+1}$.
In this region
\beq
\ii G\la_1 f+\sdd{ f}{\xi}
- f
+ \frac{G^2 \xi^2 }{4} f = O(f G^8 \log^4 G ),
\eeq
Expanding
\[ f = \sum_{k=0}^\infty G^{k} f_k,\] 
gives
\beq
\sdd{f_0}{\xi}
- f_0 = 0,
\eeq
at leading order.
Matching requires
\beqas
f &\sim& 12^{1/4} \Aa_{i+1} \ee^{\xi-X_{i+1}} \quad
\mbox{ as } \xi \ra X_{i+1}, \qquad 
f \sim -12^{1/4}   \Aa_{i}\ee^{-\xi+X_i} \quad \mbox{ as } \xi \ra X_i.
\eeqas
Thus
\[ f_0 = 12^{1/4} \Aa_{i+1} \ee^{\xi-X_{i+1}} -12^{1/4} \Aa_i \ee^{-\xi+X_i}.\]
In terms of $\xi = X_{i+1} + x$ this is
\[ f_0 = 12^{1/4} \Aa_{i+1} \ee^{x} -12^{1/4} \Aa_i \ee^{-x+X_i-X_{i+1}}.\]
Since $\ee^{X_i-X_{i+1}} = O(G^2)$ the additional term matches with $G^2 f_2$ from the $(i+1)$th hump, giving the matching condition
\[ \Aa_{i+1}f_2G^2 \sim- 12^{1/4} \Aa_i \ee^{-x+X_i-X_{i+1}},
\]
as $x \ra -\infty$ for the near-hump solution.
In terms of $\xi = X_{i} + x$ the between-hump solution is
\[ f_0 = 12^{1/4} \Aa_{i+1} \ee^{x+X_i-X_{i+1}} -12^{1/4} \Aa_i \ee^{-x}, \]
giving the matching condition
\[ \Aa_{i}f_2G^2 \sim 12^{1/4} \Aa_{i+1} \ee^{x+X_i-X_{i+1}},
\]
as $x \ra \infty$ for the near-hump solution.
Thus
\beqa
\lim_{R\ra\infty}\im\left[\fdd{f_2}{x}V_0 - f_2 \fdd{V_0}{x}\right]^R_{-R}  
& = &  0, \\
 \lim_{R\ra\infty}\re\left[\Aa_i\fdd{f_2}{x}\fdd{V_0}{x} - \Aa_i f_2 \sdd{V_0}{x}\right]^R_{-R} 
& = & -4 \sqrt{3} \Aa_{i+1} \frac{\ee^{-X_{i+1}+X_i}}{G^2}
-4 \sqrt{3} \Aa_{i-1} \frac{\ee^{-X_{i}+X_{i-1}}}{G^2}
,\\
\lim_{R\ra\infty}\left[\sdd{V_1}{x}\fdd{V_0}{x} - \fdd{V_1}{x}\sdd{V_0}{x}
\right]^R_{-R}
& = & -4 \sqrt{3}\frac{\ee^{-X_{i+1}+X_i}}{G^2}-4 \sqrt{3}\frac{\ee^{-X_{i}+X_{i-1}}}{G^2}.\label{V1xx}
\eeqa

\subsubsection{Solvability condition}

Letting $R\ra\infty$, the solvability condition (\ref{sol1ajon}) is satisfied, and,
noting that
\[ \int_{-\infty}^\infty   V_0^2\, \d x =\frac{\sqrt{3}\,
    \pi}{2},\]
  (\ref{sol2ajon})  becomes
\beq
 (\la_1^2-1)\frac{\pi}{32}\Aa_i  =  (\Aa_i-\Aa_{i+1})\frac{\ee^{X_i -
     X_{i+1}}}{G^2}+ (\Aa_i-\Aa_{i-1})\frac{\ee^{X_{i-1} -
     X_{i}}}{G^2} ,\qquad i = 1, \ldots,n.\label{Vpsol}
\eeq
Equations (\ref{Vpsol}) are a homogeneous set of linear equations for $\Aa_i$, $i = 1,\ldots,n$. For a non-trivial solution the determinant must vanish, which determines the eigenvalue $\la_1$.
Note that $\la_1=\pm 1$, $\Aa_i =\Aa$ for all $i$, is always a solution, corresponding to a global translation.
Note also that the eigenvalues again come in pairs---for every positive eigenvalue there is a negative eigenvalue of equal modulus.

\subsubsection{Examples}
\label{Vpexamples}
\paragraph{Two humps}

With $-X_1=X_2=X$ equation (\ref{Vpsol}) gives
\[ \la_1^2 = 1+  
 \left(1-\frac{\Aa_{2}}{\Aa_1}\right)X=
1+  
 \left(1-\frac{\Aa_{1}}{\Aa_2}\right)X.
\]
Thus $\Aa_1 = \pm \Aa_2$, and it appears there are solutions
\[  (\Aa_1,\Aa_2) = (1,1), \qquad (\Aa_1,\Aa_2) =(-1,1),\]
with corresponding positive eigenvalues 
\[ \la = G, \qquad \la = (1+2X)^{1/2} G.\]

\paragraph{Three humps}
For the three-bump case with $-X_1=X_3=X$, $X_2=0$, equation (\ref{Vpsol}) gives
\[
 \la_1^2 = 1+  
 \left(1-\frac{\Aa_{2}}{\Aa_1}\right) X= 1+  
 \left(1-\frac{\Aa_{3}}{\Aa_2}\right)X+
\left(1-\frac{\Aa_{1}}{\Aa_2}\right)X = 1+
 \left(1-\frac{\Aa_{2}}{\Aa_3}\right)X.
\]
Writing $(\Aa_1,\Aa_2,\Aa_3) = \mathbf{\Aa}$ the  solutions are
\[ \mathbf{\Aa} = (1,1,1), \qquad\mathbf{\Aa} = (1,0,-1) \qquad\mbox{ and } \qquad \mathbf{\Aa} = (1,-2,1),\]
with corresponding positive eigenvalues
\beq
\la=G,\qquad \la  =  (1+X)^{1/2}G,\qquad
\la  =  (1+3X)^{1/2}G.
\eeq

\paragraph{Four humps}
With  $-X_1=X_4=Y_2$, $-X_2=X_3=Y_1$, equation (\ref{Vpsol}) gives
\begin{multline*}
 \la_1^2 =  1+
 \left(1-\frac{\Aa_{2}}{\Aa_1}\right)Y_2 = 1+
\left(1-\frac{\Aa_{3}}{\Aa_2}\right)(Y_1+Y_2)+\left(1-\frac{\Aa_{1}}{\Aa_2}\right) Y_2
\\= 1+
\left(1-\frac{\Aa_{4}}{\Aa_3}\right)Y_2+
\left(1-\frac{\Aa_{2}}{\Aa_3}\right)(Y_1+Y_2)  
=  1+
\left(1-\frac{\Aa_{3}}{\Aa_4}\right)Y_2.
\end{multline*}
With $\mathbf{\Aa}=(\Aa_1,\Aa_2,\Aa_3,\Aa_4)$,  there are solutions
\beqas
\mathbf{\Aa}&=&(1,1,1,1),\\
\mathbf{\Aa} & = & (-1,1,1,-1),\\
\mathbf{\Aa} & = & \left(1+r - (r^2+2r + 2)^{1/2}, 1,-1,-1-r + (r^2+2r+
 2)^{1/2}\right),\\
\mathbf{\Aa} & = & \left(1+r + (r^2+2
r + 2)^{1/2}, 1,-1,-1-r - (r^2+2r + 2)^{1/2}\right),
\eeqas
where $r = Y_1/Y_2$,
with corresponding positive eigenvalues
\beqas
\la & = & G,\\
\la & = & (1+2 Y_2)^{1/2}G,\\
\la & = & \left(1+2Y_2+Y_1+ \left(Y_1^2+2 Y_1 Y_2
    + 2 Y_2^2\right)^{1/2}\right)^{1/2}G ,\\
\la & = &\left(1+2Y_2+Y_1- \left(Y_1^2+2 Y_1 Y_2
    + 2 Y_2^2\right)^{1/2}\right)^{1/2}G.
\eeqas

\begin{figure}[h!]
\centering
\begin{tabular}{ccc}
\includegraphics[width=0.33\textwidth]{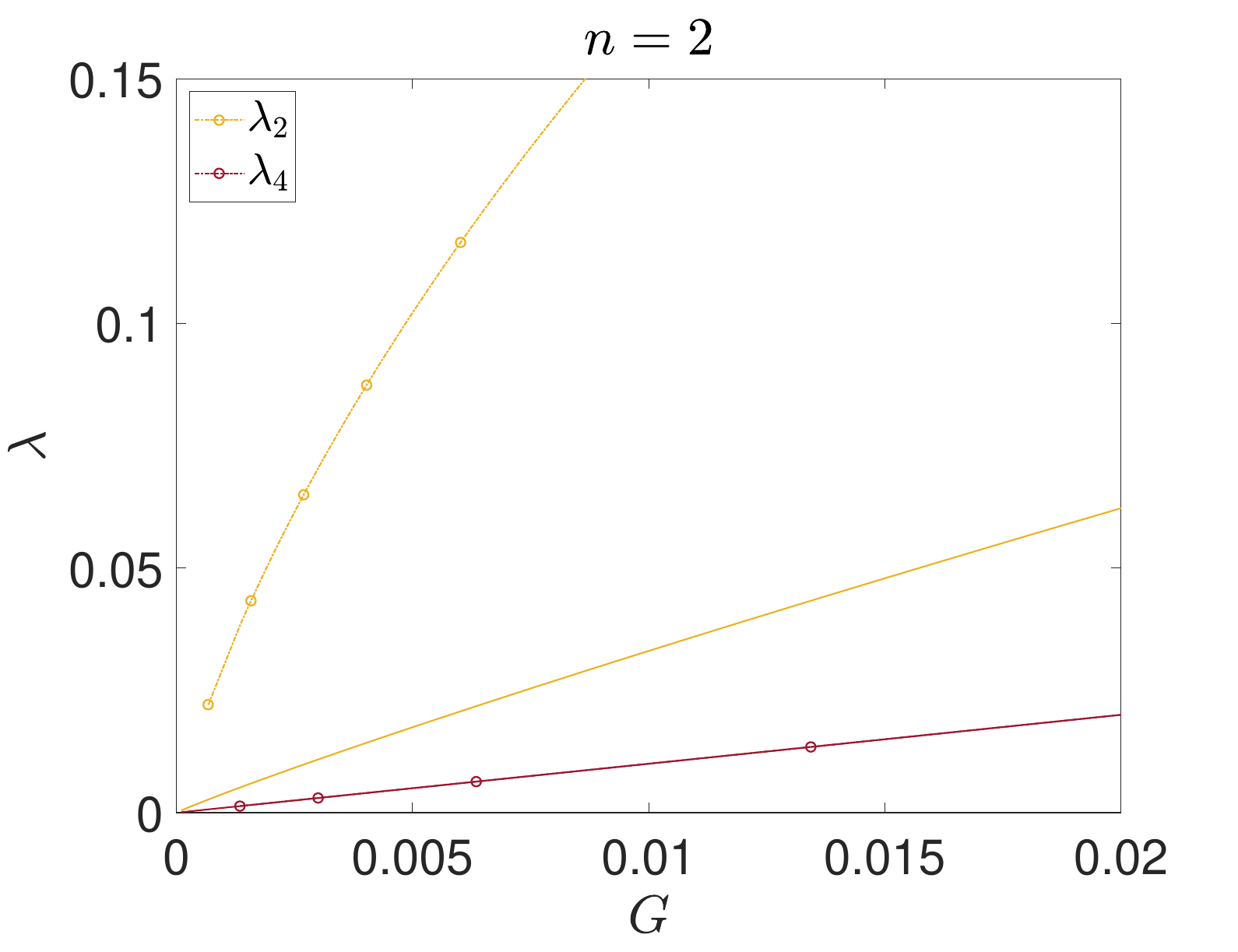}&\includegraphics[width=0.33\textwidth]{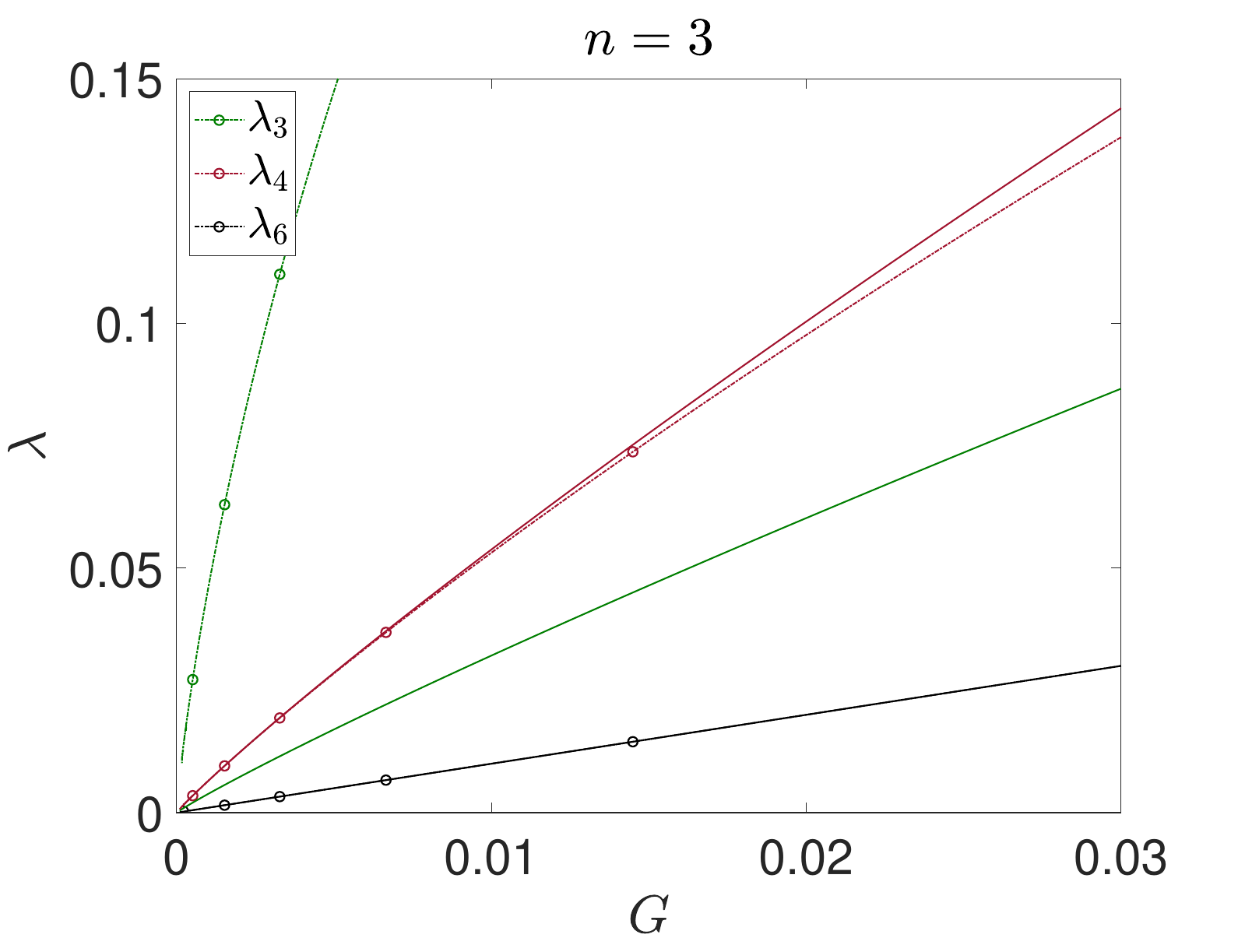} & \includegraphics[width=0.33\textwidth]{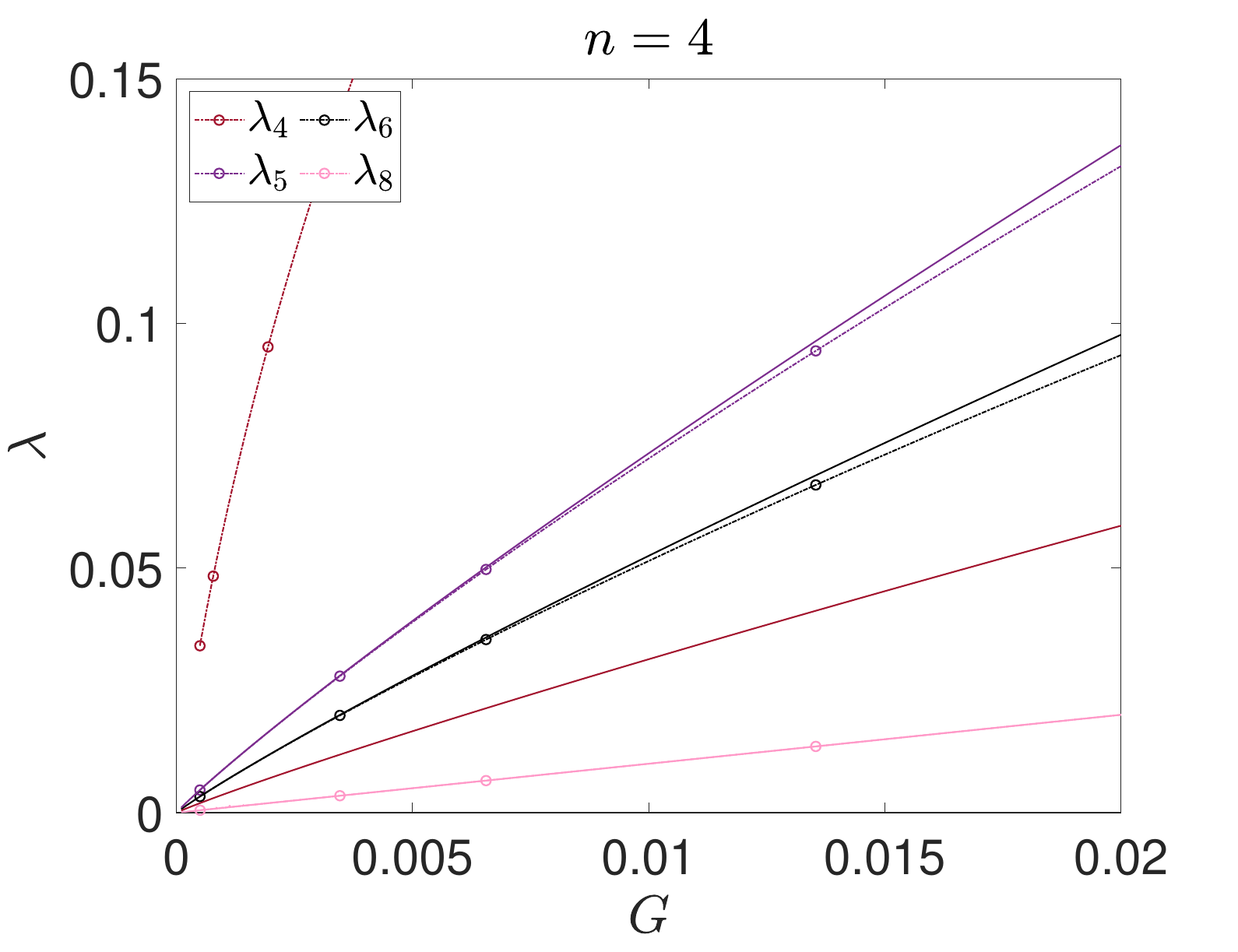}
  \\
(a)&(b)&(c)\\
\end{tabular}
\caption{Comparison of the eigenvalues proportional to $G$. The solid
  lines are the asymptotic predictions of (\ref{Vpsol}); the lines with circles correspond to the
numerically calculated values. 
  } \label{figG}

\end{figure}

In Fig. \ref{figG} we show a comparison between the predictions of equation (\ref{Vpsol}) and our numerical solutions.

We see that some of the eigenvalues are well approximated, but in each case there is a large eigenvalue that the asymptotic prediction misses, and there is a predicted eigenvalue which is not present numerically.
To understand why this is the case we have to return to our assumptions about the form of  the eigenfunctions.

\subsection{Why the naive asymptotic solution fails}
\label{sec:schematic}
Having determined that both $\ii V_0$ and $\fdd{V_0}{x}$  satisfy the leading-order problem when $\la$ is small, we should really consider a leading order solution in the near-hump region of the form
\beq
 f_0 = \Aa_i \fdd{V_0}{x} + \B_i \ii V_0,\label{combined}
 \eeq
i.e. we should combine both solutions rather than considering them separately. 
For the single-hump solution these eigenfunctions can be considered independently, but for the multi-hump solutions they interact with each other.
Let us suppose that $\la=\la_1G$ and proceed using (\ref{combined}) as the first term in the expansion near the $i$\,th hump. Then we would  find that 
there are solvability conditions at $O(G^2)$, which are of the form
\[ \mat{A(\la_1)}{0}{0}{B} \vtwo{{\bf \Aa}}{{\bf \B}} = \vtwo{{\bf
      0}}{{\bf 0}},\]
where ${\bf \Aa}$ and ${\bf \B}$ are vectors with components $\Aa_i$
and $\B_i$ respectively, $A$ is the matrix representing the linear equations (\ref{Vpsol}), and $B$ is the matrix representing the linear equations (\ref{Ghalfeigs}) but with $\la_1$ set to zero, i.e.
\beq 
 \left(\B_i-\B_{i+1}\right)\frac{\ee^{X_i - X_{i+1}}}{G^2}+
  \left(\B_i-\B_{i-1}\right)\frac{\ee^{X_{i-1} - X_{i}}}{G^2}=0.
\label{iVsol0}
 \eeq
In our analysis in \S\ref{sec:Vp} we set $\det(A)=0$ to determine the eigenvalue $\la_1$, and assumed that ${\bf \B}={\bf 0}$.
However, as we observed when writing down (\ref{Ghalfeigs}), the matrix $B$ is  singular, with null vector $(1,1,\ldots,1)$. Thus there is still a degree of freedom which is undetermined at this order, which will be determined at the next order in the expansions.
This  means that we can no longer ignore the fact that ${\bf \Aa}$,  ${\bf \B}$ and $\la$ should also be expanded in powers of $G$.

If we do so, following the scheme above, then we find that the leading-order solvability condition is 
\[ \mat{A(\la_1)}{0}{0}{B} \vtwo{{\bf \Aa}^{(0)}}{{\bf \B}^{(0)}} = \vtwo{{\bf
      0}}{{\bf 0}},\]
so that $\la_1$ is an eigenvalue of $A$, with ${\bf \Aa}^{(0)}$ the corresponding normalised eigenvector,  and ${\bf \B}^{(0)} = \B^{(0)}(1,1,\ldots,1)$ with $\B^{(0)}$ still undetermined.
Proceeding with the expansion we would find solvability conditions at the next order of the form
\beq \mat{A(\la_1)}{0}{0}{B} \vtwo{{\bf \Aa}^{(1)}}{{\bf \B}^{(1)}} = \vtwo{\la_2 C_1
    {\bf \Aa}^{(0)} + C_2{\bf \B}^{(0)} }{ C_3{\bf \Aa}^{(0)}},\label{blockeqn}
\eeq
for some matrices $C_1$, $C_2$, $C_3$. Note in particular that terms
involving ${\bf \B}^{(0)}$ appear on the right-hand side of the equation for ${\bf \Aa}^{(1)}$ while terms involving ${\bf \Aa}^{(0)}$  appear on the right-hand side  of the equation for  ${\bf \B}^{(1)}$.
Now both the matrix $A(\la_1)$ and the matrix $B$ are singular, so there are solvability conditions on the two right-hand sides. But both the degrees of freedom we have---$\la_2$ and $\B^{(0)}$---appear on the RHS of the equation for ${\bf \Aa}^{(1)}$. The solvability condition for $B {\bf \B}^{(1)} = C_3{\bf \Aa}^{(0)}$ may or may not be satisfied; we have no parameter in the equation to adjust to ensure that it is. This explains why the naive asymptotic approximation in \S\ref{sec:Vp} predicted solutions which were not found numerically---if we had continued to the next order in the expansion we would have found a solvability condition that we could not satisfy.

In order to solve the problem of both remaining degrees of freedom appearing in the same block of (\ref{blockeqn}) we need to make ${\bf \B}$ an order of $G$ larger than ${\bf \Aa}$. Suppose, therefore, that ${\bf \Aa}^{(0)} = {\bf 0}$.  
Then the leading-order solvability conditions give
\beq
B {\bf \B}^{(0)} = {\bf 0},\label{Bb0eq0}
\eeq
so that ${\bf \B}^{(0)} = b^{(0)}(1,1,\ldots,1)$.
At next order we will find
\beq
\mat{A(\la_1)}{0}{0}{B} \vtwo{{\bf \Aa}^{(1)}}{{\bf \B}^{(1)}} = \vtwo{C_2{\bf \B}^{(0)}}{ 0}.\label{Aa1eqb0}
\eeq
Since ${\bf \B}^{(0)}$ is non-zero there is a non-trivial solution ${\bf \Aa}^{(1)} = A^{-1}C_2 {\bf \B}^{(0)}$ when $A$ is non-singular. Thus $\la_1$ is still undetermined, but the relative scaling of ${\bf \Aa}^{(1)}$ and ${\bf \B}^{(0)}$ is determined.
Now, at next order, we will find
\beq
\mat{A(\la_1)}{0}{0}{B} \vtwo{{\bf \Aa}^{(2)}}{{\bf \B}^{(2)}} = \vtwo{\la_2
 C_1 {\bf \Aa}^{(1)} + C_2{\bf \B}^{(1)}}{C_3 {\bf \Aa}^{(1)} + C_4{\bf \B}^{(0)} }.\label{Bb2eqa1}
\eeq
Crucially, since ${\bf \Aa}^{(0)}$ is a function of $\la_1$, the solvability condition on the second block can be satisfied and determines the eigenvalue 
 $\la_1$.
Now that we know schematically the structure of the expansions, let us work through some of the details.

\subsection{More careful analysis of the eigenvalues $\la = O(G)$.}
We suppose that 
\beq
\la = \sum_{k=1}^\infty \la_k G^k.\label{exp1}
\eeq
Near the $i$\,th hump, with $\xi = X_i + x$, we set
\beq f = a_i U + b_i W\eeq
where
\beq f = \sum_{k=0}^\infty G^k f_k, \qquad
  U = \sum_{k=0}^\infty G^k U_k, \qquad
 W = \sum_{k=0}^\infty G^k W_k,\eeq
with  
\beq U_0 = \fdd{V_0}{x} , \qquad
  W_0 =  \ii V_0 .\eeq
We also expand $a_i$ and $b_i$ in powers of $G$, anticipating the interleaving expansions,
\beq a_i = G a_i^{(1)} + G^3 a_i^{(3)} + \cdots,
  \qquad b_i = b_i^{(0)} + G^2 b_i^{(2)} + \cdots.\label{exp2}
\eeq
Since the analysis gets now very technical, and we need to match many
orders of inner and outer expansions, we defer many of the details to
an appendix, and here simply pick out some of the key equations.
At $O(G^2)$ in the inner expansion near the $i$\,th hump we find the
solvability condition (\ref{Bb0}),
\beqa
b_i^{(0)}\left(\frac{4 \sqrt{3}}{G^2}\ee^{-X_{i+1}+X_i}+\frac{4
  \sqrt{3}}{G^2}\ee^{-X_{i}+X_{i-1}}
\right) & = & \frac{4\sqrt{3}\,b_{i+1}^{(0)}}{G^2}\ee^{X_i-X_{i+1}}
+\frac{4\sqrt{3}\,b_{i-1}^{(0)}}{G^2}\ee^{X_{i-1}-X_{i}}.\label{Bb0_main}
\eeqa
As anticipated, this is (\ref{Ghalfeigs}) with $\la_1$ set to zero, and is  equation (\ref{Bb0eq0}),  $B {\bf \B}^{(0)} = {\bf 0}$, from \S\ref{sec:schematic}.
Equation (\ref{Bb0_main}) fixes $b_i^{(0)} = b^{(0)}$ for all $i$.

At $O(G^3)$ in the inner expansion near the $i$\,th hump we find the
solvability condition (\ref{solnew2a}),
\begin{multline}
 \frac{\pi a_i^{(1)}(\la_1^2-1)}{32}   + (a_{i+1}^{(1)}- a_i^{(1)})\frac{\ee^{X_i-X_{i+1}}}{G^2}
+ (a_{i-1}^{(1)}- a_i^{(1)}) \frac{\ee^{ - X_{i}+X_{i-1}}}{G^2}
\\
=  \frac{3 \pi}{64} \la_1 b^{(0)}X_i   
- \frac{\la_1  b^{(0)}}{2}(X_i-X_{i+1})\frac{\ee^{X_i-X_{i+1}}}{G^2}
- \frac{\la_1b^{(0)}}{2} (X_{i}-X_{i-1}) \frac{\ee^{ - X_{i}+X_{i-1}}}{G^2}. 
\label{solnew2a_main}
\end{multline}

Equation (\ref{solnew2a_main}) is  equation (\ref{Aa1eqb0}), $A(\la_1) {\bf \Aa}^{(1)} = C_2{\bf \B}^{(0)}$, from \S\ref{sec:schematic}. 
When $b^{(0)}$ is non-zero it gives
 $a_i^{(1)}$ in terms of $b^{(0)}$, with no restriction on $\la_1$.
To determine $\la_1$ we need to go to one more order.
Note that there is still the possibility that $b^{(0)}=0$, in which case $\la_1$ is determined by $\det A=0$ and ${\bf \Aa}^{(1)}$ is a null vector of $A$.

At $O(G^4)$ in the inner expansion near the $i$\,th hump we find the
solvability condition (\ref{finalsol}),
\begin{multline}
\frac{ b^{(0)} \la_1^2(\la_1^2-4) \pi^3}{512} +  \frac{3\pi \la_1 X_i}{64} ( b^{(0)}\la_1  X_i-2a_{i}^{(1)}) 
+4  (b_{i+1}^{(2)} - b_i^{(2)} ) \frac{\ee^{X_i - X_{i+1}}}{G^2}
\\+4  (b_{i-1}^{(2)} - b_i^{(2)} ) \frac{\ee^{X_{i-1} - X_{i}}}{G^2}
+2  a_{i+1}^{(1)}(X_{i+1} -X_i) \la_1\frac{\ee^{X_i - X_{i+1}}}{G^2}
+2  a_{i-1}^{(1)}(X_{i-1} -X_i)\la_1\frac{\ee^{X_{i-1} - X_{i}}}{G^2}
\\ -3 a_{i+1}^{(1)} \la_1\frac{\ee^{X_i - X_{i+1}}}{G^2}
+ 3 a_{i-1}^{(1)}\la_1\frac{\ee^{X_{i-1} - X_{i}}}{G^2}
+ \frac{\la_1^2}{2}b^{(0)}\frac{\ee^{X_i - X_{i+1}}}{G^2}\left(
 -  X_i^2  + 3  X_{i+1} + 2 X_i X_{i+1} 
 -  X_{i+1}^2 \right)
\\+ \frac{\la_1^2}{2}b^{(0)}\frac{\ee^{X_{i-1} - X_{i}}}{G^2}\left(
  -  X_i^2 - 3  X_{i-1} + 2 X_i X_{i-1}  
 -  X_{i-1}^2 
\right)=0.\label{finalsol_main}
\end{multline}
This is equation (\ref{Bb2eqa1}), $B {\bf \B}^{(2)} =C_3 {\bf \Aa}^{(1)}+ C_4{\bf \B}^{(0)}$, from \S\ref{sec:schematic}.
Since $B$ has the null vector $(1,1,\ldots,1)$ these equations for $b_i^{(2)}$ have a solvability condition: if we sum over $i$ then $b_i^{(2)}$ vanishes. This sum over $i$ is the equation which determines $\la_1$.

Let us first identify which of the eigenvalues of \S\ref{sec:Vp}
survive the new solvability condition (\ref{finalsol_main}).

\subsubsection{Eigenvalues which follow the old scheme of \S\ref{sec:Vp}}
We found some numerically determined eigenvalues did match the predictions of 
\S\ref{sec:Vp}. If $b^{(0)}=0$ then the eigenvalue is determined by
requiring a non-trivial solution of (\ref{solnew2a_main}),
\beq
 \frac{\pi a_i^{(1)}(\la_1^2-1)}{32}   + (a_{i+1}^{(1)}- a_i^{(1)})\frac{\ee^{X_i-X_{i+1}}}{G^2}
+ (a_{i-1}^{(1)}- a_i^{(1)}) \frac{\ee^{ - X_{i}+X_{i-1}}}{G^2}
= 0,\label{origeiga_main}
\eeq
as in \S\ref{sec:Vp}, but the extra solvability condition given by \eqref{finalsol_main} must also be satisfied, which is 
\begin{multline}
\sum_{i=1}^n - \frac{3\pi \la_1 X_i}{32} a_{i}^{(1)} 
+2  a_{i+1}^{(1)}(X_{i+1} -X_i) \la_1\frac{\ee^{X_i - X_{i+1}}}{G^2}
\\+2  a_{i-1}^{(1)}(X_{i-1} -X_i)\la_1\frac{\ee^{X_{i-1} - X_{i}}}{G^2}
 -3 a_{i+1}^{(1)} \la_1\frac{\ee^{X_i - X_{i+1}}}{G^2}
+ 3 a_{i-1}^{(1)}\la_1\frac{\ee^{X_{i-1} - X_{i}}}{G^2}
=0.\label{extra_main}
\end{multline}
In general we would not expect this to be satisfied, but certain symmetries of the $a_i^{(1)}$ may mean it is satisfied automatically.
In particular, with some index relabelling, (\ref{extra_main}) is
\begin{multline*}
\sum_{i=1}^n   -  \frac{3\pi  X_i}{32} a_{i}^{(1)}
+2  (a_{i}^{(1)}-a_{i-1}^{(1)})X_{i} \frac{\ee^{X_{i-1} - X_{i}}}{G^2}
\\
+2 (a_{i}^{(1)}-a_{i+1}^{(1)})X_{i} \frac{\ee^{X_{i} - X_{i+1}}}{G^2}
 -3 a_{i}^{(1)}\frac{\ee^{X_{i-1} - X_{i}}}{G^2}
+ 3 a_{i}^{(1)}\frac{\ee^{X_{i} - X_{i+1}}}{G^2}
=0.
\end{multline*}
Simplifying using (\ref{origeiga_main}) and (\ref{Xeqns}) gives
\beq
\sum_{i=1}^n   - 3 X_i a_{i}^{(1)}
+  (\la_1^2-1)a_i^{(1)}
=0.\label{additional}
\eeq
Since, by symmetry, $\sum_{i=1}^n X_i=0$ we find the eigenvalue
$\la_1=\pm 1$ with eigenvector $a_i^{(1)}=1$ for $i=1,\ldots,n$ always
works as it should. 

\paragraph{Two humps}

With $-X_1=X_2=X$ we  found
\[  \mathbf{\Aa} = (-1,1), \qquad \mathbf{\Aa} =(1,1),\]
with corresponding positive eigenvalues 
\[ \la = (1+2X)^{1/2}G, \qquad \la =  G.\]
The second satisfies (\ref{additional}) but the first does not, in agreement with Fig.~\ref{figG}(a).

\paragraph{Three humps}
For the three-bump case with $-X_1=X_3=X$, $X_2=0$,
we found
\[ \mathbf{\Aa} = (1,-2,1), \qquad\mathbf{\Aa} = (1,0,-1) \qquad\mbox{ and } \qquad \mathbf{\Aa} = (1,1,1),\]
with corresponding positive eigenvalues 
\beq
\la=(1+3X)^{1/2}G,\qquad \la  =  (1+X)^{1/2}G,\qquad
\la  =  G.
\eeq
The first and third satsify (\ref{additional}), but the second does not, in agreement with  Fig.~\ref{figG}(b).

\paragraph{Four humps}
With  $-X_1=X_4=Y_2$, $-X_2=X_3=Y_1$, we found
\beqas
\mathbf{\Aa} & = & \left(1+r - (r^2+2r + 2)^{1/2}, 1,-1,-1-r + (r^2+2r+
 2)^{1/2}\right),\\
\mathbf{\Aa} & = & (-1,1,1,-1),\\
\mathbf{\Aa} & = & \left(1+r + (r^2+2
r + 2)^{1/2}, 1,-1,-1-r - (r^2+2r + 2)^{1/2}\right),\\
\mathbf{\Aa}&=&(1,1,1,1),
\eeqas
where $r = Y_1/Y_2$,
with corresponding positive eigenvalues
\beqas
\la & = & \left(1+2Y_2+Y_1+ \left(Y_1^2+2 Y_1 Y_2
    + 2 Y_2^2\right)^{1/2}\right)^{1/2}G ,\\
\la & = & (1+2 Y_2)^{1/2}G,\\
\la & = &\left(1+2Y_2+Y_1- \left(Y_1^2+2 Y_1 Y_2
    + 2 Y_2^2\right)^{1/2}\right)^{1/2}G,\\
\la & = & G.
\eeqas
The  second and fourth satisfy (\ref{additional}), but the first and third do  not.
It appears in Fig.~\ref{figG}(c) that we correctly predicted the first  eigenvalue also. However, this was merely a coincidence, and the first eigenfunction in fact follows the new scheme with nonzero $b^{(0)}$.

\subsubsection{Eigenvalues which follow the new scheme}

\paragraph{Two humps}
With $-X_1=X_2=X$ the sum over $i$ of (\ref{finalsol}) is
\beq
\frac{\la_1^2(\la_1^2-4)\pi(\pi^2(\la_1^2-1-2X) + 8 X^2(3-4X))}
{256(\la_1^2-1-2X)}=0.
\eeq
We see the positive eigenvalue $\la_1=2$, and also a singularity at the false eigenvalue $\la_1=(1+2X)^{1/2}$ of \S\ref{sec:Vp}. There is one other positive eigenvalue given by
\beq
\la_1 = \left(1+2X + \frac{8 X^2(4X-3)}{\pi^2}\right)^{1/2}. 
\eeq
The corresponding eigenvector is
\[ {\bf b}^{(0)} = (1,1), \qquad {\bf a}^{(1)} =
\frac{\la_1 X(2X-3)}{2(\la_1^2-1-2X)}
    (1,-1).\]

\paragraph{Three humps}
With $-X_1=X_3=X$, $X_2=0$, the sum over $i$ of (\ref{finalsol}) is
\beq
\frac{\la_1^2(\la_1^2-4)\pi(3\pi^2(\la_1^2-1-X) + 16 X^2(3-2X))}
{512(\la_1^2-1-X)}=0.
\eeq
We see the positive eigenvalue $\la_1=2$, and also a singularity at the false eigenvalue $\la_1=(1+X)^{1/2}$ of \S\ref{sec:Vp}. There is one other positive eigenvalue given by
\beq
\la_1 = \left(1+X + \frac{16 X^2(2X-3)}{3\pi^2}\right)^{1/2}. 
\eeq
The corresponding eigenvector is
\[ {\bf b}^{(0)} = (1,1,1), \qquad {\bf a}^{(1)} =
\frac{\la_1 X(X-3)}{2(\la_1^2 -1 -X)}
    (1,0,-1).\]

\paragraph{Four humps}
With $-X_1=X_4=Y_2$, $-X_2=X_3=Y_1$, the sum over $i$ of  (\ref{finalsol})
has two  positive solutions in addition to $\la_1=2$.

\section{Conclusions and Future Challenges}
\label{sec:conclusions}

In the present work we have revisited the topic of multi-pulse
collapsing solutions in the context of the one-dimensional
nonlinear Schr{\"o}dinger equation with a power-law nonlinearity. Returning to the results of~\cite{budd:1999}
from a novel perspective, we have showcased not only in
a qualitative, but also in a quantitative way, how 
{\it countably infinitely many} branches of such 
solutions bifurcate from the critical point, along
with their most well-known cousin, namely the stationary
single collapse point waveform. We have shown that all
of these solutions emerge from an infinite distance between
the pulses in an example of ``bifurcation from infinity''
with the distance between the pulses shrinking as the 
growth rate of the collapse increases. We have
provided quantitative characterizations both for
the bifurcation curve of the relevant waveforms for
the blowup rate $G$ as a function of $\sigma$, as well
as for the separation between their peaks as a function 
of $G$. Importantly, we have also tackled both qualitatively
and quantitatively the spectral stability analysis of
the relevant states. Indeed, we have found all of them 
to be unstable, bearing $n-1$ pairs of eigenvalues of
practically of size $G^{1/2}$, another $n+1$ pairs of 
size practically proportional to $G$, while $n-1$
pairs emerge on the imaginary axis, and one remains
in the vicinity of the origin. We have verified the relevant
count for cases of $n=1$ to $n=4$ and indeed examined
cases even up to $n=6$. While some of the O$(G^{1/2})$
eigenvalues were less accurately captured when $G$ was large,
generally our predictions were found to be in 
excellent qualitative and good quantitative agreement
with the numerical observations. Finally, some select
dynamical simulations gave us the opportunity to 
observe the symmetry breaking events leading to
the instability of the multi-pulse configurations
and the eventual predominance of a single pulse,
in line with the expectations associated with stability
based on our
eigenvector analysis.

\begin{figure}[!pt]  
\centering
\includegraphics[width=0.5\textwidth]{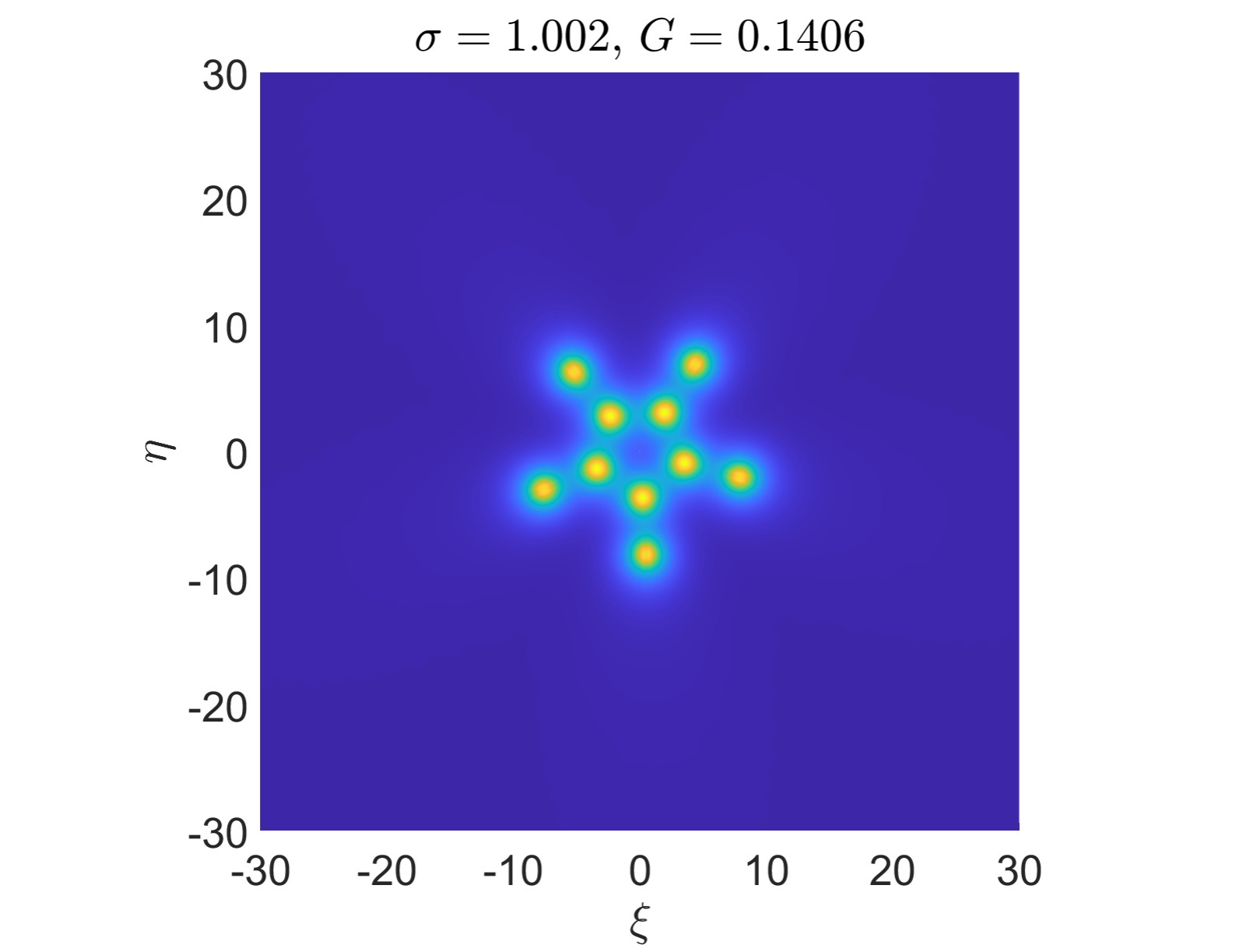}
\caption{Star-shaped self-similar profile, $|v(\xi,\eta)|$, for the 2D NLS equation. 
Here, $\sigma=1.002$ and the computed blow-up rate is $G=0.1406$.
}
\label{fig:star_shaped}
\end{figure}

Naturally, this study paves the way for a number of
directions of future study.
Arguably, the most canonical of these concerns whether
such multi-bump collapsing solutions can be systematically
found in higher dimensions. 
Our preliminary considerations along this vein suggest
the existence of a wide range of possible
higher-dimensional
waveforms. 
For instance, in Fig.~\ref{fig:star_shaped}, we illustrate a 
prototypical star-shaped self-similar waveform computed in the rescaled $(\xi,\eta) \equiv (x/L, y/L)$ space for the 2D NLS equation. Generalizations of the analytical
considerations herein enable the consideration of
multiple collapse ``spots'' as an intriguing
interacting particle system in the higher-dimensional
setting meriting investigation (in terms of its equilibria)
in its own right.
These can then potentially be associated with the recent
experiments of~\cite{banerjee2024collapse} and offer
a systematic explanation for the observations therein.
These can also potentially be connected
to the stability analysis of the {\em ring type solutions}
that have been discovered in the work of~\cite{FIBICH200755,BARUCH20101968}. It would be 
particularly interesting to explore the spectral stability of
the ring in terms of its eigenvalues in the co-exploding
frame and the conditions under which these may potentially
give rise to multi-pulse configurations as the parameters
of the system (such as the nonlinearity exponent) may vary.%
~From a computational perspective, and especially for such higher- dimensional PDEs, domain decomposition methods~\cite{doi:10.1137/1.9781611974065,widlund_book} together with the parallelization features provided by the Portable, Extensible Toolkit for Scientific Computation (PETSc)~\cite{petsc} already embedded in FreeFEM~\cite{SADAKA2025109378} (and references therein) have provided us with the necessary computational environment to efficiently and reliably study these types of problems. 

It is also relevant to point out here that recent experimental
developments in optics have enabled the realization of
fractional derivatives (in the form of derivatives
of the Riesz type) which, in turn, allow for the manipulation
of dispersion~\cite{hoang2024observationfractionalevolutionnonlinear}. Considering the  fractional Laplacian
operator $(-\Delta)^{\alpha/2}$, with the L{\'e}vy index 
$0 < \alpha < 2$, this experimental realization has 
made considerations in the vicinity of $\alpha=1$ possible.
This constitutes (even for the cubic nonlinearity) a critical
point for transition to collapse-type solutions and hence
represents an additional area of possible future research
for the waveforms and phenomena identified herein.~Such
studies are currently underway and the reporting of
associated findings is deferred to future publications.

\section*{Acknowledgements}
This work has been supported by the U.S. National Science Foundation
under Grants DMS-2204782 (EGC), and PHY-2110030, PHY-2408988 and DMS-2204702
(PGK), and by the US National Science Foundation (IGK). ~EGC expresses his gratitude to Prof. Hannes Uecker (University of Oldenburg) for fruitful discussions related to the finite-element implementation of phase conditions in \texttt{pde2path}~\cite{uecker_book}.~Also, EGC thanks Profs. Christopher Douglas (Duke), Pierre Jolivet (Sorbonne Universit\'e), Georges Sadaka (University of Rouen Normandie) for their continuing 
support and encouragement on technical aspects on \texttt{FreeFEM}.

\bibliographystyle{apsrev4-1}
\bibliography{bibliography}

\appendix
\renewcommand{\thefigure}{A\arabic{figure}}

\setcounter{figure}{0}

\section{Comparison with finite element computations}
\label{app:num_comp}
\begin{figure}[h!]
\centering
\begin{tabular}{cc}
\includegraphics[width=0.49\textwidth]{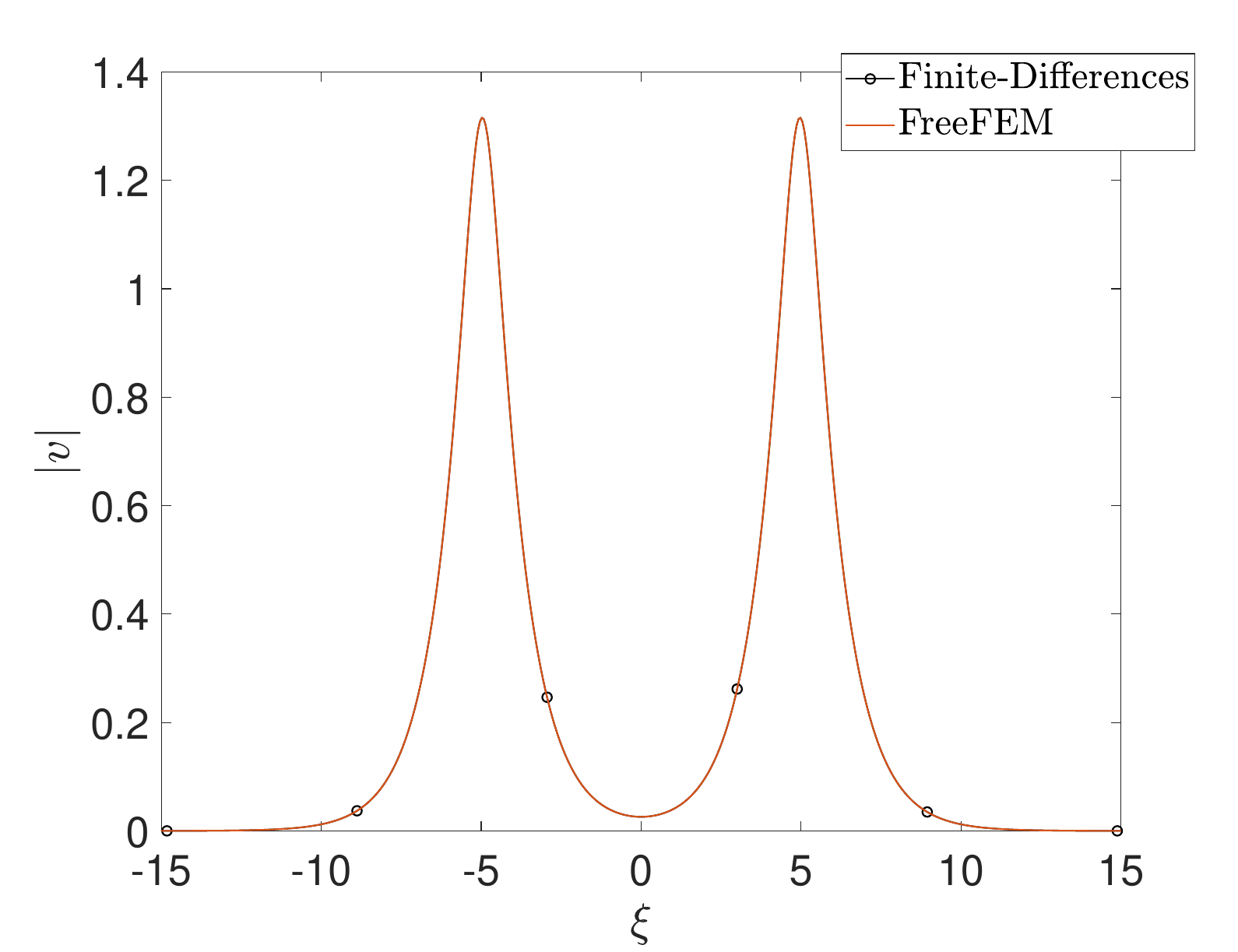}  &
\includegraphics[width=0.49\textwidth]{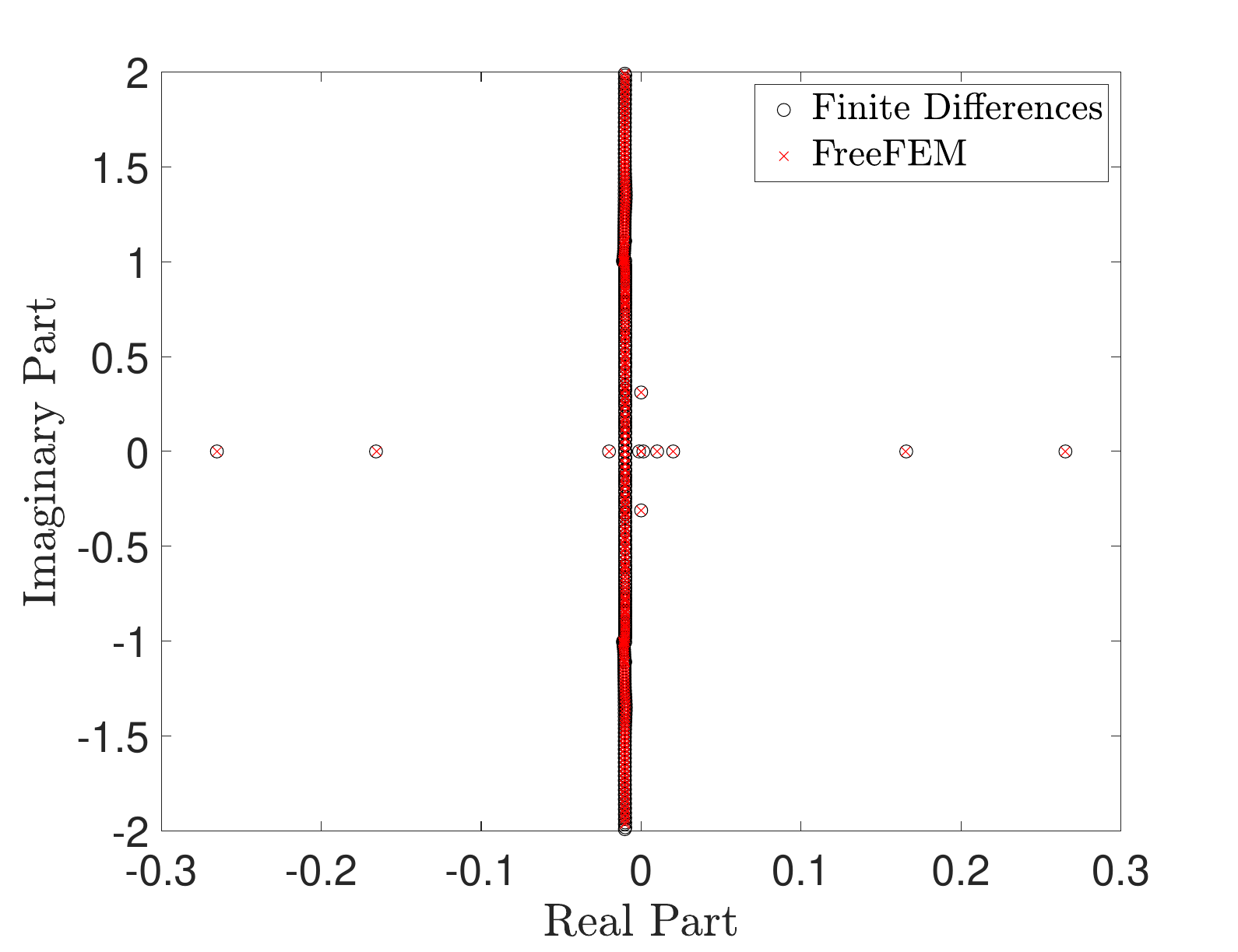} \\
(a) & (b)
\end{tabular}
\caption{
  (a) Finite-difference (black line with open circles) and finite-element (red line) two-humped solutions with a blow-up rate of $G=0.01$. 
  The two solutions are visually indistinguishable.
  The computational domain $\xi \in [-300,300]$ is discretized using equidistant nodes with a spacing of $\Delta \xi=0.06$ (10,000 nodes). 
 (b) Spectra comparison between the finite-difference discretization and MATLAB's \texttt{eigs} function (black circles) and \texttt{FreeFEM}'s ARPACK eigensolver (red crosses).
  } \label{fig:FD_FEM_comparison}

\end{figure}

In this section, we compare numerical computations presented in the main text of our manuscript with those obtained using the open-source, finite-element software, \texttt{FreeFEM}~\cite{freefem}.
Figure~\ref{fig:FD_FEM_comparison} shows a comparison of two-humped solutions computed numerically with finite differences and P3 finite elements.
The blow-up rate is $G=0.01$ and the computational domain: $\xi \in [-300,300]$.
Both solutions appear visually indistinguishable. 
For the finite-difference computation, the domain is discretized using equidistant nodes with a nodal spacing of $\Delta \xi=0.06$ (i.e., $10,000$ points) whereas for the finite-element, $10,000$ nodes were considered on $[-300,300]$.
In addition to the solution comparison, we also present the spectra computed using MATLAB's \texttt{eigs} function and \texttt{FreeFEM}'s eigensolver (ARPACK).
This comparison demonstrates that both eigensolvers yield nearly identical eigenvalues.

\section{Full domain dynamical simulations}
\label{app:fullMN}
\begin{figure}[H]
\centering
\begin{tabular}{c}
\includegraphics[width=0.75\textwidth]{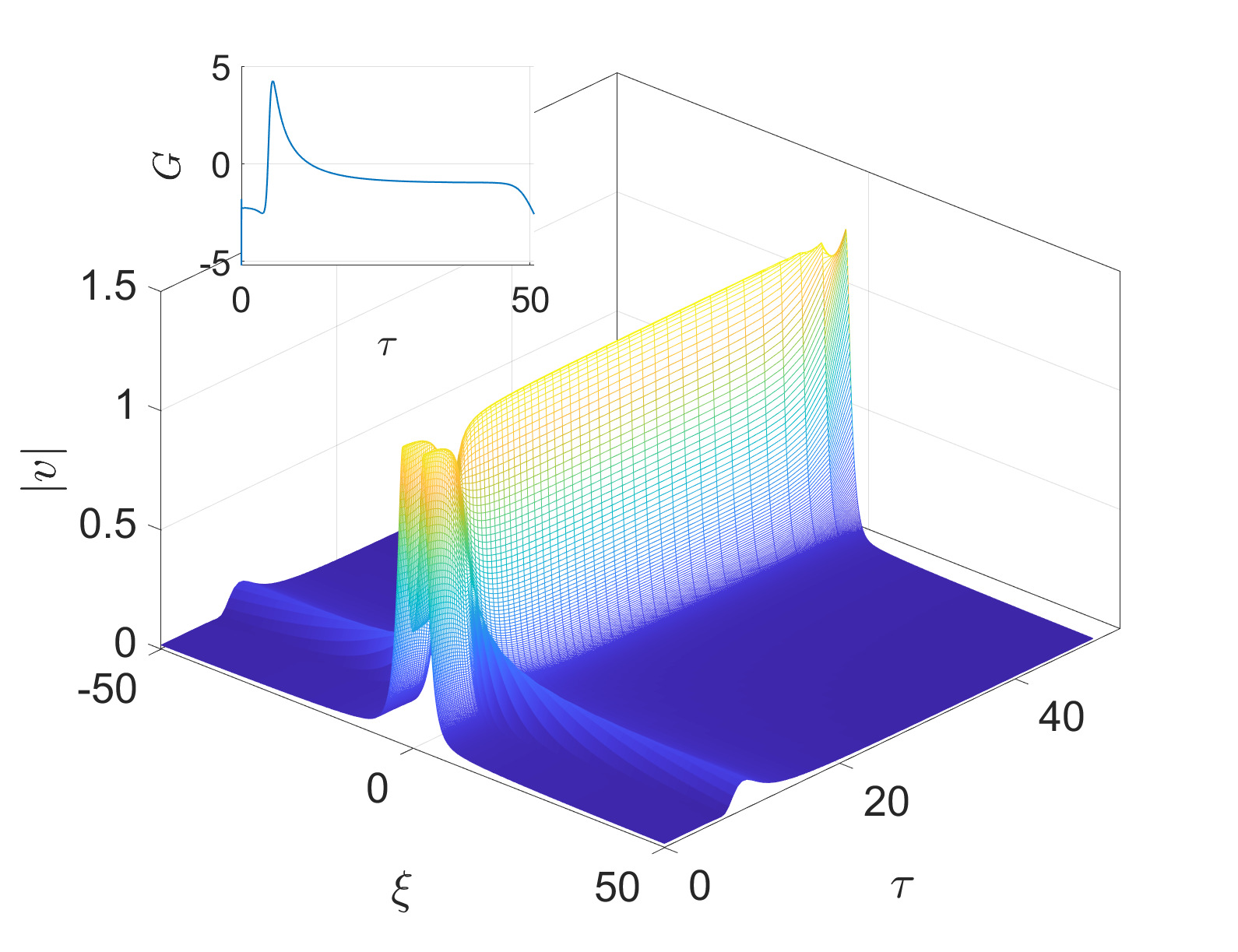}  \\
(a)\\
\includegraphics[width=0.75\textwidth]{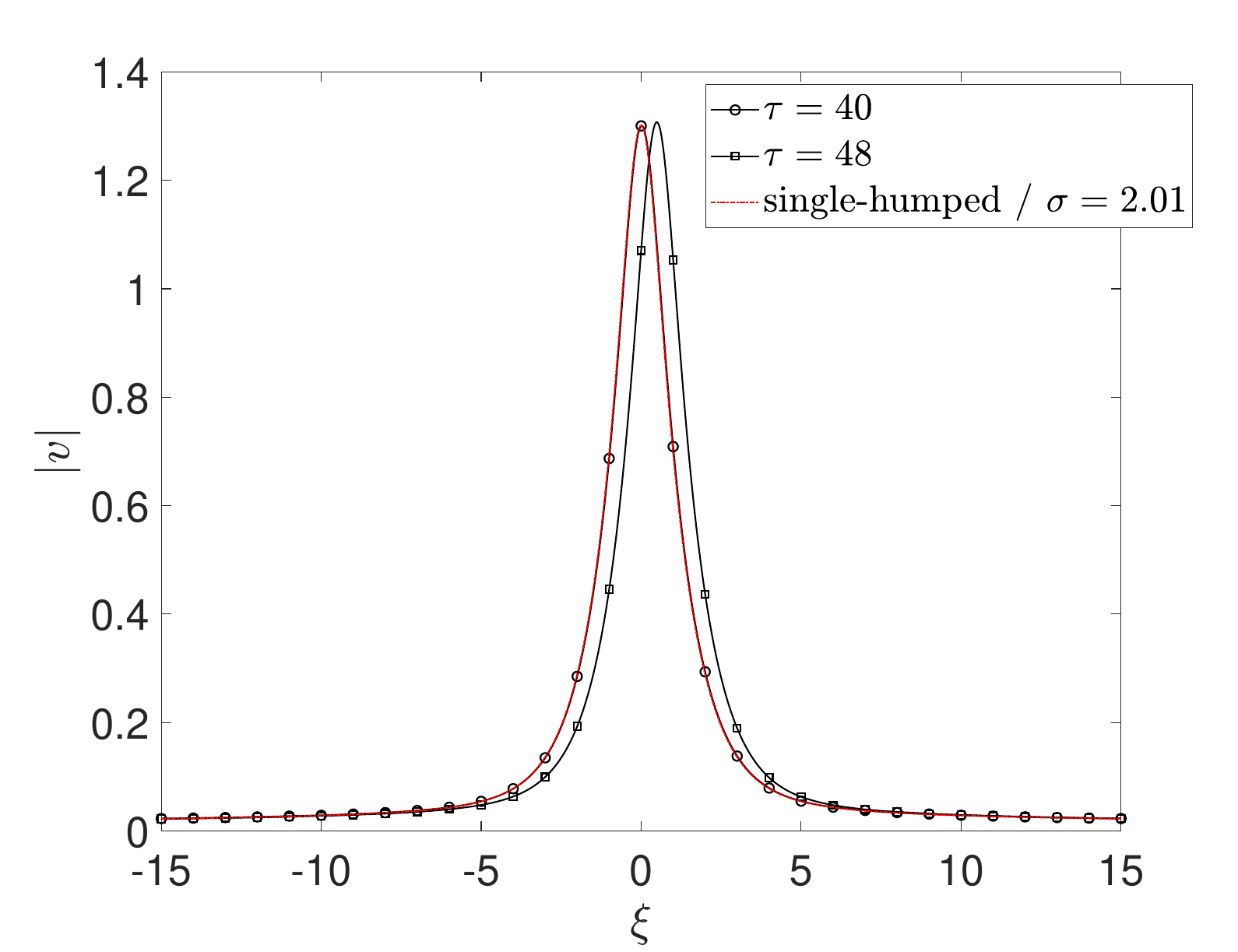}  \\
(b)
\end{tabular}
\caption{
  (a) Dynamics in the renormalized/rescaled spatio-temporal framework $\xi-\tau$: Starting from a two-humped solution at $\sigma=2.01$ and perturbing along the eigenfunction that corresponds to the second largest real eigenvalue the solution evolves toward the corresponding single-humped solution after an initial ``restructuring'' $\tau$-period.
  Convergence is lost at around $\tau=50$.
  The inset illustrates the evolution of blow-up rate, $G$.
 (b) To enhance visualization of the shift from the origin at later stages of the simulation, we plot two snapshots: one at $\tau=40$ (solid line with open circles), where the solution is visually indistinguishable from the single-humped self-similar solution (red dashed line), and one at $\tau=48$, where the solution has visibly shifted from the origin.
 $\xi \in [-50,50]$, nodal distance: $\delta x = 0.001$, and $dt=0.01$.
  } \label{fig:fulldomain_single}

\end{figure}

\begin{figure}{H}
\centering
\begin{tabular}{cc}
\includegraphics[width=0.49\textwidth]{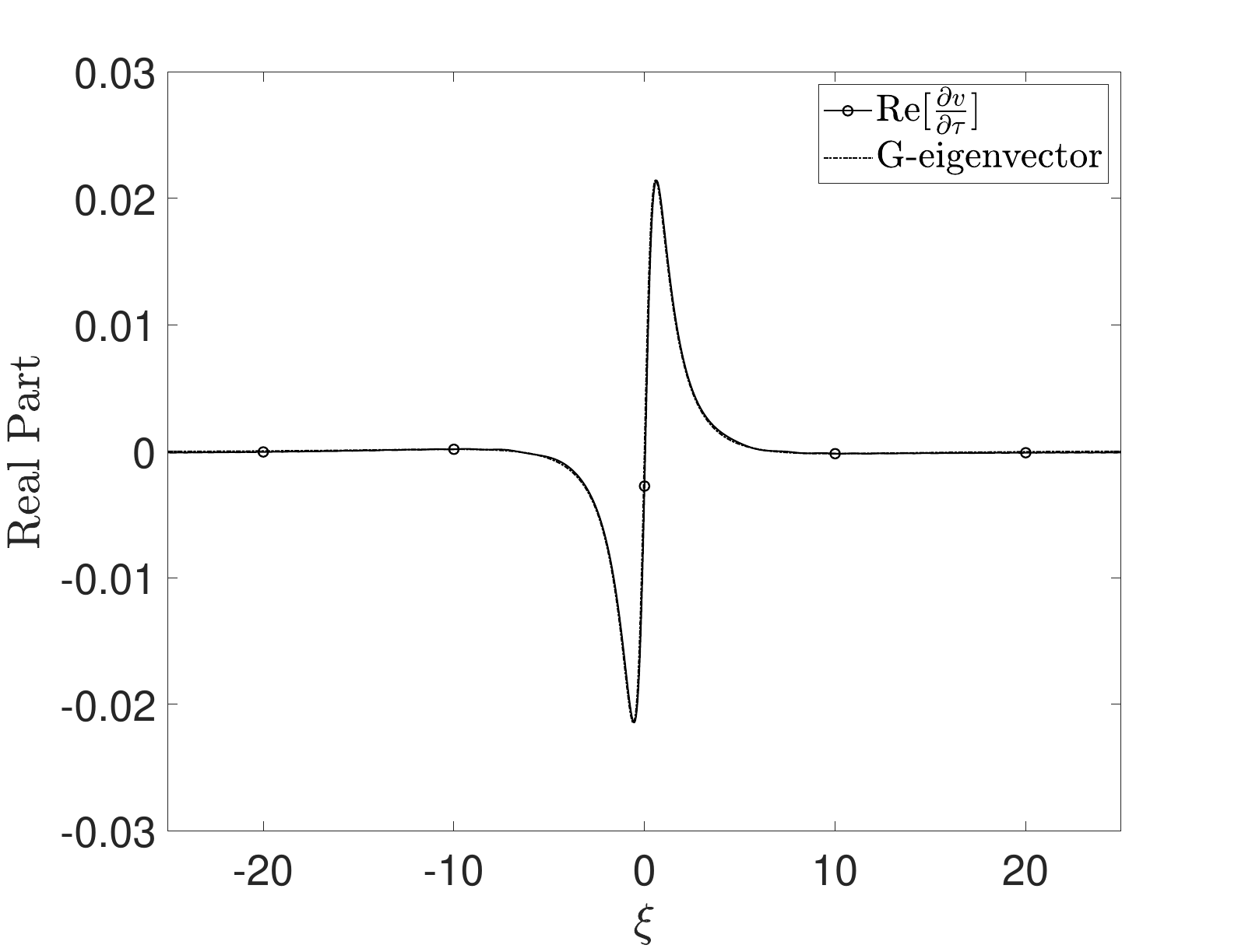}&\includegraphics[width=0.49\textwidth]{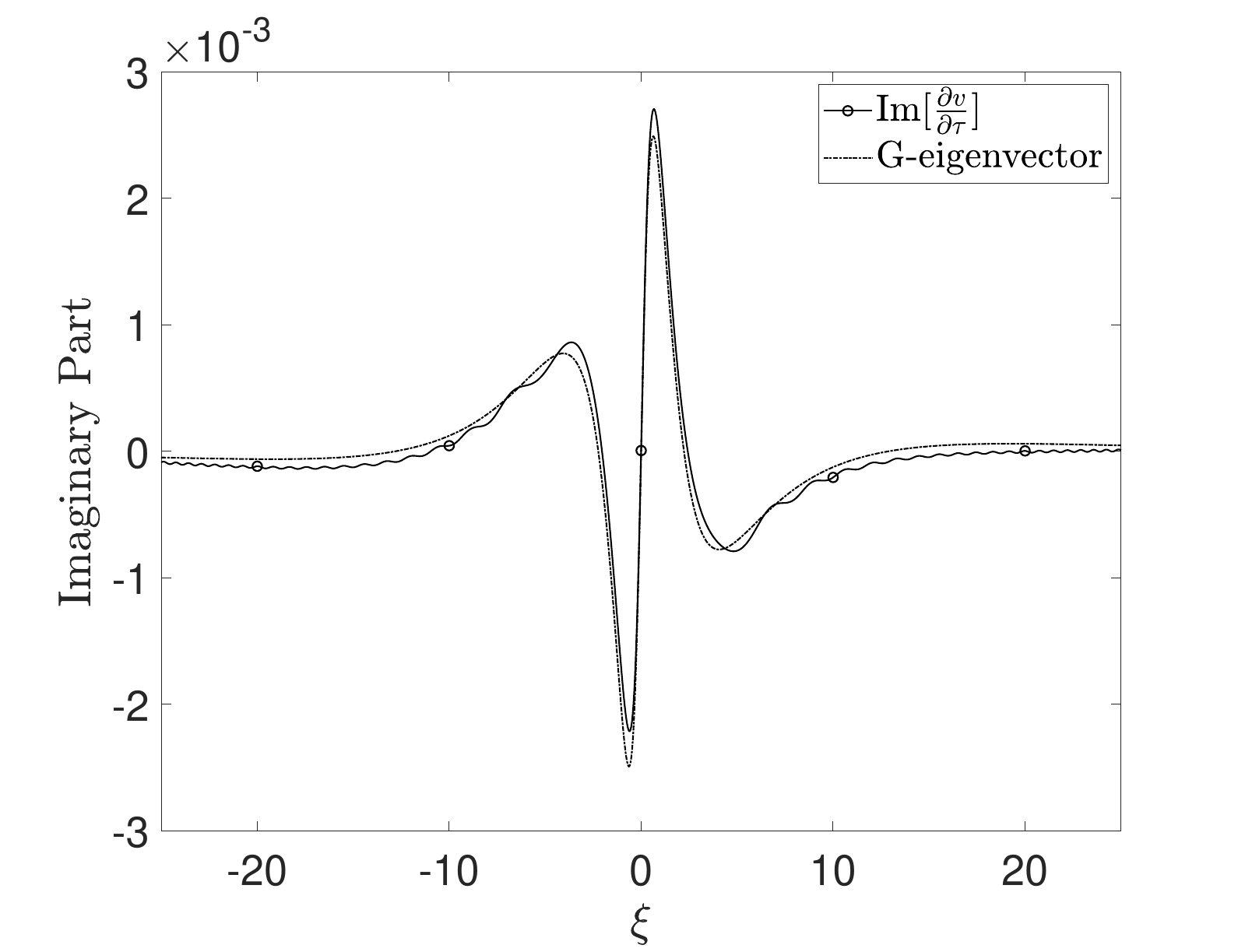}
  \\
(a)&(b)
\end{tabular}
\begin{tabular}{c}
\includegraphics[width=0.49\textwidth]{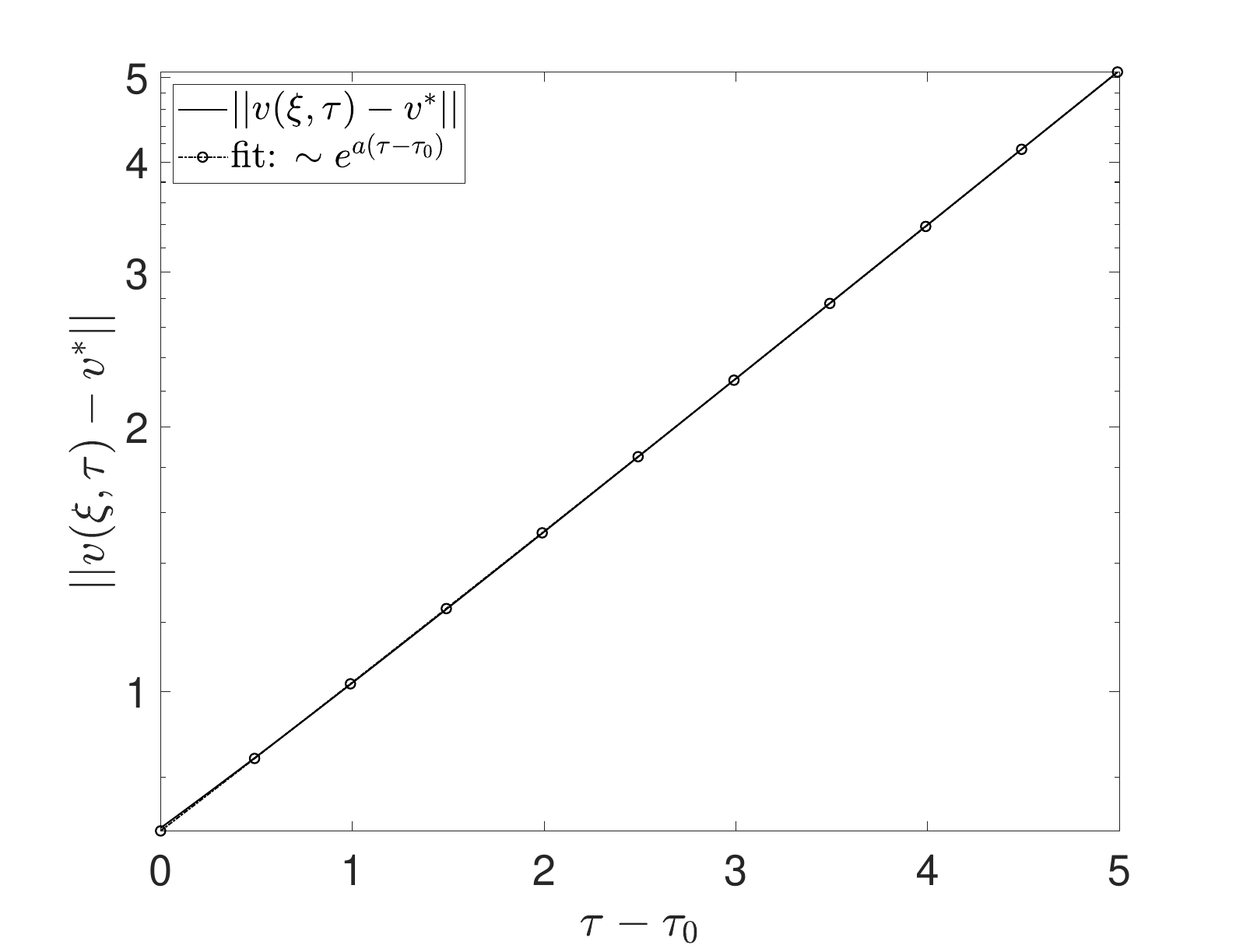}
  \\
(c)
\end{tabular}
\caption{Comparison of solution's (a) real part and (b) imaginary part, partial $\tau$ derivative (computed at $\tau=42$) with the real and imaginary part of the $G$-eigenvector indicating that the instability presented in Fig.~\ref{fig:fulldomain_single} leads to growth along the direction of the $G$-eigenvector.
 (c) Exponential evolution of solution's $v$ deviation from the single-humped solution, $v^*$, $||v(\xi,\tau)-v^*||$.
 The starting time to measure deviation is $\tau_0=40$.
 Here, $a \approx 0.4085$, which is close to the blow-up rate of the single-humped solution: $G = 0.403$.
  } \label{fig:instability}

\end{figure}

In this section, we present dynamical simulations performed on the full domain, i.e., $\xi \in [-K,K]$.
When simulations are performed on $[-K,K]$, convergence
of an unstable multi-hump state to a steady-state corresponding to a single-humped solution is not observed. 
The initial data evolves toward the corresponding single-humped self-similar state; however the $\lambda=G$ associated eigenmode drives the solution away from this final state.
Figure~\ref{fig:fulldomain_single} illustrates the $\xi-\tau$ spatiotemporal evolution of an initial two-humped self-similar solution computed at $\sigma=2.01$, slightly perturbed along the symmetric eigenmode corresponding to the second largest eigenvalue.
After an initial $\tau$ period spanning approximately $[0,6]$, where the solution visibly restructures toward a single-humped state, then at $\tau \approx 40$, $v$ starts shifting from the origin to positive $\xi$ values. 
This shift is attributed to the $\lambda=G$ eigenvalue of the single-humped self-similar solution. 
In Fig.~\ref{fig:instability}(a)-(b), we compare the numerically approximated value of $\frac{\partial v}{\partial \tau}$ at $\tau = 48$ with the $\lambda=G$-associated eigenmode of the single-humped self-similar solution ((a) real part and (b) imaginary part). 
The comparison shows good agreement, indicating that $v$ shifts in the direction of the $\lambda=G$-associated eigenmode. 
The blow-up rate for a single-humped solution at $\sigma=2.01$ is $G=0.403$.
In Fig.~\ref{fig:instability}(c), we plot the evolution of the deviation $||v(\xi,\tau)-v^*||$, with $v^*$ the single-humped self-similar solution at $\sigma=2.01$.
Our starting point for measuring the deviation is $\tau_0=40$.
This deviation is expected to grow exponentially in rescaled time, following: $||v(\xi,\tau)-v^*|| \sim  e^{a(\tau-\tau_0)}$, where $a \approx G$. 
Indeed, fitting yields $a \approx 0.4085$ which is close to the expected value of the blow-up rate for a single-humped solution at $\sigma=2.01$ (which, as mentioned above, is $G = 0.403$).

\begin{figure}{H}
\centering
\begin{tabular}{c}
\includegraphics[width=0.5\textwidth]{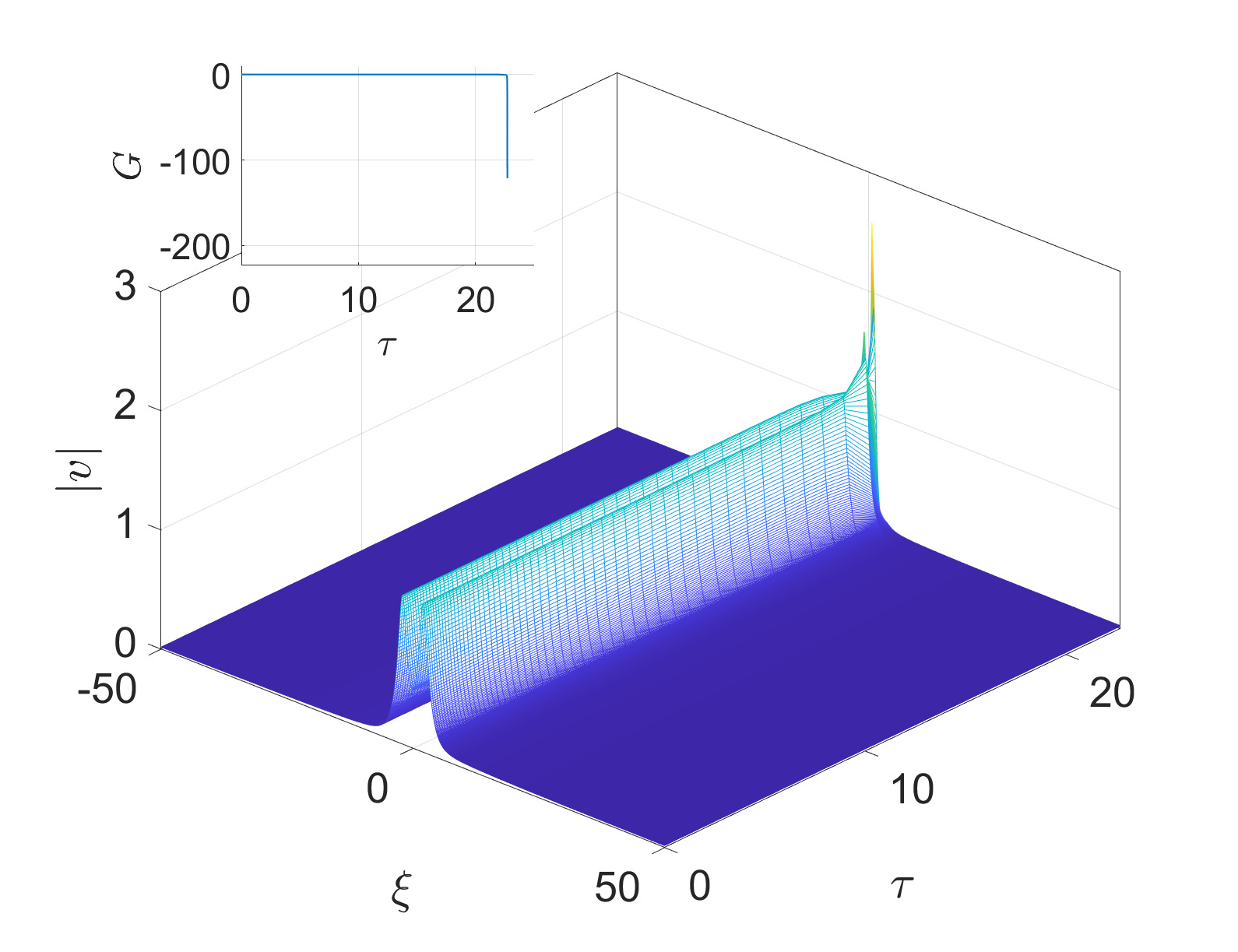}\\
(a) \\
\end{tabular}
\begin{tabular}{cc}
\includegraphics[width=0.4\textwidth]{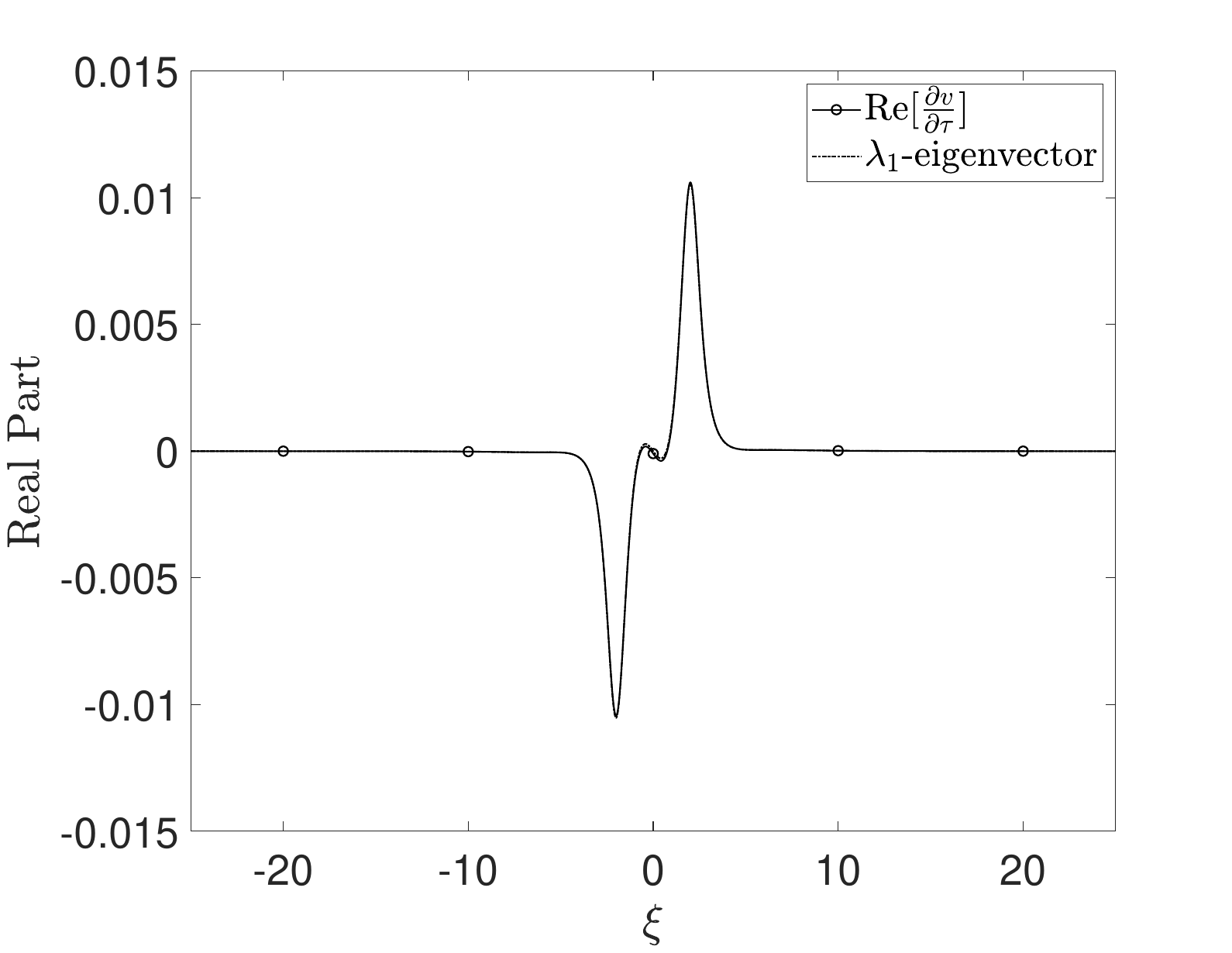} &  \includegraphics[width=0.4\textwidth]{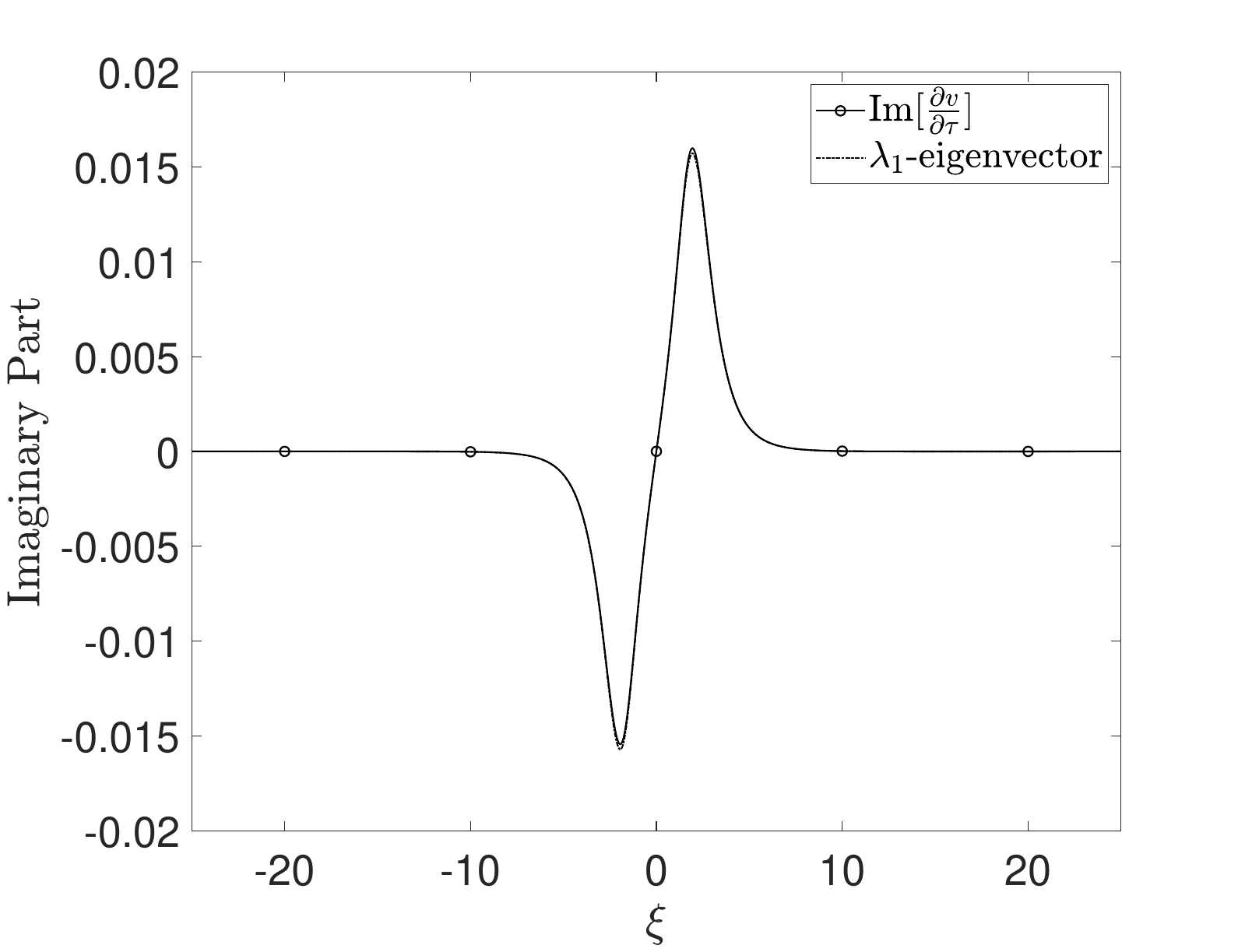} \\
(b) & (c) \\
\end{tabular}
\begin{tabular}{c}
\includegraphics[width=0.4\textwidth]{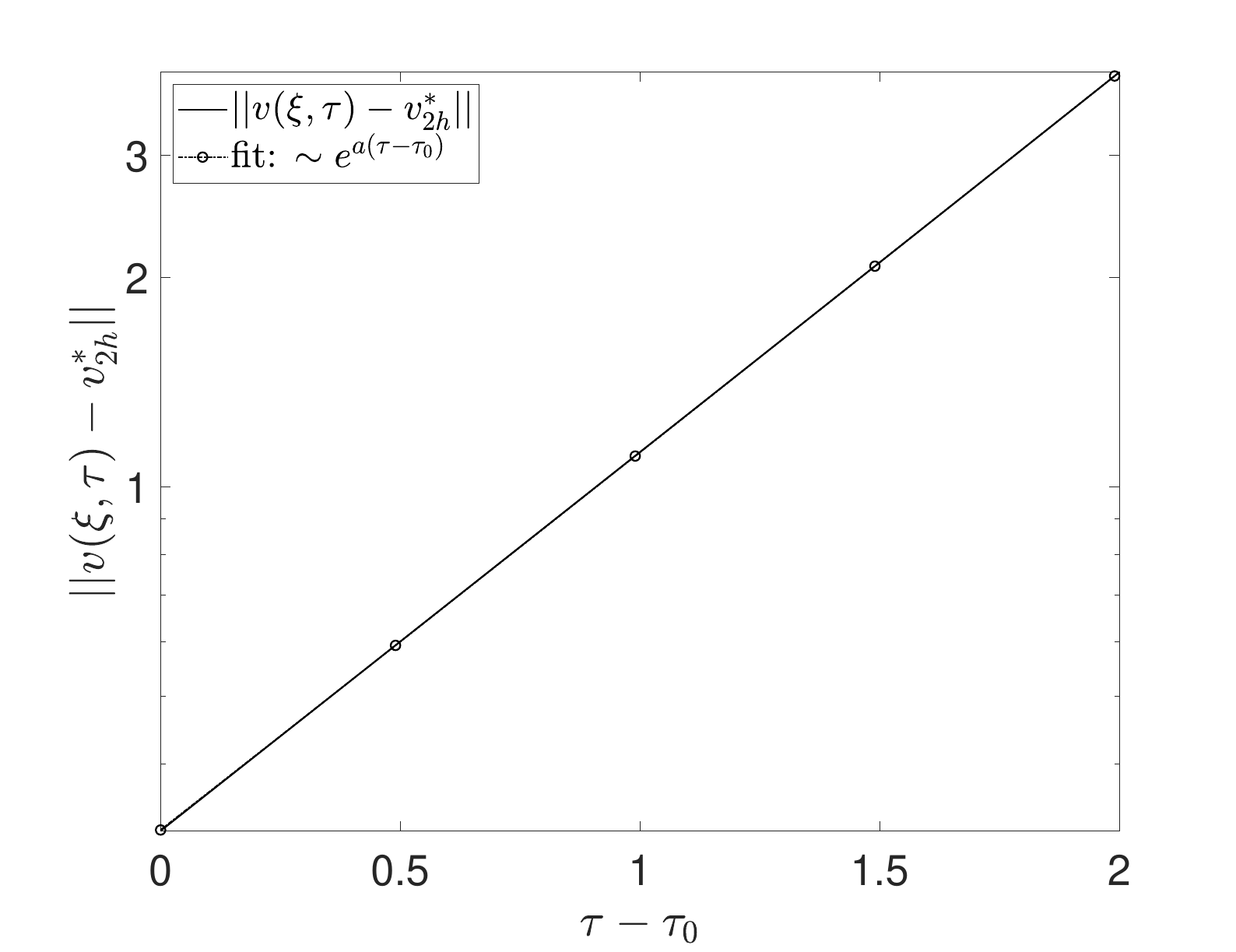} \\
(d) 
\end{tabular}
\caption{(a) Dynamics in the renormalized/rescaled spatio-temporal framework $\xi-\tau$: Starting from a two-humped solution at $\sigma=2.01$, we observe that the dynamics for the time interval $[0,18]$ are slow.
The accumulation of numerical errors drives the solution to asymmetric profiles for $\tau>20$, while the blow-up rate, $G$ (shown in the inset) shifts towards large negative values, and convergence is lost at around $\tau \approx 23 $ [in line
with the blow-up of such solutions also reported in~\cite{budd:1999}].
(b)-(c) Comparison of solution's (rescaled-$\tau$) time derivative ((b) real part and (c) imaginary part) computed at $\tau=18$  with the real and imaginary part of the eigenvector associated with the largest real eigenvalue of the two-humped solution, $\lambda_1=1.245$.
The agreement indicates that the deviation grows along the direction of the $\lambda_1$ eigenvector.
(d) Exponential evolution of solution's $v$ deviation from the two-humped solution, $v^*_{2h}$, $||v(\xi,\tau)-v^*_{2h}||$.
 The starting time to measure deviation is $\tau_0=18$.
 Here, $a \approx 1.26$, which is close to the largest real eigenvalue, $\lambda_1$ of the two-humped solution.
  } \label{fig:asymmetry}

\end{figure}

In Fig.~\ref{fig:asymmetry}(a), we present a second simulation initialized with the two-humped solution for $\sigma=2.01$.
This time, no perturbation is directly applied to the initial condition, and we perform a direct numerical
simulation with the roundoff
error providing the initial perturbation. 
The unstable self-similar solution retains its initial shape for a time interval $\tau \in [0,18]$ before gradually losing its symmetry in the later stages of the simulation.
At this point, we observe an asymmetric shape for $v$ and the blow-up rate that trends toward large negative values. 
The solution begins to deviate in the direction of the largest real eigenvalue of the two-humped solution.
This behavior is illustrated in Fig.~\ref{fig:asymmetry}(b)-(c), where we plot the time-derivative of $v$ (real and imaginary part) computed at $\tau=18$ and compare against the eigenmode which is associated with the largest real eigenvalue of the two-humped solution.
If we fit the exponential growth of the deviation from the two-humped solution $v^*_{2h}$, $E \equiv || v(\xi,\tau - v^*_{2h} ||\sim e^{a (\tau- \tau_0)}$, then $a$ is expected to be close to $\lambda_1$.
This fitting is shown in Fig.~\ref{fig:asymmetry}(d), where $a=1.2606$ close enough to $\lambda_1=1.245$.
We start measuring the deviation at $\tau_0 = 18$.

\section{Higher order terms in the bifurcation diagram}

To get a good quantitative analysis in Figure \ref{fig:bifdiagram}
when $G$ is not so small we need to 
include higher-order terms in equation (\ref{sigeqn2secIV}).
We can do so by following a similar procedure to that by which we
determined the higher-order terms for the one-hump solution in \cite{jon1}.

\subsection{Higher-order terms in the far field}
We proceed to calculate more terms in the expansion (\ref{farfieldexp}) of \S \ref{sec:farfield}. The calculation is similar to that in \cite{jon1}.
The equation for $A_1$ is
\[ 2 A_1' \phi' + A_0'' + A_1 \phi''=0,\]
i.e.,
\[  \fdd{}{\rho}A_1(-\phi')^{1/2} =  \frac{A_0''}{2(-\phi')^{1/2}} = \frac{a_0(8+3 \rho^2)}{4(4-\rho^2)^{5/2}}.\]
Thus,
\[ A_1 =  \frac{a_0 \rho(24-\rho^2)}{24 \sqrt{2}\,(4-\rho^2)^{7/4}} + \frac{2^{1/2}a_1}{(4-\rho^2)^{1/4}}.
\]
Matching the near field gives $a_1=0$.
At the next order
\[ 2 A_2' \phi' + A_1'' + A_2 \phi''=0,\]
i.e.
\[  \fdd{}{\rho}A_2(-\phi')^{1/2} =  \frac{A_1''}{2(-\phi')^{1/2}} = \frac{a_0 \rho(3648+640 \rho^2-3 \rho^4)}{192(\rho^2-4)^4}.\]
Thus
\[ A_2 = \frac{a_0(2320-996 \rho^2+9 \rho^4)}{576\sqrt{2}\,(4-\rho^2)^{13/4}} + \frac{2^{1/2}a_2}{(4-\rho^2)^{1/4}}.
\]
As $\rho\ra0$,
\beq
A_2 \ra \frac{145 a_0}{4608} + a_2.\label{A2inner}
\eeq

\subsection{Higher-order terms in outer limit of the inner}
From \S\ref{app:V1} as $x \ra \infty$,
the asymptotic behaviour of $V_1=V_o+V_e$ is given by
\beqa
V_o 
&\sim&  3^{1/4}\frac{X_i}{16}\sqrt{2}\, \ee^{-x}\left(2x^2+2x-2\log 2 + 1\right)
,\\
V_e & \sim & 12^{1/4}(\al_0+\al_1x+\al_2 x^2+\al_3 x^3)\ee^{-x} - \frac{X_i^2}{16}3^{1/4}\sqrt{2}\left(1 - 2x \right)\ee^{-x},
\eeqa
with $\alpha_i$, $i=0,1,2,3$ given by (\ref{alpha0123}).
With $x = \xi-X_i$, The coefficient of $\ee^{-\xi}$ is
\[ \left(3^{1/4}\frac{X_i}{16}\sqrt{2}\,(2X_i^2-2X_i-2\log2+1) + 12^{1/4}(\alpha_0 - \alpha_1 X_i + \alpha_2 X_i^2 - \alpha_3 X_i^3)- \frac{X_i^2}{16}3^{1/4}\sqrt{2}(1+2X_i)\right)\ee^{X_i}.
\]
Matching with (\ref{A2inner}) gives
\begin{multline*}
  a_2 =  \left(3^{1/4}\frac{X_i}{16}\sqrt{2}\,(2X_i^2-2X_i-2\log2+1) + 12^{1/4}(\alpha_0 - \alpha_1 X_i + \alpha_2 X_i^2 - \alpha_3 X_i^3)
    \right. \\
    \left. - \frac{X_i^2}{16}3^{1/4}\sqrt{2}(1+2X_i)\right)\ee^{X_i}
  -\frac{145 a_0}{4608}.
\end{multline*}

\subsection{Higher-order terms in the integral of $\V_s^2$}
The final step in finding higher-order terms in (\ref{sigsol}) is to get a better approximation to the right-hand side, for which we need the integral of $\V_s^2$.
If we spilt the integral we have
\beqas
\int_{-\infty}^\infty & = & \int_{-\infty}^{X_1-R}
+\int_{X_1-R}^{X_1+R}+\int_{X_1+R}^{X_2-R} + \cdots
+ \int_{X_n+R}^\infty
\eeqas
Now, using the inner expansion,
\beqas
\int_{X_i-R}^{X_i+R} \V_s^2\, \d \xi &=& \int_{-R}^R (V_0^2 + 2 G^2 V_0 V_1 + \cdots)\, \d x\\
& \sim & \frac{\sqrt{3}\pi}{2}+ 2 G^2 \int_{-R}^R V_{\mathrm{even}} V_0\, \d x\\
& \sim & \frac{\sqrt{3}\pi}{2}+ 2 G^2 \int_{-R}^R \left(\frac{3^{1/4}}{2}\left(\frac{\ee^{X_{i}-X_{i+1}}}{G^2}+
 \frac{\ee^{X_{i-1}-X_i}}{G^2} \right)v_2(x) + \hat{V}_1(x)- \frac{X_i^2}{4} \left(\frac{V_0}{4} + \frac{x}{2}\fdd{V_0}{x}\right)\right) V_0\, \d x\\
&= & \frac{\sqrt{3}\pi}{2}+ 2 G^2 \frac{\sqrt{3} \pi^3}{256}+ 3^{1/4}\left(\ee^{X_{i}-X_{i+1}}+
 \ee^{X_{i-1}-X_i} \right) \int_{-R}^R v_2 V_0\, \d x\\
&= & \frac{\sqrt{3}\pi}{2}+ 2 G^2 \frac{\sqrt{3} \pi^3}{256}+ 3^{1/4}\left(\ee^{X_{i}-X_{i+1}}+
 \ee^{X_{i-1}-X_i} \right) \left[ 2 \times 3^{1/4}(x-\tanh 2x) \right]_{-R}^R\\
&= & \frac{\sqrt{3}\pi}{2}+  G^2 \frac{\sqrt{3} \pi^3}{128}+ 4\sqrt{3}\left(\ee^{X_{i}-X_{i+1}}+
 \ee^{X_{i-1}-X_i} \right) \left( R-1 \right).
\eeqas
using the outer (between humps) expansion,
\beqas
\int_{X_i+R}^{X_{i+1}-R} V_s^2\, \d \xi & = &  \int_{X_i+R}^{X_{i+1}-R} (V_0^2 + 2 G^2V_1 V_0+\cdots )\, \d \xi\\
& = & \int_{X_i+R}^{X_{i+1}-R} 2 \sqrt{3}(\ee^{-\xi+X_i} + \ee^{\xi - X_{i+1}})^2  + 2 G^2 12^{1/4}V_1 (\ee^{-\xi+X_i} + \ee^{\xi - X_{i+1}})+\cdots )\, \d \xi\\
& \sim & - 8 \sqrt{3}\,R \ee^{X_i-X_{i+1}} + 4 \sqrt{3} (X_{i+1}-X_i)\ee^{X_i-X_{i+1}}  +\cdots
\eeqas
Thus, together
\beqas
\int_{-\infty}^\infty V_s^2\, \d x & \sim & \frac{\sqrt{3}\pi}{2}\left(1 + \frac{G^2\pi^2}{64}\right)n + 4 \sqrt{3}(R-1)\sum_{i=1}^n\left(\ee^{X_{i}-X_{i+1}}+
 \ee^{X_{i-1}-X_i} \right) \\
&& \mbox{ } - 8 \sqrt{3}\,R\sum_{i=1}^{n-1} \ee^{X_i-X_{i+1}} + 4 \sqrt{3}\sum_{i=1}^{n-1} (X_{i+1}-X_i)\ee^{X_i-X_{i+1}} \\
& \sim & \frac{\sqrt{3}\pi}{2}\left(1 + \frac{G^2\pi^2}{64}\right)n + 4 \sqrt{3}(R-1)\sum_{i=1}^{n-1}\ee^{X_{i}-X_{i+1}}+4 \sqrt{3}(R-1)\sum_{i=1}^{n-1}
 \ee^{X_{i}-X_{i+1}}  \\
&& \mbox{ } - 8 \sqrt{3}\,R\sum_{i=1}^{n-1} \ee^{X_i-X_{i+1}} + 4 \sqrt{3}\sum_{i=1}^{n-1} (X_{i+1}-X_i)\ee^{X_i-X_{i+1}} 
\\
& \sim & \frac{\sqrt{3}\pi}{2}\left(1 + \frac{G^2\pi^2}{64}\right)n 
 + 4 \sqrt{3}\sum_{i=1}^{n-1} (X_{i+1}-X_i-2)\ee^{X_i-X_{i+1}} 
\eeqas
\subsection{Normal form}
With $X_1 = - X_n$ equation \eqref{sigsol} now becomes
\begin{multline}
4 \sqrt{3}\ee^{2X_n}\ee^{-\pi/G}\left(1+
  G^2\left(\frac{X_i}{16}(2X_i^2-2X_i-2\log2+1) \right.\right. \\ \left.\left.+ (\alpha_0 - \alpha_1 X_i + \alpha_2 X_i^2 - \alpha_3 X_i^3)
     - \frac{X_i^2}{16}(1+2X_i)\right)\ee^{X_i}
    -\frac{145 G^2}{4608}\right)^2\\  =\frac{(\sigma-2)G}{2 \sigma}\left( \frac{\sqrt{3}\pi}{2}\left(1 + \frac{G^2\pi^2}{64}\right)n 
 + 4 \sqrt{3}\sum_{i=1}^{n-1} (X_{i+1}-X_i-2)\ee^{X_i-X_{i+1}} \right)
\end{multline}
This is what is plotted in Figure \ref{fig:bifdiagram}.

\section{Detailed asymptotic analysis of the eigenvalues $\la = O(G)$.}
With the expansions \eqref{exp1}-\eqref{exp2}, 
equating coefficients at $O(G)$ we have
\[f_1  =  a_i^{(1)} U_0 + b_i^{(0)} W_1,
\]
where
\beqas
\sdd{W_1}{x} 
+  2 V_0^{4} W_1^* 
+ 3V_0^{4} W_1 
-  W_1 
  &=&  \la_1 V_0.
\eeqas
The solvability conditions on $W_1$ are satisfied automatically, and
\[   W_1 =\la_1\left(\frac{V_0}{4} + \frac{x}{2}\fdd{V_0}{x}\right).
\]
Equating coefficients at $O(G^2)$ we have
\[f_2  =  a_i^{(1)} U_1  + b_i^{(2)} W_0+ b_i^{(0)} W_2,
\]
where
\beqas
\sdd{ U_1}{x} +
2 V_0^{4}U_1^* +3 V_0^{4} U_1 
- U_1
 &=& -\ii \la_1 \fdd{V_0}{x} ,\\
 \eq{W_2}{W^*_2}
& = &  -\ii \la_1^2\left(\frac{V_0}{4} +
  \frac{x}{2}\fdd{V_0}{x}\right)+ \la_2 V_0 - \ii\frac{ (x+X_i)^2}{4}
V_0
- 4 \ii V_1V_0^4.
\eeqas
The solvability conditions on $U_1$ are automatically satisfied, while those on $W_2$ are
\beqa
 \im \left[\fdd{W_2}{x}V_0 - W_2  \fdd{V_0}{x}
\right]^R_{-R} & =& \left[\fdd{V_1}{x} V_0 - V_1 \fdd{V_0}{x}
\right]^R_{-R}
\label{sol3a},\\
  \re\left[\fdd{W_2}{x}\fdd{V_0}{x} - W_2\sdd{V_0}{x}\right]^R_{-R} &
  =&  0. \label{sol4a}
\eeqa
The solutions are
\beq
U_1 = - \frac{\ii \la_1 x }{2}V_0,
  \qquad
W_2  =  -\frac{\ii \la_1^2 x^2V_0}{8}+\ii V_1 +\la_2\left(\frac{V_0}{4}+\frac{x}{2}\fdd{V_0}{x}\right).\label{U1W2}
\eeq
Note that $W_2$ grows at infinity, as does $V_1$, so that to evaluate (\ref{sol3a}) we  need to match with the region between humps.
In preparation for matching we list the limiting behaviour of the near-hump solutions as $x \ra\pm\infty$:
\begin{align*}
U_0&\sim    \mp 12^{1/4} \ee^{\mp x} \quad \mbox{ as }x \ra \pm\infty,&
W_0 &\sim  12^{1/4}\ii  \ee^{\mp x} \quad \mbox{ as }x \ra \pm\infty,\\
U_1 & \sim  - \frac{12^{1/4}\ii}{2} x  \ee^{\mp x} \quad \mbox{ as }x \ra \pm\infty,& 
W_1  &\sim  
  12^{1/4}  \left(\frac{1}{4} \mp  \frac{x}{2} \right)\ee^{\mp x}\quad \mbox{ as }x \ra \mp \infty,
\end{align*}
Thus,
\[
f_0  \sim 12^{1/4}\ii b_i^{(0)} \ee^{\mp x}\quad \mbox{ as }x \ra
\pm\infty,\qquad  
f_1   \sim 12^{1/4} b_i^{(0)}\left(\frac{1}{4} \mp \frac{x}{2}  \right)\ee^{\mp x}\mp 12^{1/4} a_i^{(1)}\ee^{\mp x}\quad \mbox{ as }x \ra \pm \infty.
\]

\subsection{In between the humps}
As usual, away from the humps,
\beqas
\ii  \la f+\sdd{ f}{\xi} 
- f
+ \frac{G^2 \xi^2 }{4} f &=& O(G^{8}\log G^4 f).
\eeqas
Consider the gap between $X_i$ and $X_{i+1}$.
At leading order
\[ \sdd{ f_0}{\xi} 
- f_0=0.\]
Matching requires
\[
f_0 \sim 12^{1/4} \ii b_{i+1}^{(0)} \ee^{\xi-X_{i+1}}\quad
\mbox{as } \xi \ra X_{i+1},\qquad
f_0 \sim 12^{1/4} \ii  b_i^{(0)}\ee^{-\xi+X_i}\quad\mbox{as }
\xi \ra X_i. 
\]
Thus,
\[ f_0 = 12^{1/4} \ii b_{i+1}^{(0)} \ee^{\xi-X_{i+1}} +12^{1/4}
  \ii b_i^{(0)} \ee^{-\xi+X_i}.\]
In terms of $\xi = X_{i+1} + x$ this is
\[ f_0 = 12^{1/4} \ii b_{i+1}^{(0)}\ee^{x} +12^{1/4} 
  \ii b_i^{(0)} \ee^{-x+X_i-X_{i+1}},\]
giving the matching condition (translating to $i+1 \rightarrow i$)
\beq
 G^2f_2 \sim 12^{1/4} \ii b_{i-1}^{(0)}
  \ee^{-x+X_{i-1}-X_{i}} \qquad \mbox{ as }x \ra -\infty\label{match1}
\eeq
on the local solution near the $i$\,th hump.
Similarly we find 
\beq
G^2 f_2  \sim 12^{1/4} \ii b_{i+1}^{(0)}
 \ee^{x+X_i-X_{i+1}} 
\qquad \mbox{ as }x \ra \infty.
\label{match2}
\eeq
on the local solution near the $i$\,th hump.
Since
\[f_2  =  a_i^{(1)} U_1  + b_i^{(2)} W_0+ b_i^{(0)} W_2,
\]
and the local hump solutions for  $U_1$ and $W_0$ decay at infinity,  matching gives
\beqa
\im \left[b_i^{(0)}\fdd{W_2}{x}V_0 -b_i^{(0)}W_2  \fdd{V_0}{x}
\right]^R_{-R} & =&  \frac{4\sqrt{3}\,b_{i+1}^{(0)}}{G^2}\ee^{X_i-X_{i+1}} +\frac{4\sqrt{3}\,b_{i-1}^{(0)}}{G^2}\ee^{X_{i-1}-X_{i}}
,\label{m3}\\
  \re\left[b_i^{(0)}\fdd{W_2}{x}\fdd{V_0}{x} -
    b_i^{(0)}W_2\sdd{V_0}{x}\right]^R_{-R} & =&0
.\qquad\label{m4}
\eeqa
Using (\ref{m3}), (\ref{m4}) and (\ref{V1x}) in (\ref{sol3a})-(\ref{sol4a}) gives
\beqa
b_i^{(0)}\left(\frac{4 \sqrt{3}}{G^2}\ee^{-X_{i+1}+X_i}+\frac{4
  \sqrt{3}}{G^2}\ee^{-X_{i}+X_{i-1}}
\right) & = & \frac{4\sqrt{3}\,b_{i+1}^{(0)}}{G^2}\ee^{X_i-X_{i+1}}
+\frac{4\sqrt{3}\,b_{i-1}^{(0)}}{G^2}\ee^{X_{i-1}-X_{i}}.\label{Bb0}
\eeqa
This is equation \eqref{Bb0_main}  from the main text.
Equation (\ref{Bb0}) fixes $b_i^{(0)} = b^{(0)}$ for all $i$.

\subsection{The outer limit of $f_2$}
We will need to know the outer limit of $f_2$, for which we need to
know the behaviour of $W_2$ as $x \ra \pm \infty$. We have just
evaluated the exponentially growing component by matching, but we will
find that unfortunately we need also the exponentially decaying component.
This requires the asumptotic behaviour of $V_1$.
Note that  there are also higher-order terms arising from the  expansion of $f_0$ and $f_1$ as $x \ra \pm \infty$, but these don't contribute until $O(\ee^{-5x}) = O(G^{10})$ so we will ignore them.

We show in \S\ref{app:V1} that, as $x \ra - \infty$,
\begin{multline}
V_1 
\sim  3^{1/4}\frac{X_i}{16}\sqrt{2}\, \ee^{x}\left(-2x^2+2x+2\log 2 - 1\right)
+
3^{1/4} \frac{\ee^{X_{i-1}-X_i}}{G^2} \sqrt{2}\, \ee^{-x} \\
+12^{1/4}(\al_0-\al_1x+\al_2 x^2-\al_3 x^3)\ee^{x} - \frac{X_i^2}{16}3^{1/4}\sqrt{2}\left(1 + 2x\right) \ee^{x},\label{V1minffull}
\end{multline}
while, as  $x \ra \infty$,
\begin{multline}
V_1
\sim  3^{1/4}\frac{X_i}{16}\sqrt{2}\, \ee^{-x}\left(2x^2+2x-2\log 2 + 1\right)
+3^{1/4} \sqrt{2}\frac{\ee^{X_{i}-X_{i+1}}}{G^2}\ee^{x}\\
+ 12^{1/4}(\al_0+\al_1x+\al_2 x^2+\al_3 x^3)\ee^{-x} - \frac{X_i^2}{16}3^{1/4}\sqrt{2}\left(1 - 2x \right)\ee^{-x}.\label{V1inffull}
\end{multline}
Thus
\beqas
W_2 & \sim & 
- \frac{\ii \la_1^2 x^2}{8} 12^{1/4} \ee^{-x} + 12^{1/4}\ii\frac{X_i}{16}\, \ee^{-x}\left(2x^2+2x-2\log 2 + 1\right)
+12^{1/4} \ii\frac{\ee^{X_{i}-X_{i+1}}}{G^2}\ee^{x}\\
&& \mbox{ }+ 12^{1/4} \ii(\al_0+\al_1x+\al_2 x^2+\al_3 x^3)\ee^{-x} - \ii\frac{X_i^2}{16}12^{1/4}\left(1 - 2x \right)\ee^{-x}+ \frac{\la_2}{4}12^{1/4} \left(1-2x\right)\ee^{-x} \qquad \mbox{ as } x \ra \infty,\\
W_2 & \sim & 
- \frac{\ii \la_1^2 x^2}{8} 12^{1/4} \ee^{x} + 12^{1/4}\ii\frac{X_i}{16}\, \ee^{x}\left(-2x^2+2x+2\log 2 - 1\right)
+12^{1/4} \ii\frac{\ee^{X_{i-1}-X_{i}}}{G^2}\ee^{-x}\\
&& \mbox{ }+ 12^{1/4} \ii(\al_0-\al_1x+\al_2 x^2-\al_3 x^3)\ee^{x} - \ii\frac{X_i^2}{16}12^{1/4}\left(1 + 2x \right)\ee^{x} + \frac{\la_2}{4}12^{1/4} \left(1+2x\right)\ee^{x} \qquad \mbox{ as } x \ra -\infty.
\eeqas
Thus
\begin{multline}
f_2 \sim 12^{1/4}b_i^{(2)} \ii \ee^{-x} -  12^{1/4}a_i^{(1)} \frac{\ii \la_1 x}{2}\ee^{-x}
-b^{(0)} \frac{\ii \la_1^2 x^2}{8} 12^{1/4} \ee^{-x} +  12^{1/4}\ii b^{(0)}\frac{X_i}{16}\, \ee^{-x}\left(2x^2+2x-2\log 2 + 1\right)\\
+12^{1/4} \ii b^{(0)}\frac{\ee^{X_{i}-X_{i+1}}}{G^2}\ee^{x}
+ 12^{1/4} \ii b^{(0)}(\al_0+\al_1x+\al_2 x^2+\al_3 x^3)\ee^{-x} - \ii\frac{X_i^2}{16}12^{1/4} b^{(0)}\left(1 - 2x \right)\ee^{-x}\\
 + \frac{\la_2}{4}12^{1/4} b^{(0)}\left(1-2x\right)\ee^{-x} \qquad \mbox{ as } x \ra \infty,\label{f2inf}
\end{multline}
\begin{multline}
f_2 \sim  12^{1/4}b_i^{(2)} \ii \ee^{x}-  12^{1/4}a_i^{(1)} \frac{\ii \la_1 x}{2}\ee^{x}- 
\frac{\ii \la_1^2 x^2}{8} 12^{1/4} b^{(0)}\ee^{x} + 12^{1/4}b^{(0)}\ii\frac{X_i}{16}\, \ee^{x}\left(-2x^2+2x+2\log 2 - 1\right)\\
+12^{1/4} b^{(0)}\ii\frac{\ee^{X_{i-1}-X_{i}}}{G^2}\ee^{-x}
+ 12^{1/4} b^{(0)}\ii(\al_0-\al_1x+\al_2 x^2-\al_3 x^3)\ee^{x} - b^{(0)}\ii\frac{X_i^2}{16}12^{1/4}\left(1 + 2x \right)\ee^{x}\\+ \frac{\la_2}{4}12^{1/4}b^{(0)} \left(1+2x\right)\ee^{x} \qquad \mbox{ as } x \ra -\infty.\label{f2minf}
\end{multline}

\subsection{$O(G^3)$ terms in the inner expansion}

At $O(G^3)$ we find
\[ f_3 = a_i^{(3)} U_0 + a_i^{(1)} U_2 + b_i^{(2)} W_1 + b_i^{(0)} W_3,
\]
where
\beqas
\eq{U_2}{U^*_2}
& = &- \frac{\la_1^2 x}{2}V_0- \ii \la_2 \fdd{V_0}{x}- \frac{ (x+X_i)^2}{4}\fdd{V_0}{x} - 20 V_1V_0^3\fdd{V_0}{x},\\
\eq{W_3}{W^*_3}
&=& -\frac{\la_1^3 x^2V_0}{8}+ \la_1V_1 -2\ii \la_2 \la_1\left(\frac{V_0}{4} +
  \frac{x}{2}\fdd{V_0}{x}\right)+ \la_3 V_0 \\
&& \mbox{ }- \frac{
   (x+X_i)^2}{4}\la_1\left(\frac{V_0}{4} + \frac{x}{2}\fdd{V_0}{x}\right) - 20 V_1V_0^3\la_1\left(\frac{V_0}{4} + \frac{x}{2}\fdd{V_0}{x}\right) .
\eeqas
The solvability conditions are 
\beqa
 \im \left[\fdd{U_2}{x}V_0 - U_2  \fdd{V_0}{x}
\right]^R_{-R} & =&  0
\label{sol1a},\\
  \re\left[\fdd{U_2}{x}\fdd{V_0}{x} - U_2\sdd{V_0}{x}\right]^R_{-R} &
  =&    \frac{\sqrt{3}\, \pi(\la_1^2-1)}{8}  +\left[\sdd{V_1}{x}\fdd{V_0}{x} - \fdd{V_1}{x}\sdd{V_0}{x}
\right]^R_{-R}\label{sol2a}\\
 \im \left[\fdd{W_3}{x}V_0 - W_3  \fdd{V_0}{x}
 \right]^R_{-R} & =&
0 
\label{sol3b},\\
  \re\left[\fdd{W_3}{x}\fdd{V_0}{x} - W_3\sdd{V_0}{x}\right]^R_{-R} &
  =& \la_1 \int_{-R}^R \left(
    V_1 \fdd{V_0}{x} - \left(\frac{ x X_i}{2}+  20 V_1 V_0^3\right)\left(\frac{V_0}{4} +
  \frac{x}{2}\fdd{V_0}{x}\right)\fdd{V_0}{x} 
    \right)\, \d x, \label{sol4b}\qquad
    \eeqa
where we have taken advantage of the evenness of $V_0$ to simplify.
The terms forced by $\la_2$ and $\la_3$ can be evaluated, so that we can write
\beqas
U_2 & = & U_2^R - \frac{\ii \la_2 x}{2} V_0,\\
W_3 & = & W_3^R - \ii \la_2 \la_1 \frac{x^2 V_0}{4} + \la_3 \left(\frac{V_0}{4} + \frac{x}{2}\fdd{V_0}{x}\right),
\eeqas
where
\beqa
\eqr{U_2^R} 
& = &- \frac{\la_1^2 x}{2}V_0- \frac{ (x+X_i)^2}{4}\fdd{V_0}{x} - 20 V_1V_0^3\fdd{V_0}{x},\label{eq:U2R}\\
\eqr{W_3^R}
&=& -\frac{\la_1^3 x^2V_0}{8}+ \la_1V_1
- \frac{(x+X_i)^2}{4}\la_1\left(\frac{V_0}{4} + \frac{x}{2}\fdd{V_0}{x}\right) - 20 V_1V_0^3\la_1\left(\frac{V_0}{4} + \frac{x}{2}\fdd{V_0}{x}\right) .\label{eq:W3R}\qquad
\eeqa

\subsection{$O(G)$ terms in the outer expansion}
At $O(G)$ in the outer we find
\[\ii \la_1 f_0 +  \sdd{ f_1}{\xi} 
  - f_1=0,
\]
with
\[ f_0 = 12^{1/4} \ii b^{(0)} \ee^{\xi-X_{i+1}} +12^{1/4}
  \ii b^{(0)} \ee^{-\xi+X_i}.\]
Matching requires
\beqas
f_1 &\sim&12^{1/4} a_{i+1}^{(1)} \ee^{\xi-X_{i+1}}
+ 12^{1/4} b^{(0)}\left( \frac{\la_1 (\xi-X_{i+1})}{2} +\frac{\la_1}{4}\right)\ee^{\xi-X_{i+1}}\qquad
\mbox{ as } \xi \ra X_{i+1},\\
f_1 &\sim&  -12^{1/4}a_{i}^{(1)} \ee^{-\xi+X_i}+12^{1/4}b^{(0)}  \left(-
  \frac{\la_1 (\xi-X_{i})}{2} +\frac{\la_1}{4}\right)\ee^{-\xi+X_i}\qquad\mbox{ as } \xi \ra X_{i}.
\eeqas
Thus
\begin{multline*}
f_1  =  12^{1/4} \left(a_{i+1}^{(1)}+ 
 \frac{\la_1  b^{(0)}(\xi-X_{i+1})}{2}
+\frac{\la_1 b^{(0)}}{4}\right)\ee^{\xi-X_{i+1}}\\ +
12^{1/4}  \left(-a_{i}^{(1)} -
  \frac{\la_1  b^{(0)}(\xi-X_{i})}{2}+\frac{\la_1 b^{(0)}}{4}\right)\ee^{-\xi+X_i}.
\end{multline*}
In terms of $\xi = X_{i+1} + x$ this is
\begin{multline*}
f_1  =  12^{1/4} \left(a_{i+1}^{(1)} +
 \frac{\la_1 x}{2}
 b^{(0)}+\frac{\la_1}{4}b^{(0)}\right)\ee^{x}\\ +
 12^{1/4}  \left(-a_{i}^{(1)} 
 -
  \frac{\la_1 (x+X_{i+1}-X_{i})}{2} b^{(0)}+\frac{\la_1}{4}b^{(0)}\right)\ee^{-x 
  - X_{i+1}+X_i}.
\end{multline*}
giving the matching condition
\beq
f_3G^2 \sim 12^{1/4}  \left(-a_{i-1}^{(1)}-
  \frac{\la_1 (x+X_{i}-X_{i-1})}{2} b^{(0)}+\frac{\la_1}{4}b^{(0)}\right)\ee^{-x
  - X_{i}+X_{i-1}},\label{f3minf}
\eeq
as $x \ra -\infty$ on the inner solution.
In terms of $\xi = X_{i} + x$ we have
\begin{multline*}
  f_1  =  12^{1/4} \left(a_{i+1}^{(1)}
  + \frac{\la_1 (x+X_i-X_{i+1})}{2}
 b^{(0)}+\frac{\la_1}{4}b^{(0)}\right)\ee^{x+X_i-X_{i+1}}\\ +
12^{1/4}  \left(-a_{i}^{(1)} -
  \frac{\la_1 x}{2} b^{(0)}+\frac{\la_1}{4}b^{(0)}\right)\ee^{-x},
\end{multline*}
giving the matching condition
\beq
f_3G^2 \sim  12^{1/4} \left(a_{i+1}^{(1)}+
 \frac{\la_1 (x+X_i-X_{i+1})}{2}
 b^{(0)}+\frac{\la_1}{4}b^{(0)}\right)\ee^{x+X_i-X_{i+1}},\label{f3inf}
\eeq
as $x \ra \infty$ on the inner solution.

\subsection{Solvability condition on $f_3$}
The matching conditions (\ref{f3minf}), (\ref{f3inf})  give
\beqas
G^2\left[ \fdd{f_3}{x} V_0 - f_3 \fdd{V_0}{x}  \right]^{R}_{-R} & = &  2 \sqrt{3}\, \left(2a_{i+1}^{(1)} +
 \la_1 (1+R+X_i-X_{i+1}) b^{(0)} \right)\ee^{X_i-X_{i+1}}\\
&& \mbox{ }
- 2 \sqrt{3}\, \left(2a_{i-1}^{(1)} +
  \la_1 (-1-R+X_{i}-X_{i-1}) b^{(0)}\right)
\ee^{ - X_{i}+X_{i-1}}\\
G^2\left[ \fdd{f_3}{x} \fdd{V_0}{x} - f_3 \sdd{V_0}{x}  \right]^{R}_{-R} & = &
-  2 \sqrt{3}\, \left(2a_{i+1}^{(1)} +
 \la_1 (1+R+X_i-X_{i+1}) b^{(0)} \right)\ee^{X_i-X_{i+1}}\\
&& \mbox{ }
- 2 \sqrt{3}\, \left(2a_{i-1}^{(1)} +
  \la_1 (-1-R+X_{i}-X_{i-1}) b^{(0)}\right)
\ee^{ - X_{i}+X_{i-1}}.
\eeqas
The solvability conditions  (\ref{sol1a}), (\ref{sol2a}), (\ref{sol3b}), (\ref{sol4b}) then give
\beqa
\lefteqn{ \frac{\sqrt{3}\, \pi a_i^{(1)}(\la_1^2-1)}{8}  +a_i^{(1)}\left[\sdd{V_1}{x}\fdd{V_0}{x} - \fdd{V_1}{x}\sdd{V_0}{x}
\right]^R_{-R} }\hspace{5cm}&&\non\\ 
\lefteqn{\mbox{ }+
\la_1 b^{(0)}\int_{-R}^R \left(
     V_1 \fdd{V_0}{x} - \left(\frac{ x X_i}{2}+  20 V_1 V_0^3\right)\left(\frac{V_0}{4} +
  \frac{x}{2}\fdd{V_0}{x}\right)\fdd{V_0}{x} 
    \right)\, \d x}\hspace{4cm}&&\non\\
 & = & - 2\sqrt{3}\, \left(2a_{i+1}^{(1)}+
 \la_1 (1+R+X_i-X_{i+1}) b^{(0)} \right)\frac{\ee^{X_i-X_{i+1}}}{G^2}\non\\
&& \mbox{ }
- 2 \sqrt{3}\, \left(2a_{i-1}^{(1)}  +
  \la_1 (-1-R+X_{i}-X_{i-1}) b^{(0)}\right)
\frac{\ee^{ - X_{i}+X_{i-1}}}{G^2}.\qquad \label{solnew2}
\eeqa
We show in \S\ref{app:W2I} that the terms proportional to $R$ vanish as they should, and  (\ref{solnew2}) simplifies to
\begin{multline}
 \frac{\pi a_i^{(1)}(\la_1^2-1)}{32}   + (a_{i+1}^{(1)}- a_i^{(1)})\frac{\ee^{X_i-X_{i+1}}}{G^2}
+ (a_{i-1}^{(1)}- a_i^{(1)}) \frac{\ee^{ - X_{i}+X_{i-1}}}{G^2}
\\
=  \frac{3 \pi}{64} \la_1 b^{(0)}X_i   
- \frac{\la_1  b^{(0)}}{2}(X_i-X_{i+1})\frac{\ee^{X_i-X_{i+1}}}{G^2}
- \frac{\la_1b^{(0)}}{2} (X_{i}-X_{i-1}) \frac{\ee^{ - X_{i}+X_{i-1}}}{G^2}. 
\label{solnew2a}
\end{multline}
Equation (\ref{solnew2a}) is  equation (\ref{solnew2a_main}) from the
main text.
When $b^{(0)}$ is non-zero it gives
 $a_i^{(1)}$ in terms of $b^{(0)}$, with no restriction on $\la_1$.
To determine $\la_1$ we need to go to one more order.

\subsection{$O(G^4)$ terms in the inner expansion}
At $O(G^4)$ in the inner expansion near hump $i$ we find
\[
f_4 = a_i^{(3)} U_1 + a_i^{(1)} U_3 + b_i^{(4)} W_0 + b_i^{(2)} W_2 + b_i^{(0)} W_4,
\]
where
\beqas
\eq{U_3}{U^*_3}
&=&  -\ii \la_1U_2^R - \la_1 \la_2 xV_0 -\ii \la_3 V_0' + \frac{
  \ii\la_1 x (x+X_i)^2}{8}V_0 + 2  \ii \la_1xV_1V_0^4  ,
\eeqas
\begin{multline*}
\eq{W_4}{W^*_4}
= -\ii \la_1 \left( W_3^R -  \ii \la_1 \la_2\frac{x^2 V_0}{4}\right)
-\ii \la_2\left(  -\frac{\ii \la_1^2 x^2V_0}{8}+\ii V_1 + \la_2 \left(\frac{V_0}{4} + \frac{x}{2} \fdd{V_0}{x}\right)\right)\\
-\ii \la_3 \la_1\left(\frac{V_0}{4} + \frac{x}{2} \fdd{V_0}{x} \right) + \la_4 V_0
- \frac{
   (x+X_i)^2}{4}\left(  -\frac{\ii \la_1^2 x^2V_0}{8}+\ii V_1 + \la_2 \left(\frac{V_0}{4} + \frac{x}{2} \fdd{V_0}{x}\right)\right)\\
 - 4 V_1V_0^3 \left(  -\frac{\ii \la_1^2 x^2V_0}{8}+\ii V_1+ 5\la_2 \left(\frac{V_0}{4} + \frac{x}{2} \fdd{V_0}{x}\right)\right)  
- 2 V_0^2(3 V_1^2+2 V_0V_2) \ii V_0 .
\end{multline*}
The relevant solvability conditions are
\beqa
 \im \left[\fdd{U_3}{x}V_0 - U_3  \fdd{V_0}{x}
\right]^R_{-R} & =&  \int_{-R}^R \left(- \la_1 U_2^R V_0 + \frac{\la_1}{4}
x^2 X_i V_0^2 + 2 \la_1 x V_1 V_0^5\right)\, \d x
\label{sol1b},\\
 \im \left[\fdd{W_4}{x}V_0 - W_4  \fdd{V_0}{x}
 \right]^R_{-R}  &=& -\int_{-R}^R \left(\la_1 W_3^R V_0 +
 \left(\frac{(x+X_i)^2 V_0}{4} + 4 V_1 V_0^4\right) \left(V_1 -
   \frac{\la_1^2 x^2 V_0}{8}\right)\right.
 \non \\ && \mbox{ } \hspace{3cm}
\left.\frac{}{} + 2 V_0^4(3 V_1^2+2 V_0V_2)\right) \, \d x
\label{sol3c}.
\eeqa

\subsection{$O(G^2)$ terms in the outer expansion}
At $O(G^2)$ in the outer
\[\ii \la_1 f_1 + \ii \la_2 f_0 +  \sdd{ f_2}{\xi} 
  - f_2=- \frac{\xi^2 f_0}{4}.
\]
Matching with (\ref{f2inf}), (\ref{f2minf}) gives the solution as
\begin{multline}
f_2 = 12^{1/4}b_i^{(2)} \ii \ee^{-\xi+X_i} -  12^{1/4}a_i^{(1)} \frac{\ii \la_1 (\xi-X_i)}{2}\ee^{-\xi+X_i}
-b^{(0)} \frac{\ii \la_1^2 (\xi-X_i)^2}{8} 12^{1/4} \ee^{-\xi+X_i} \\+  12^{1/4}\ii b^{(0)}\frac{X_i}{16}\, \ee^{-\xi+X_i}\left(2(\xi-X_i)^2+2(\xi-X_i)-2\log 2 + 1\right)
+12^{1/4} \ii b^{(0)}\frac{\ee^{X_{i}-X_{i+1}}}{G^2}\ee^{\xi-X_i}\\
+ 12^{1/4} \ii b^{(0)}(\al_0+\al_1(\xi-X_i)+\al_2 (\xi-X_i)^2+\al_3 (\xi-X_i)^3)\ee^{-\xi+X_i} - \ii\frac{X_i^2}{16}12^{1/4} b^{(0)}\left(1 - 2(\xi-X_i) \right)\ee^{-\xi+X_i}\\
 + \frac{\la_2}{4}12^{1/4} b^{(0)}\left(1-2(\xi-X_i)\right)\ee^{-\xi+X_i}\\
+  12^{1/4}b_{i+1}^{(2)} \ii \ee^{\xi-X_{i+1}}-  12^{1/4}a_{i+1}^{(1)} \frac{\ii \la_1 (\xi-X_{i+1})}{2}\ee^{\xi-X_{i+1}}- 
\frac{\ii \la_1^2 (\xi-X_{i+1})^2}{8} 12^{1/4} b^{(0)}\ee^{\xi-X_{i+1}} \\+ 12^{1/4}b^{(0)}\ii\frac{X_{i+1}}{16}\, \ee^{\xi-X_{i+1}}\left(-2(\xi-X_{i+1})^2+2(\xi-X_{i+1})+2\log 2 - 1\right)
+12^{1/4} b^{(0)}\ii\frac{\ee^{X_{i-1}-X_{i}}}{G^2}\ee^{-\xi+X_{i+1}}\\
+ 12^{1/4} b^{(0)}\ii(\al_0-\al_1(\xi-X_{i+1})+\al_2 (\xi-X_{i+1})^2-\al_3 (\xi-X_{i+1})^3)\ee^{\xi-X_{i+1}} - b^{(0)}\ii\frac{X_{i+1}^2}{16}12^{1/4}\left(1 + 2(\xi-X_{i+1}) \right)\ee^{\xi-X_{i+1}}\\+ \frac{\la_2}{4}12^{1/4}b^{(0)} \left(1+2(\xi-X_{i+1})\right)\ee^{\xi-X_{i+1}}.
\end{multline}
This then gives the matching conditions 
\begin{multline}
G^2 f_4 \sim 12^{1/4}b_{i-1}^{(2)} \ii \ee^{-x - X_{i}+X_{i-1}} -  12^{1/4}a_{i-1}^{(1)} \frac{\ii \la_1 (x+X_{i}-X_{i-1})}{2}\ee^{-x-X_{i}+X_{i-1}}\\
-b^{(0)} \frac{\ii \la_1^2 (x+X_{i}-X_{i-1})^2}{8} 12^{1/4} \ee^{-x-X_{i}+X_{i-1}} \\+  12^{1/4}\ii b^{(0)}\frac{X_{i-1}}{16}\, \ee^{-x-X_{i}+X_{i-1}}\left(2(x+X_{i}-X_{i-1})^2+2(x+X_{i}-X_{i-1})-2\log 2 + 1\right)\\
+ 12^{1/4} \ii b^{(0)}(\al_0+\al_1(x+X_{i}-X_{i-1})+\al_2 (x+X_{i}-X_{i-1})^2+\al_3 (x+X_{i}-X_{i-1})^3)\ee^{-x-X_{i}+X_{i-1}} \\- \ii\frac{X_{i-1}^2}{16}12^{1/4} b^{(0)}\left(1 - 2(x+X_{i}-X_{i-1}) \right)\ee^{-x-X_{i}+X_{i-1}}\\
 + \frac{\la_2}{4}12^{1/4} b^{(0)}\left(1-2(x+X_{i}-X_{i-1})\right)\ee^{-x-X_{i}+X_{i-1}}\qquad \mbox{ as } x \ra -\infty\label{G2f4minf}
\end{multline}
\begin{multline}
G^2 f_4 \sim 12^{1/4}b_{i+1}^{(2)} \ii \ee^{x+X_i-X_{i+1}}-  12^{1/4}a_{i+1}^{(1)} \frac{\ii \la_1 (x+X_i-X_{i+1})}{2}\ee^{x+X_i-X_{i+1}}\\- 
\frac{\ii \la_1^2 (x+X_i-X_{i+1})^2}{8} 12^{1/4} b^{(0)}\ee^{x+X_i-X_{i+1}} \\+ 12^{1/4}b^{(0)}\ii\frac{X_{i+1}}{16}\, \ee^{x+X_i-X_{i+1}}\left(-2(x+X_i-X_{i+1})^2+2(x+X_i-X_{i+1})+2\log 2 - 1\right)
\\
+ 12^{1/4} b^{(0)}\ii(\al_0-\al_1(x+X_i-X_{i+1})+\al_2 (x+X_i-X_{i+1})^2-\al_3 (x+X_i-X_{i+1})^3)\ee^{x+X_i-X_{i+1}} \\- b^{(0)}\ii\frac{X_{i+1}^2}{16}12^{1/4}\left(1 + 2(x+X_i-X_{i+1}) \right)\ee^{x+X_i-X_{i+1}}\\+ \frac{\la_2}{4}12^{1/4}b^{(0)} \left(1+2(x+X_i-X_{i+1})\right)\ee^{x+X_i-X_{i+1}}\qquad \mbox{ as } x \ra \infty.\label{G2f4inf}
\end{multline}
on the inner solution near hump $i$.

\subsection{Final solvability condition}
We have, from (\ref{sol1b}), (\ref{sol3c})
\beqa
\lefteqn{ \im \left[\fdd{f_4}{x}V_0 - f_4  \fdd{V_0}{x}
\right]^R_{-R}  = 
a_i^{(1)} \int_{-R}^R \left(- \la_1 U_2^R V_0 + \frac{\la_1}{4}
x^2 X_i V_0^2 + 2 \la_1 x V_1 V_0^5\right)\, \d x } \qquad&& \non \\
&& \mbox{ }
 -b^{(0)}\int_{-R}^R \left(\la_1 W_3^R V_0 + \left(\frac{(x+X_i)^2 V_0}{4} +
   4 V_1 V_0^4\right) \left(V_1 - \frac{\la_1^2 x^2 V_0}{8}\right)
 \non
 + 2 V_0^4(3 V_1^2+2 V_0V_2) \right) \, \d x \non\\
&&\mbox{ } \qquad+b_i^{(2)}\left(\frac{4 \sqrt{3}}{G^2}\ee^{-X_{i+1}+X_i}+\frac{4
  \sqrt{3}}{G^2}\ee^{-X_{i}+X_{i-1}}
\right)\label{finalint}
\eeqa
Evaluating the LHS by matching using (\ref{G2f4minf}), (\ref{G2f4inf})  gives
\beqas
 \im \left[\fdd{f_4}{x}V_0 - f_4  \fdd{V_0}{x}
\right]^R_{-R} & =& \frac{4 \sqrt{3}}{G^2} \ee^{X_i-X_{i+1}} b_{i+1}^{(2)} -
\frac{\sqrt{3}}{G^2} \ee^{X_i-X_{i+1}}\la_1 a_{i+1}^{(1)} (1+2 R + 2 X_i - 2 X_{i+1}) \\ &&\mbox{ }-
\frac{b^{(0)}}{32 \sqrt{3} G^2} \ee^{X_i-X_{i+1}}
\left(\pi^2 + 48 \la_1^2 R + 48 \la^2 R^2 + 16 R^3 + 48(\la_1^2+R)X_i^2\right.\\
&& \mbox{ }\qquad \qquad 
\left. + 16 X_i^3 - 48(\la_1^2+2\la_1^2 R + \log 2) X_{i+1} + 48(1+\la_1^2) X_{i+1}^2 - 16 X_{i+1}^3\right.\\
&& \mbox{ }\qquad \qquad
\left. + 48 X_i(\la_1^2+2\la_1^2 R + R^2 - 2 \la_1^2 X_{i+1})\right)\\
&&\mbox{ }
+ \frac{4 \sqrt{3}}{G^2} \ee^{X_{i-1}-X_i} b_{i-1}^{(2)}
- \frac{\sqrt{3} }{G^2} \ee^{X_{i-1}-X_i} \la_1 a_{i-1}^{(1)}(-1-2R-2X_{i-1}+2X_i) \\
&& \mbox{ } 
- \frac{b^{(0)}}{32 \sqrt{3}G^2}  \ee^{X_{i-1}-X_i}\left(
\pi^2 + 48 \la_1^2 R + 48 \la_1^2 R^2 + 16 R^3 + 48(1+\la_1^2)X_{i-1}^2
\right.\\
&& \mbox{ }\qquad \qquad
\left. +
16 X_{i-1}^3 - 48(R^2 + \la_1^2(1+2R))X_i + 48(\la_1^2+R) X_i^2 - 16 X_i^3 \right.\\
&& \mbox{ }\qquad \qquad
\left.- 48 X_{i-1}(\la_1^2(-1-2R) - \log 2 + 2\la_1^2 X_i)\right).
\eeqas
We show in \S\ref{sec:simplifyfinal} that all the terms in $R$  cancel as they should, and  (\ref{finalint}) simplifies to  
\begin{multline}
\frac{ b^{(0)} \la_1^2(\la_1^2-4) \pi^3}{512} +  \frac{3\pi \la_1 X_i}{64} ( b^{(0)}\la_1  X_i-2a_{i}^{(1)}) 
\\
+4  (b_{i+1}^{(2)} - b_i^{(2)} ) \frac{\ee^{X_i - X_{i+1}}}{G^2}
+4  (b_{i-1}^{(2)} - b_i^{(2)} ) \frac{\ee^{X_{i-1} - X_{i}}}{G^2}
\\
+2  a_{i+1}^{(1)}(X_{i+1} -X_i) \la_1\frac{\ee^{X_i - X_{i+1}}}{G^2}
+2  a_{i-1}^{(1)}(X_{i-1} -X_i)\la_1\frac{\ee^{X_{i-1} - X_{i}}}{G^2}
\\
 -3 a_{i+1}^{(1)} \la_1\frac{\ee^{X_i - X_{i+1}}}{G^2}
+ 3 a_{i-1}^{(1)}\la_1\frac{\ee^{X_{i-1} - X_{i}}}{G^2}
\\
+ \frac{\la_1^2}{2}b^{(0)}\frac{\ee^{X_i - X_{i+1}}}{G^2}\left(
 -  X_i^2  + 3  X_{i+1} + 2 X_i X_{i+1} 
 -  X_{i+1}^2 \right)\\
+ \frac{\la_1^2}{2}b^{(0)}\frac{\ee^{X_{i-1} - X_{i}}}{G^2}\left(
  -  X_i^2 - 3  X_{i-1} + 2 X_i X_{i-1}  
 -  X_{i-1}^2 
\right)=0.\label{finalsol}
\end{multline}
Equation (\ref{finalsol}) is equation (\ref{finalsol_main}) from the main text.
The sum over $i$ is the equation which determines $\la_1$.

\section{Analysis of $V_1$}
\label{app:V1}
We have equation (\ref{inner1}), which we restate for convenience
\beq
\eqr{V_1}  =  -\frac{(x+X_i)^2V_0}{4}.\label{V1eqn}
\eeq
The exponential growth of $V_1$ at infinity arises due to the term proportional to $xV_0$ on the RHS, which is not orthogonal to $\fdd{V_0}{x}$.
We divide $V_1$ into terms forced by the odd and even components on the RHS by setting 
\[ V_1 = V_o + V_e,\]
where
\beqa
\eqr{V_o} & = & -\frac{xX_iV_0}{2},\label{Voeqn}\\
\eqr{V_e} & = & -\frac{(x^2+X_i^2)V_0}{4},\label{Veeqn}
\eeqa
and $V_e \ra 0$ as $x \ra \pm \infty$. Note that $V_e$ is even but this last boundary condition means that $V_o$ is not odd but has an even component (which satisfies the homogeneous equation). 

Since the homogeneous version of (\ref{V1eqn}) has linearly independent solutions
\beq
v_1=  \frac{\sinh (2 \xi)}{\cosh^{3/2} (2 \xi)}, \qquad v_2 =
  \frac{\cosh(4 \xi) - 3}{\cosh^{3/2}(2 \xi)},\label{v1v2}
\eeq
the general solution of (\ref{Voeqn}) may be written as
\[ V_o = \frac{X_i}{2}\frac{v_1(x)}{4} \int_a^x v_2(\bar{x})\bar{x}V_0(\bar{x})\, \d \bar{x} -
\frac{X_i}{2}\frac{v_2(x)}{4}\int_{b}^x v_1(\bar{x}) \bar{x}V_0(\bar{x})\, \d \bar{x},\]
for some constants $a$ and $b$.
Varying the constant $a$ corresponds to translating the position of the inner region. To fix it we need to be more specific about how the $X_i$ are defined. If we define $X_i$ to be the position of the local maximum of the modulus of $V$, then we have $\fdd{V}{\xi}=0$ at $\xi = X_i$, so that  in the inner coordinate $\fdd{V}{x}=0$ at $x=0$. This fixes $a=0$. To determine $b$ we use the matching conditions (\ref{V1minffull}), (\ref{V1inffull}), which give
\[
V_o  \sim   12^{1/4}  \frac{\ee^{x + X_i - X_{i+1}}}{G^2} \quad \mbox{as } x \ra \infty,\qquad
V_o  \sim   12^{1/4}  \frac{\ee^{-x + X_{i-1} - X_{i}}}{G^2} \quad \mbox{as } x \ra -\infty.
\]
We can set $b=-\infty$ if we add
on the multiple of $v_2'(x)$ we expect there. This gives
\beqa
V_o &=& \frac{X_i}{2}\frac{v_1(x)}{4} \int_0^x v_2(\bar{x})\bar{x}V_0(\bar{x})\, \d \bar{x} -
\frac{X_i}{2}\frac{v_2(x)}{4}\int_{-\infty}^x v_1(\bar{x})
\bar{x}V_0(\bar{x})\, \d \bar{x} +
3^{1/4} \frac{\ee^{X_{i-1}-X_i}}{G^2} v_2(x)\non \\
 &=& 3^{1/4}\frac{X_i}{2}\frac{v_1(x)}{4}\left( \log (1+\ee^{4x}) + x^2-2x - 2x \tanh 2x - \log 2
 \right) \non \\
&& \mbox{ }+
\frac{3^{1/4}}{4}\frac{X_i}{2}\frac{v_2(x)}{4}\left(
2 x \sech 2x - \sin^{-1} \sech 2x
\right)
 +
3^{1/4} \frac{\ee^{X_{i-1}-X_i}}{G^2} v_2(x).\label{Vo}
\eeqa
Note the branch of $\sin^{-1}$ is such that
\[  \sin^{-1}\sech 2x = \begin{cases}
\sin^{-1} \sech 2x & x<0,\\
\pi - \sin^{-1} \sech 2x & x>0.
\end{cases}
\]
The odd part of (\ref{Vo}) is
\begin{multline}
V_{\mathrm{odd}} 
 = 3^{1/4}\frac{X_i}{2}\frac{v_1(x)}{4}\left( \log (1+\ee^{4x}) + x^2-2x - 2x \tanh 2x - \log 2
 \right)  \\
+
\frac{3^{1/4}}{4}\frac{X_i}{2}\frac{v_2(x)}{4}\left(
2 x \sech 2x +\frac{\pi}{2}- \sin^{-1} \sech 2x
\right)\label{Vodd}
\end{multline}
leaving an even multiple of $v_2$,
\beqas
V_{o}^e & = & 3^{1/4}\left(-\frac{X_i \pi}{64}+
 \frac{\ee^{X_{i-1}-X_i}}{G^2} \right)v_2(x) =\frac{3^{1/4}}{2}\left(\frac{\ee^{X_{i}-X_{i+1}}}{G^2}+
 \frac{\ee^{X_{i-1}-X_i}}{G^2} \right)v_2(x).
\eeqas
As  $x \ra -\infty$,
\beqa
V_o  
&\sim&  3^{1/4}\frac{X_i}{16}\sqrt{2}\, \ee^{x}\left(-2x^2+2x+2\log 2 - 1\right)
+
3^{1/4} \frac{\ee^{X_{i-1}-X_i}}{G^2} \sqrt{2}\, \ee^{-x}, \label{Vominf}
\eeqa
while as  $x \ra \infty$,
\beqa
V_o 
&\sim&  3^{1/4}\frac{X_i}{16}\sqrt{2}\, \ee^{-x}\left(2x^2+2x-2\log 2 + 1\right)
+3^{1/4} \sqrt{2}\frac{\ee^{X_{i}-X_{i+1}}}{G^2}\ee^{x},\label{Voinf}
 \eeqa
where we have used (\ref{Xeqns}) to simplify.
Let us now turn to (\ref{Veeqn}). 
The solution may be written
\[ V_e = \hat{V}_1- \frac{X_i^2}{4} \left(\frac{V_0}{4} + \frac{x}{2}\fdd{V_0}{x}\right),
\]
where 
\[ \hat{V}_1(x) = v_1(x) \int_0^x \frac{v_2(\bar{x}) \bar{x}^2 V_0(\bar{x})\, \d \bar{x}}{16} + v_2(x) \int_{\bar{x}}^\infty \frac{v_1(\bar{x}) \bar{x}^2 V_0(\bar{x})\, \d \bar{x}}{16},
\]
is the perturbation to the single hump solution (i.e. without the change of origin).
As $x \ra \infty$,
\[ \hat{V}_1 \sim 12^{1/4}(\al_0+\al_1x+\al_2 x^2+\al_3 x^3)\ee^{-x}, \]
with
\beq
\al_0 =  \frac{1}{32 }\left(1 - \frac{\pi^2}{12}\right), \qquad
\al_1  =  \frac{1}{16},\qquad
\al_2  =  \frac{1}{16},\qquad
\al_3  =  \frac{1}{24}.\label{alpha0123}
\eeq
Since $\hat{V}_1$ is even, as $x\ra-\infty$, 
\[ \hat{V}_1 \sim 12^{1/4}(\al_0-\al_1x+\al_2 x^2-\al_3 x^3)\ee^{x}. \]
Thus
\beqa
V_e & \sim & 12^{1/4}(\al_0+\al_1x+\al_2 x^2+\al_3 x^3)\ee^{-x} - \frac{X_i^2}{16}3^{1/4}\sqrt{2}\left(1 - 2x \right)\ee^{-x} \qquad \mbox{ as } x \ra \infty,\label{Veinf}\\
V_e & \sim & 12^{1/4}(\al_0-\al_1x+\al_2 x^2-\al_3 x^3)\ee^{x} - \frac{X_i^2}{16}3^{1/4}\sqrt{2}\left(1 + 2x\right) \ee^{x} \qquad \mbox{ as } x \ra -\infty.\label{Veminf}\qquad
\eeqa
Together (\ref{Vominf}), (\ref{Voinf}), (\ref{Veinf}), and
(\ref{Veminf})  give (\ref{V1minffull}) and (\ref{V1inffull}).

{\bf PGK: here there is a citation missing.}

\section{Analysis of $V_2$}
\label{app:V2}
If we proceed to $O(G^4)$ in the local analysis of \S\ref{sec:nearhump}, \S\ref{sec:nearhump1} we find the equation for $V_2$ is
\beqas
\eqr{V_2} & = & -\frac{(x+X_i)^2V_1}{4} -10 V_0^3 V_1^2.
\eeqas
Multiplying by $V_0$ and integrating gives
\beqas
- \int_{-R}^R \frac{(x+X_i)^2V_1V_0}{4} +10 V_0^4 V_1^2\, \d x
 & = &\int_{-R}^R V_0\left(\eqr{V_2}\right)\, \d x\\
& = & \int_{-R}^R \left(V_2 \sdd{V_0}{x} + 5 V_0^5 V_2 - V_0 V_2  \right)\, \d x 
+ \left[V_0 \fdd{V_2}{x} - V_2 \fdd{V_0}{x}\right]^R_{-R}\\
& = & \int_{-R}^R 4 V_0^5 V_2 \, \d x 
+ \left[V_0 \fdd{V_2}{x} - V_2 \fdd{V_0}{x}\right]^R_{-R}
\eeqas
Thus
\beqa
 \int_{-R}^R 4 V_0^5 V_2 + 10V_0^4 V_1^2  +\frac{(x+X_i)^2V_1V_0}{4} \, \d x  & = & -\left[V_0 \fdd{V_2}{x} - V_2 \fdd{V_0}{x}\right]^R_{-R}.\label{V2Id1}
\eeqa
To evaluate the right-hand side we need to go to one more term in the outer region between humps than we did in \S\ref{sec:gaps}.
Equating coefficients of $G^2$ in the region between humps gives
\beq
\sdd{V_1}{\xi} - V_1 = -\frac{\xi^2}{4}V_0 = -\frac{12^{1/4}\xi^2}{4}\left(\ee^{-\xi+X_i} + \ee^{\xi - X_{i+1}}\right).
\eeq
Thus
\beq
V_1 = 12^{1/4}\left(C_1-\frac{\xi}{16} +  \frac{\xi^2}{16} - \frac{\xi^3}{24}\right)\ee^{\xi - X_{i+1}} + 
12^{1/4}\left(C_2+\frac{\xi}{16} +  \frac{\xi^2}{16} +\frac{\xi^3}{24}\right)\ee^{-\xi + X_{i}}.
\eeq
The coefficients $C_1$ and $C_2$ are determined by matching with the subdominant exponentials in the inner solution $V_1$.
As  $x \ra -\infty$ in the inner solution, from (\ref{Vominf}), (\ref{Veminf}),
\begin{multline*}
V_1 
\sim  3^{1/4}\frac{X_i}{16}\sqrt{2}\, \ee^{x}\left(-2x^2+2x+2\log 2 - 1\right)
+
3^{1/4} \frac{\ee^{X_{i-1}-X_i}}{G^2} \sqrt{2}\, \ee^{-x} \\
+12^{1/4}(\al_0-\al_1x+\al_2 x^2-\al_3 x^3)\ee^{x} - \frac{X_i^2}{16}3^{1/4}\sqrt{2}\left(1 + 2x\right) \ee^{x}.
\end{multline*}
With $x = \xi - X_{i+1}$ (and $i \ra i+1$) this is
\begin{multline*}
V_1 
\sim  3^{1/4}\frac{X_{i+1}}{16}\sqrt{2}\, \ee^{\xi - X_{i+1}}\left(-2(\xi - X_{i+1})^2+2(\xi - X_{i+1})+2\log 2 - 1\right)
+
3^{1/4} \frac{\ee^{X_{i}-X_{i+1}}}{G^2} \sqrt{2}\, \ee^{-(\xi - X_{i+1})} \\
+12^{1/4}(\al_0-\al_1(\xi - X_{i+1})+\al_2 (\xi - X_{i+1})^2-\al_3 (\xi - X_{i+1})^3)\ee^{\xi - X_{i+1}} - \frac{X_{i+1}^2}{16}3^{1/4}\sqrt{2}\left(1 + 2(\xi - X_{i+1})\right) \ee^{\xi - X_{i+1}}.
\end{multline*}
Removing the terms which match with $V_0$ in the outer leaves
\beqas
V_1 
&\sim&  12^{1/4}\ee^{\xi - X_{i+1}}\left(
\frac{X_{i+1}}{16}\, \left(-2(\xi - X_{i+1})^2+2(\xi - X_{i+1})+2\log 2 - 1\right)
\right. \\ && \mbox{ }\left.
+(\al_0-\al_1(\xi - X_{i+1})+\al_2 (\xi - X_{i+1})^2-\al_3 (\xi - X_{i+1})^3)
- \frac{X_{i+1}^2}{16}\left(1 + 2(\xi - X_{i+1})\right) \right)\\
&\sim&  12^{1/4}\ee^{\xi - X_{i+1}}\left(
\frac{1}{32}\left(1-\frac{\pi^2}{12}\right) - \frac{\xi}{16} + \frac{\xi^2}{16} - \frac{\xi^3}{24} - \frac{X_{i+1}^2}{8} + \frac{X_{i+1}^3}{24} + \frac{X_{i+1}\log 2}{8}
\right).
\eeqas
Thus 
\[ C_1 = 
\frac{1}{32}\left(1-\frac{\pi^2}{12}\right)  - \frac{X_{i+1}^2}{8} + \frac{X_{i+1}^3}{24} + \frac{X_{i+1}\log 2}{8}.\]
As  $x \ra \infty$ in the inner solution, from (\ref{Voinf}), (\ref{Veinf}),
\begin{multline*}
V_1
\sim  3^{1/4}\frac{X_i}{16}\sqrt{2}\, \ee^{-x}\left(2x^2+2x-2\log 2 + 1\right)
+3^{1/4} \sqrt{2}\frac{\ee^{X_{i}-X_{i+1}}}{G^2}\ee^{x}\\
+ 12^{1/4}(\al_0+\al_1x+\al_2 x^2+\al_3 x^3)\ee^{-x} - \frac{X_i^2}{16}3^{1/4}\sqrt{2}\left(1 - 2x \right)\ee^{-x}.
\end{multline*}
With $ x = \xi - X_i$ this is
\beqas
V_1
&\sim&  3^{1/4}\frac{X_i}{16}\sqrt{2}\, \ee^{-(\xi - X_i)}\left(2(\xi - X_i)^2+2(\xi - X_i)-2\log 2 + 1\right)
+3^{1/4} \sqrt{2}\frac{\ee^{X_{i}-X_{i+1}}}{G^2}\ee^{(\xi - X_i)}\\
&&\mbox{ }+ 12^{1/4}(\al_0+\al_1(\xi - X_i)+\al_2 (\xi - X_i)^2+\al_3 (\xi - X_i)^3)\ee^{-(\xi - X_i)} \\ && \mbox{ }- \frac{X_i^2}{16}3^{1/4}\sqrt{2}\left(1 - 2(\xi - X_i) \right)\ee^{-(\xi - X_i)}.
\eeqas
Removing the terms which match with $V_0$ in the outer leaves
\beqas
V_1
&\sim&  12^{1/4}\, \ee^{-(\xi - X_i)}
\left(
\frac{X_i}{16}\left(2(\xi - X_i)^2+2(\xi - X_i)-2\log 2 + 1\right)\right.
\\
&&\left.\mbox{ }+ (\al_0+\al_1(\xi - X_i)+\al_2 (\xi - X_i)^2+\al_3 (\xi - X_i)^3)- \frac{X_i^2}{16}\left(1 - 2(\xi - X_i) \right)\right)\\
&\sim&12^{1/4}\ee^{-\xi + X_{i}}\left(
\frac{1}{32}\left(1-\frac{\pi^2}{12}\right) + \frac{\xi}{16} + \frac{\xi^2}{16} + \frac{\xi^3}{24} - \frac{X_{i}^2}{8} - \frac{X_{i}^3}{24} - \frac{X_{i}\log 2}{8}
\right).
\eeqas
Thus
\[ C_2  = \frac{1}{32}\left(1-\frac{\pi^2}{12}\right) - \frac{X_{i}^2}{8} - \frac{X_{i}^3}{24} - \frac{X_{i}\log 2}{8}.
\]
Thus the outer solution for $V_1$ when $X_i<\xi<X_{i+1}$ is
\begin{multline}
V_1 = 12^{1/4}\ee^{-\xi + X_{i}}\left(
\frac{1}{32}\left(1-\frac{\pi^2}{12}\right) + \frac{\xi}{16} + \frac{\xi^2}{16} + \frac{\xi^3}{24} - \frac{X_{i}^2}{8} - \frac{X_{i}^3}{24} - \frac{X_{i}\log 2}{8}
\right)\\+ 12^{1/4}\ee^{\xi - X_{i+1}}\left(
\frac{1}{32}\left(1-\frac{\pi^2}{12}\right) - \frac{\xi}{16} + \frac{\xi^2}{16} - \frac{\xi^3}{24} - \frac{X_{i+1}^2}{8} + \frac{X_{i+1}^3}{24} + \frac{X_{i+1}\log 2}{8}
\right).
\label{outerV1}
\end{multline}
Thus the matching condition on $V_2$ is
\beqas
G^2 V_2 & \sim & 12^{1/4}\ee^{x+X_i - X_{i+1}}\left(
\frac{1}{32}\left(1-\frac{\pi^2}{12}\right) - \frac{x+X_i}{16} +
\frac{(x+X_i)^2}{16} - \frac{(x+X_i)^3}{24}
  - \frac{X_{i+1}^2}{8} + \frac{X_{i+1}^3}{24} + \frac{X_{i+1}\log 2}{8}\right)\\
& \sim & 12^{1/4}\ee^{x+X_i - X_{i+1}}\left(
\frac{1}{32}\left(1-\frac{\pi^2}{12}\right) + \frac{x(-1+2X_i-2X_i^2)}{16} + \frac{x^2(1-2X_i)}{16} - \frac{x^3}{24}\right.\\
&& \mbox{ }\hspace{6cm}  \left. -\frac{X_i}{16}+\frac{X_i^2}{16}-\frac{X_i^3}{24} -\frac{X_{i+1}^2}{8} + \frac{X_{i+1}^3}{24} + \frac{X_{i+1}\log 2}{8}\right),
\eeqas
as $x \ra \infty$, and
\beqas
G^2 V_2 & \sim & 12^{1/4}\ee^{-(x+X_i) + X_{i-1}}\left(
\frac{1}{32}\left(1-\frac{\pi^2}{12}\right) + \frac{(x+X_i)}{16} +
\frac{(x+X_i)^2}{16} + \frac{(x+X_i)^3}{24}
  - \frac{X_{i-1}^2}{8} - \frac{X_{i-1}^3}{24} - \frac{X_{i-1}\log 2}{8}
\right)\\ 
& \sim & 12^{1/4}\ee^{-x-X_i + X_{i-1}}\left(\frac{1}{32}\left(1-\frac{\pi^2}{12}\right) + \frac{x(1+2X_i+2X_i^2)}{16} + \frac{x^2(1+2X_i)}{16}+ \frac{x^3}{24}\right.\\
&& \mbox{ }\hspace{6cm}  \left. +\frac{X_i}{16}+\frac{X_i^2}{16}+\frac{X_i^3}{24} -\frac{X_{i-1}^2}{8} - \frac{X_{i-1}^3}{24} - \frac{X_{i-1}\log 2}{8}\right),
\eeqas
as $x \ra -\infty$.
Thus, finally,
\begin{multline}
\left[V_0 \fdd{V_2}{x} - V_2 \fdd{V_0}{x}\right]^R_{-R}  \\=  
\frac{\sqrt{3}}{96}\ee^{X_i - X_{i+1}}\left(-\pi^2 - 16R^3-48X_iR^2
  -  48 X_i^2 R - 16 X_i^3-48 X_{i+1}^2 + 16 X_{i+1}^3 + 48 \log 2\,
  X_{i+1} \right)\\
+ \frac{\sqrt{3}}{96}\ee^{-X_i + X_{i-1}}\left(-\pi^2 - 16R^3+48X_iR^2
  - 48 X_i^2 R + 16 X_i^3 -48 X_{i-1}^2 - 16 X_{i-1}^3 - 48 \log 2\, X_{i-1} \right).\label{V2V0}
\end{multline}

\section{Derivation of (\ref{solnew2a})}
\label{app:W2I}

The solvability condition (\ref{solnew2}) involves the integral
\begin{multline*}
\lim_{R\ra\infty}\int_{-R}^R  \left(
    V_1 \fdd{V_0}{x} - \left(\frac{ x X_i}{2}+  20 V_1 V_0^3\right)\left(\frac{V_0}{4} +
  \frac{x}{2}\fdd{V_0}{x}\right)\fdd{V_0}{x} 
    \right)\, \d x \\
 =   -\frac{ \sqrt{3}\,\pi^3  X_i}{256} +\lim_{R\ra\infty} \int_{-R}^R   V_{\mathrm{odd}} 
 \left(
    \fdd{V_0}{x}
     -   20  V_0^3\left(\frac{V_0}{4} +
  \frac{x}{2}\fdd{V_0}{x}\right)\fdd{V_0}{x} 
    \right)\, \d x
\end{multline*}
where we have used the fact that
\beqas
\int_{-\infty}^\infty \frac{x}{2}\left(\frac{V_0}{4} +
  \frac{x}{2}\fdd{V_0}{x}\right)\fdd{V_0}{x} 
   \, \d x & = & \frac{\sqrt{3} \pi^3}{256},
\eeqas
and that only the odd part of $V_1$ contributes to the integral.
Using (\ref{Vodd}) we find numerically that
\beqas
\int_{-\infty}^\infty 20 V_{\mathrm{odd}}  V_0^3\left(\frac{V_0}{4} +
  \frac{x}{2}\fdd{V_0}{x}\right)\fdd{V_0}{x} 
    \, \d x & \approx & -0.0731175 X_i.
    \eeqas
Thus we are left with evaluating
\[ \int_{-R}^R V_{\mathrm{odd}}  \fdd{V_0}{x} \, \d x.\]
Now
\beqas
\int^x v_2(\bar{x}) V_0'(\bar{x})\, \d \bar{x} & = & - 3^{1/4} (\log \cosh 2x +  \sech^2 2x).
\eeqas
Thus, integrating by parts,
\beqas
\lefteqn{\lim_{R\ra\infty}\int_{-R}^R  v_2(\bar{x}) V_0'(\bar{x}) 3^{1/4}
 \frac{ X_i}{32} \left(
2 \bar{x} \sech 2\bar{x}+\frac{\pi}{2} - \sin^{-1} \sech 2\bar{x}
 \right) \, \d \bar{x} }\qquad \qquad \qquad &&\\
& = & \sqrt{3}\lim_{R\ra\infty}\left[ -(\log \cosh 2x + \sech^2 2x) \frac{X_i}{32}\left(
  2 {x} \sech 2{x} +\frac{\pi}{2}- \sin^{-1} \sech 2{x}\right)\right]^R_{-R} \\
&&\qquad  \mbox{ }+ \frac{\sqrt{3}X_i}{32}\int_{-\infty}^\infty (\log \cosh 2\bar{x} + \sech^2 2\bar{x})(-4 \bar{x} \sech 2 \bar{x} \tanh 2 \bar{x}) \, \d \bar{x}\\ 
& = & \sqrt{3} (2R - \log 2) \frac{X_i \pi}{32} - 0.31625 X_i,
\eeqas
where we have used the fact that
\[
\frac{\sqrt{3}}{32}\int_{-\infty}^\infty (\log \cosh 2\bar{x} + \sech^2 2\bar{x})(-4 \bar{x} \sech 2 \bar{x} \tanh 2 \bar{x}) \, \d \bar{x} \approx -0.31625.
\]
Since, numerically, 
\[
\int_{-\infty}^\infty \frac{3^{1/4}}{8}V_0'(x) v_1(x)\left( \log (1+\ee^{4x}) + x^2-2x - 2x \tanh 2x - \log 2
\right) \, \d x \approx -0.109394, \]
we have
\beqa
\lim_{R\ra\infty}\int_{-R}^R V_{\mathrm{odd}} \fdd{V_0}{x}\, \d x & = & 
\sqrt{3} (2R - \log 2) \frac{X_i \pi}{32} -0.425644 X_i.\label{VoddV0pint}
\eeqa
Thus, finally,
\beqa
\lefteqn{\lim_{R\ra\infty} \int_{-R}^R \left(
     V_1 \fdd{V_0}{x} - \left(\frac{ x X_i}{2}+  20 V_1 V_0^3\right)\left(\frac{V_0}{4} +
  \frac{x}{2}\fdd{V_0}{x}\right)\fdd{V_0}{x} 
    \right)\, \d x }\hspace{4cm} && \non \\
& = & -\frac{ \sqrt{3}\,\pi^3  X_i}{256} +0.0731175X_i+
\sqrt{3} (2R - \log 2) \frac{X_i \pi}{32}-0.425644 X_i  \non \\
& = &    \frac{\sqrt{3}\, R \pi X_i}{16}-0.680174 X_i.\label{solnew2int}
\eeqa
Using (\ref{solnew2int}) and (\ref{V1xx}) in (\ref{solnew2}) gives
\begin{multline*}
 \frac{\sqrt{3}\, \pi a_i^{(1)}(\la_1^2-1)}{8}  +a_i^{(1)}\left(-\frac{4 \sqrt{3}}{G^2}\ee^{-X_{i+1}+X_i}-\frac{4 \sqrt{3}}{G^2}\ee^{-X_{i}+X_{i-1}}
\right) +
\la_1 b^{(0)} \left( \frac{\sqrt{3}\, R \pi X_i}{16}-0.680174 X_i
\right)\\
  =  - 2\sqrt{3}\, \left(2a_{i+1}^{(1)}+
 \la_1 (1+R+X_i-X_{i+1}) b^{(0)} \right)\frac{\ee^{X_i-X_{i+1}}}{G^2}
\\
- 2 \sqrt{3}\, \left(2a_{i-1}^{(1)}  +
  \la_1 (-1-R+X_{i}-X_{i-1}) b^{(0)}\right)
\frac{\ee^{ - X_{i}+X_{i-1}}}{G^2}.
\end{multline*}
Using (\ref{Xeqns}) the terms proportional to $R$ cancel as they should, leaving
\begin{multline*}
 \frac{\sqrt{3}\, \pi a_i^{(1)}(\la_1^2-1)}{8}   + 4\sqrt{3}\,(a_{i+1}^{(1)}- a_i^{(1)})\frac{\ee^{X_i-X_{i+1}}}{G^2}
+ 4 \sqrt{3}\,(a_{i-1}^{(1)}- a_i^{(1)}) \frac{\ee^{ - X_{i}+X_{i-1}}}{G^2}
\\
 = 1.02026\,\la_1 b^{(0)}X_i
- 2\sqrt{3}\, \la_1  b^{(0)}\left((X_i-X_{i+1})\frac{\ee^{X_i-X_{i+1}}}{G^2}
+ (X_{i}-X_{i-1}) \frac{\ee^{ - X_{i}+X_{i-1}}}{G^2}\right).
\end{multline*}
Dividing by $4 \sqrt{3}$ gives
\begin{multline}
 \frac{\pi a_i^{(1)}(\la_1^2-1)}{32}   + (a_{i+1}^{(1)}- a_i^{(1)})\frac{\ee^{X_i-X_{i+1}}}{G^2}
+ (a_{i-1}^{(1)}- a_i^{(1)}) \frac{\ee^{ - X_{i}+X_{i-1}}}{G^2}
\\
 = 0.147262\,\la_1 b^{(0)}X_i
-  \frac{\la_1  b^{(0)}}{2}\left((X_i-X_{i+1})\frac{\ee^{X_i-X_{i+1}}}{G^2}
+ (X_{i}-X_{i-1}) \frac{\ee^{ - X_{i}+X_{i-1}}}{G^2}\right).\label{solnew2temp}
\end{multline}
We can check our analysis, and identify the constant $0.147262$, by considering the exact eigenfunction corresponding to $\la=2G$.
For this eigenfunction
\beqas 
f &=& \ii V_s + G\left( \frac{V_s}{2} + \xi \fdd{V_s}{\xi} - \frac{\ii G \xi V_s}{2}\right)
\eeqas
so that with $\xi = x + X_i$,
\beqas
f & = & \ii V_0 + G\left(\frac{V_0}{2} + (x+X_i) \fdd{V_0}{\xi} \right) 
+ \ii G^2 \left(  V_1 - \frac{ (x+X_i) V_0}{2}\right) + G^3   \left( \frac{V_1}{2} + \xi \fdd{V_1}{\xi}\right)+\cdots.
\eeqas
Comparing to our expansion, with $b^{(0)}=1$, and $\la_1=2$,
\beqas
f & = & \ii V_0 + G \left(a_i^{(1)} \fdd{V_0}{x} + \frac{V_0}{2} + x \fdd{V_0}{x}\right) +\ii  G^2\left(- a_i^{(1)} x V_0 + b_i^{(2)} V_0  
- \frac{x^2 V_0}{2} + V_1 
\right)  + G^3\left(\cdots\right)
\eeqas
 we have 
\[ a_i^{(1)} = X_i, \qquad b_i^{(2)} = -\frac{X_i^2}{2}.\]
Substituting this into (\ref{solnew2temp}) gives 
\begin{multline}
 \frac{3\pi X_i}{32}   + (X_{i+1}- X_i)\frac{\ee^{X_i-X_{i+1}}}{G^2}
+ (X_{i-1}- X_i) \frac{\ee^{ - X_{i}+X_{i-1}}}{G^2}
\\
 = 2\times 0.147262\, X_i
-  \left((X_i-X_{i+1})\frac{\ee^{X_i-X_{i+1}}}{G^2}
+ (X_{i}-X_{i-1}) \frac{\ee^{ - X_{i}+X_{i-1}}}{G^2}\right).\label{solnew2temp2}
\end{multline}
This identifies the constant as 
\beq
 0.147262 = \frac{3 \pi}{64},\label{0.14}
\eeq
so that  (\ref{solnew2temp}) becomes (\ref{solnew2a}).

\section{Derivation of (\ref{finalsol})}
\label{sec:simplifyfinal}
Let us evaluate the integrals on the right-hand side of (\ref{finalint}), namely 
\beq
I_1 =\lim_{R\ra\infty} \int_{-R}^R \left(- \la_1 U_2^R V_0 + \frac{\la_1}{4}
x^2 X_i V_0^2 + 2 \la_1 x V_1 V_0^5\right)\, \d x \label{U2Rint} 
\eeq
and
\beq
I_2=-\lim_{R\ra\infty} \int_{-R}^R \la_1 W_3^R V_0 + \left(\frac{(x+X_i)^2 V_0}{4} + 4 V_1 V_0^4\right) \left(V_1 - \frac{\la_1^2 x^2 V_0}{8}\right)  + 2 V_0^4(3 V_1^2+2 V_0V_2) \, \d x. \label{W3Iint}
\eeq
For (\ref{U2Rint}) we require the odd part of $V_1$. Using (\ref{Vodd}) gives, numerically
\beq
\int_{-\infty}^\infty \left( \frac{\la_1}{4}
x^2 X_i V_0^2 + 2  \la_1 x V_1 V_0^5\right)\, \d x  \approx 0.373466 \la_1 X_i.
\label{Apart2}
\eeq
From (\ref{V2Id1}) we have
\beqa
 \int_{-R}^R 4 V_0^5 V_2 + 10V_0^4 V_1^2  +\frac{(x+X_i)^2V_1V_0}{4} \, \d x  & = & -\left[V_0 \fdd{V_2}{x} - V_2 \fdd{V_0}{x}\right]^R_{-R}.\label{V2Id1copy}
\eeqa
so that (\ref{W3Iint}) is 
\beq
I_2 = \left[V_0 \fdd{V_2}{x} - V_2 \fdd{V_0}{x}\right]^R_{-R}-\int_{-R}^R \la_1 W_3^R V_0 + \left(\frac{(x+X_i)^2 V_0}{4} + 4 V_1 V_0^4\right) \left( - \frac{\la_1^2 x^2 V_0}{8}\right)   \, \d x \label{W3Iint2}
\eeq
Also
\beqa
\int_{-\infty}^\infty \frac{(x+X_i)^2 V_0}{4} \left(- \frac{\la_1^2 x^2 V_0}{8}\right) \, \d x &=& -\la_1^2 \frac{\sqrt{3}\, \pi^3(5 \pi^2 + 16 X_i^2)}{16384},\\
\int_{-\infty}^\infty4 V_1 V_0^4\left(- \frac{\la_1^2 x^2 V_0}{8}\right)  \, \d x &= &- \frac{\la_1^2}{2}\int_{-\infty}^\infty x^2V_{\mathrm{even}} V_0^5   \, \d x,
\eeqa
where 
\beq
V_{\mathrm{even}} = \frac{3^{1/4}}{2}\left(\frac{\ee^{X_{i}-X_{i+1}}}{G^2}+
 \frac{\ee^{X_{i-1}-X_i}}{G^2} \right)v_2(x) + \hat{V}_1(x)- \frac{X_i^2}{4} \left(\frac{V_0}{4} + \frac{x}{2}\fdd{V_0}{x}\right).\label{Veven}
\eeq
Since
\beqas
\int_{-\infty}^\infty v_2(x) x^2 V_0(x)^5  \, \d x & = & \frac{3^{1/4}(24-\pi^2)}{24} ,\\
\int_{-\infty}^\infty  \hat{V}_1(x) x^2 V_0(x)^5  \, \d x & \approx & 0.0227007 ,\\
\int_{-\infty}^\infty  \left(\frac{V_0}{4} + \frac{x}{2}\fdd{V_0}{x}\right)x^2 V_0(x)^5  \, \d x & = &  0,
\eeqas
\beqa
- \frac{\la_1^2}{2}\int_{-\infty}^\infty x^2V_{\mathrm{even}} V_0^5   \, \d x
& = & - \la_1^2\left(\frac{\sqrt{3}(24-\pi^2)}{96} \left(\frac{\ee^{X_{i}-X_{i+1}}}{G^2}+
 \frac{\ee^{X_{i-1}-X_i}}{G^2} \right)+ 0.0113504 
\right).\qquad\quad
\eeqa
Collecting all this together gives
\beqa
I_1 & = & - \la_1 \lim_{R\ra\infty} \int_{-R}^R  U_2^R V_0 \, \d x +  0.373466 \la_1 X_i, \label{I1} 
\\
I_2 & = &-\la_1\int_{-R}^R  W_3^R V_0 \, \d x + \left[V_0 \fdd{V_2}{x} - V_2 \fdd{V_0}{x}\right]^R_{-R}
+\la_1^2 \frac{\sqrt{3}\, \pi^3(5 \pi^2 + 16 X_i^2)}{16384}\non \\&& \mbox{ }
+ \la_1^2\left(\frac{\sqrt{3}(24-\pi^2)}{96} \left(\frac{\ee^{X_{i}-X_{i+1}}}{G^2}+
 \frac{\ee^{X_{i-1}-X_i}}{G^2} \right)+ 0.0113504 
\right).\label{I2}
\eeqa
The term in square brackets involving $V_0$ and $V_2$ is evaluated in \S\ref{app:V2} and given by (\ref{V2V0}). We are left with the integrals involving $U_2^R$ and $W_3^R$.

\subsection{Analysis of $U_2^R$ and $W_3^R$}
Differentiating (\ref{inner1}) gives
\beqas
\frac{\rd^3 V_1}{\rd x^3} + 5 V_0^4 \fdd{V_1}{x} - \fdd{V_1}{x} & = &
-20 V_0^3 V_1 \fdd{V_0}{x} -\frac{(x+X_i)V_0}{2} -\frac{(x+X_i)^2}{4}\fdd{V_0}{x}.
\eeqas
Thus, recalling (\ref{eq:U2R}), (\ref{eq:W3R})   we may write
\beqa
 U_2^R &=& \fdd{V_1}{x} + U,\label{Ueqn}\\
 W_3^R &=& \la_1\left(\frac{V_1}{4}+\frac{x}{2}\fdd{V_1}{x}\right)+W,\label{Weqn}
\eeqa
where 
\beqas
 \eqr{U} &=&  \frac{(1-\la_1^2)}{2}xV_0 + \frac{X_i}{2}V_0,\\
\eqr{W} &=& \frac{\la_1(2x^2+3x X_i + X_i^2)}{4}  V_0  -\frac{\la_1^3 x^2}{8}V_0.
\eeqas
Integrating by parts
 \beq
\int_{-R}^R \fdd{V_1}{x} V_0 \, \d x = \left[ V_1 V_0 \right]^R_{-R} -\int_{-R}^R V_1 \fdd{V_0}{x} \, \d x = \left[ V_1 V_0 \right]^R_{-R} - \int_{-R}^R V_{\mathrm{odd}} \fdd{V_0}{x} \, \d x,\label{V1pV0}
\eeq
where we evaluated the final integral already in (\ref{VoddV0pint}).
Note that
\beqa
\left[ V_1 V_0 \right]^R_{-R}& = & 2\sqrt{3}\frac{\ee^{X_i-X_{i+1}}}{G^2} - 2\sqrt{3}\frac{\ee^{-X_i+X_{i-1}}}{G^2}  = -\frac{\sqrt{3}\, \pi X_i}{16}.\label{V1V0}
\eeqa
Also
\[ \int_{-R}^R \left(\frac{V_1}{4}+\frac{x}{2}\fdd{V_1}{x}\right)V_0\,  \d x = \int_{-R}^R \left(\frac{V_{\mathrm{even}}}{4}+\frac{x}{2}\fdd{V_{\mathrm{even}}}{x}\right)V_0\,  \d x,\]
where $V_{\mathrm{even}}$ is given by (\ref{Veven}).
Since
\beqas
\lim_{R\ra\infty}\int_{-R}^R \left(\frac{v_2}{4}+\frac{x}{2}\fdd{v_2}{x}\right)V_0\,  \d x & = &  
3^{1/4} \left(\frac{1}{2}-\frac{\pi^2}{48} +R^2+R\right)
,\\
\int_{-\infty}^\infty \left(\frac{\hat{V}_1}{4}+\frac{x}{2}\fdd{\hat{V}_1}{x}\right)V_0\,  \d x & \approx & 0.173106 ,\\
\int_{-\infty}^\infty V_0\left(\frac{1}{4}+\frac{x}{2}\fdd{}{x}\right)^2V_0\,  \d x & = & -\frac{\sqrt{3}\,\pi^3}{256} ,
\eeqas
we find
\beqa
\lim_{R \ra\infty}\int_{-R}^R\left(\frac{V_1}{4}+\frac{x}{2}\fdd{V_1}{x}\right)V_0\,  \d x & = & 
\frac{\sqrt{3}}{2}\left(\frac{\ee^{X_{i}-X_{i+1}}}{G^2}+
 \frac{\ee^{X_{i-1}-X_i}}{G^2} \right) \left(\frac{1}{2}-\frac{\pi^2}{48} +R^2+R\right)\non \\&& \mbox{ }
+  0.173106
+  \frac{\sqrt{3}\,\pi^3}{1024} X_i^2.\label{V1xV1p}
\eeqa
This leaves us just with the integrals involving $U$ and $W$.
To determine $U$ and $W$ we will need to match with the solution in between the humps, which means we have to reintroduce the constant multipliers.
The actual term in the integral (\ref{finalint}) is
\beq
I_3 = \lim_{R \ra \infty} \int_{-R}^R a_i^{(1)} U + b^{(0)} W \, \d x .
\eeq
The general solution to
\[ \eqr{f} = g,\]
is
\[ f = -\frac{v_1(x)}{4} \int_a^x v_2(\bar{x})g(\bar{x})\, \d \bar{x} +
\frac{v_2(x)}{4}\int_{b}^x v_1(\bar{x}) g(\bar{x})\, \d \bar{x},\]
where $v_1$ and $v_2$ are the homogeneous solutions given by (\ref{v1v2}).
With $f = U$, we have
\[ g = g_U = \frac{(1-\la_1^2)}{2}x V_0 + \frac{X_i}{2} V_0,\]
while with $f = W$, we have
\[ g =g_W=   \frac{\la_1(2x^2+3x X_i + X_i^2)}{4}  V_0  -\frac{\la_1^3 x^2}{8}V_0
.\]
Note that 
\beqas
 \int^x V_0(\bar{x}) v_1(\bar{x})\, \d \bar{x} &=&-\frac{V_0(x)^2}{2\times3^{1/4}} ,\\
 \int^x V_0(\bar{x}) v_2(\bar{x})\, \d \bar{x} &=& 2\times 3^{1/4}(x - \tanh 2x).
\eeqas
Then, integrating by parts,
\beqas
\lefteqn{\lim_{R\ra\infty}\int_{-R}^R f V_0 \, \d x = \frac{1}{4\times 3^{1/4}} \lim_{R\ra\infty}\int_{-R}^R V_0(x) V_0'(x) \int_a^x v_2(\bar{x})g(\bar{x})\, \d \bar{x} \, \d x } && \\
&&\hspace{4cm}\mbox{ }-
\frac{1}{4\times 3^{1/4}}\lim_{R\ra\infty} \int_{-R}^RV_0(x)v_2(x) \int_{b}^x V_0'(\bar{x}) g(\bar{x})\, \d \bar{x} \, \d x\\
&=& \frac{1}{4\times 3^{1/4}} \lim_{R\ra\infty}\left[ \frac{V_0(x)^2}{2} \int_a^x v_2(\bar{x})g(\bar{x}) \, \d \bar{x}\right]^R_{-R}-\frac{1}{4\times 3^{1/4}}\lim_{R\ra\infty} \int_{-R}^R \frac{V_0(x)^2}{2} v_2(x)g(x) \, \d x\\&& \mbox{ } -
\frac{1}{2} \lim_{R\ra\infty}\left[(x-\tanh 2x) \int_{b}^x V_0'(\bar{x}) g(\bar{x})\, \d \bar{x} \right]^R_{-R} +
\frac{1}{2} \lim_{R\ra\infty}\int_{-R}^R(x-\tanh 2x) V_0'(x) g(x) \, \d x\\
&= & -\frac{1}{4\times 3^{1/4}} \int_{-\infty}^\infty \frac{V_0(x)^2}{2} v_2(x)g(x) \, \d x -
\frac{1}{2} (R-1)  \int_{b}^\infty V_0'(\bar{x}) g(\bar{x})\, \d \bar{x}   \\&& \mbox{ }-
\frac{1}{2} (R-1) \int_{b}^{-\infty} V_0'(\bar{x}) g(\bar{x})\, \d \bar{x}  +
\frac{1}{2} \int_{-\infty}^\infty(x-\tanh 2x) V_0'(x) g(x) \, \d x.
\eeqas
Evaluating the full range integrals we find
\beqas
\int_{-\infty}^\infty  V_0(x)^2 v_2(x) g_U(x) \, \d x & = &0,\\
\int_{-\infty}^\infty  (x-\tanh 2x) V_0'(x) g_U(x)  \, \d x & = &0,\\
\int_{-\infty}^\infty  V_0(x)^2 v_2(x) g_W(x) \, \d x & = & \frac{\la_1(4-\la_1^2)}{8}\frac{3^{3/4} \pi}{2},\\
\int_{-\infty}^\infty  (x-\tanh 2x) V_0'(x) g_W(x)  \, \d x & = &\frac{\la_1(4-\la_1^2)}{8}\frac{\sqrt{3}\, \pi(4-\pi^2)}{32} .
\eeqas
Thus
\beqa
\lim_{R\ra\infty}\int_{-R}^R U V_0 & = &  -
\frac{1}{2} (R-1)  \int_{b}^\infty V_0'(\bar{x}) g_U(\bar{x})\, \d \bar{x}   
-
\frac{1}{2} (R-1) \int_{b}^{-\infty} V_0'(\bar{x}) g_U(\bar{x})\, \d \bar{x} ,\label{Uint}\\
\lim_{R\ra\infty}\int_{-R}^R W V_0 & = &  -
\frac{1}{2} (R-1)  \int_{b}^\infty V_0'(\bar{x}) g_W(\bar{x})\, \d \bar{x} \non  \\&& \mbox{ }-
\frac{1}{2} (R-1) \int_{b}^{-\infty} V_0'(\bar{x}) g_W(\bar{x})\, \d \bar{x}  -
\frac{\sqrt{3}\,\pi^3\la_1(4-\la_1^2)}{512}
.\label{Wint}
\eeqa
To evaluate the reminaing integrals we need to determine $b$ using 
the matching conditions (\ref{f3minf}), (\ref{f3inf}) on $f_3$, which give
\beqas
(a_i^{(1)}U_2 + b^{(0)} W_3)G^2 &\sim& 12^{1/4}  \left(-a_{i-1}^{(1)}-
  \frac{\la_1 (x+X_{i}-X_{i-1})}{2} b^{(0)}+\frac{\la_1}{4}b^{(0)}\right)\ee^{-x
  - X_{i}+X_{i-1}}\quad \mbox{ as }x \ra -\infty,\\
(a_i^{(1)}U_2 + b^{(0)} W_3)G^2 &\sim&  12^{1/4} \left(a_{i+1}^{(1)}+
 \frac{\la_1 (x+X_i-X_{i+1})}{2}
 b^{(0)}+\frac{\la_1}{4}b^{(0)}\right)\ee^{x+X_i-X_{i+1}}\quad \mbox{ as } x \ra \infty.
\eeqas
Using (\ref{V1inffull}), (\ref{V1minffull}), (\ref{Ueqn}), (\ref{Weqn}), gives the matching conditions
\beqas
a_i^{(1)}U + b^{(0)}W & \sim & 12^{1/4}  \left(a_i^{(1)}-a_{i-1}^{(1)}-
  \frac{\la_1 (X_{i}-X_{i-1})}{2} b^{(0)}\right)\frac{\ee^{-x
  - X_{i}+X_{i-1}}}{G^2}\qquad \mbox{ as }x \ra -\infty,\\
a_i^{(1)}U + b^{(0)} W &\sim&  12^{1/4} \left(-a_i^{(1)}+a_{i+1}^{(1)}+
 \frac{\la_1 (X_i-X_{i+1})}{2}
 b^{(0)}\right)\frac{\ee^{x+X_i-X_{i+1}}}{G^2}\qquad \mbox{ as } x \ra \infty.
\eeqas
We can satisfy the conditions at $-\infty$ by setting $b=-\infty$ in both $U$ and $W$ and adding an appropriate multiple of $v_2$ corresponding to the matching condition.
Since
\[ v_2 \sim \sqrt{2}\, \ee^{-x} \mbox{ as }x \ra -\infty,\]
this multiple is
\beq
 3^{1/4}  \left(a_i^{(1)}-a_{i-1}^{(1)}-
  \frac{\la_1 (X_{i}-X_{i-1})}{2} b^{(0)}\right)\frac{\ee^{
  - X_{i}+X_{i-1}}}{G^2}v_2(x).\label{v2mult}
\eeq
As a check, at infinity then
\beqas
a_i^{(1)}U + b^{(0)}W & \sim & 3^{1/4}  \left(a_i^{(1)}-a_{i-1}^{(1)}-
  \frac{\la_1 (X_{i}-X_{i-1})}{2} b^{(0)}\right)\frac{\ee^{
  - X_{i}+X_{i-1}}}{G^2}v_2(x)\\
&&\qquad\mbox{ }+\frac{v_2}{4}\int_{-\infty}^\infty v_1(\bar{x})\left(a_i^{(1)}g_U + b^{(0)}g_W\right)\, \d\bar{x}\\
 & = & 3^{1/4}  \left(a_i^{(1)}-a_{i-1}^{(1)}-
  \frac{\la_1 (X_{i}-X_{i-1})}{2} b^{(0)}\right)\frac{\ee^{
  - X_{i}+X_{i-1}}}{G^2}v_2(x)\\
&&\qquad\mbox{ }-\frac{v_2}{4\times 3^{1/4}}\left(a_i^{(1)}\frac{(1-\la_1^2)}{2} + b^{(0)}\frac{\la_1 3 X_i}{4}\right)\int_{-\infty}^\infty V_0'(\bar{x})\bar{x}V_0(\bar{x})\, \d\bar{x}\\
 & = & 12^{1/4}  \left(a_i^{(1)}-a_{i-1}^{(1)}-
  \frac{\la_1 (X_{i}-X_{i-1})}{2} b^{(0)}\right)\frac{\ee^{
  - X_{i}+X_{i-1}}}{G^2}\ee^x\\
&&\qquad\mbox{ }+12^{1/4}\pi\left(a_i^{(1)}\frac{(1-\la_1^2)}{32} + b^{(0)}\frac{\la_1 3 X_i}{64}\right)\ee^x\\
& = &  12^{1/4}  \left(a_{i+1}^{(1)}-a_{i}^{(1)}+
  \frac{\la_1 (X_{i}-X_{i+1})}{2} b^{(0)}\right)\frac{\ee^{x+
  X_{i}-X_{i+1}}}{G^2}
\eeqas
as required, using (\ref{solnew2a}) (which also provides proof of \eqref{0.14}).
With $b=-\infty$ we have two further integrals to evaluate.
We have
\beqa
-\frac{1}{2}(R-1)\int_{-\infty}^\infty V_0' (a_i^{(1)}g_U + b^{(0)} g_W)\, \d x & = & -\frac{1}{2}(R-1) \left(a_i^{(1)}\frac{(1-\la_1^2)}{2} + b^{(0)} \frac{3\la_1 X_i}{4}\right) \int_{-\infty}^\infty V_0' x V_0\, \d x \non \\
& = & \frac{1}{2}(R-1) \left(a_i^{(1)}\frac{(1-\la_1^2)}{2} + b^{(0)} \frac{3\la_1 X_i}{4}\right) \frac{\sqrt{3}\, \pi}{4}\label{binfintegrals}
\eeqa
and we also need the integral of the extra term (\ref{v2mult}), 
\begin{multline}
 3^{1/4}  \left(a_i^{(1)}-a_{i-1}^{(1)}-
  \frac{\la_1 (X_{i}-X_{i-1})}{2} b^{(0)}\right)\frac{\ee^{
  - X_{i}+X_{i-1}}}{G^2} \int_{-R}^R v_2 V_0\, \d x \\ =  
4  \sqrt{3}  \left(a_i^{(1)}-a_{i-1}^{(1)}-
  \frac{\la_1 (X_{i}-X_{i-1})}{2} b^{(0)}\right)\frac{\ee^{
  - X_{i}+X_{i-1}}}{G^2} (R - 1).\label{extraterm}
\end{multline}
Collecting together the results of this section we have, using (\ref{Ueqn}), (\ref{V1V0}),
 and \eqref{VoddV0pint},
\beq
\lim_{R\ra\infty}\int_{-R}^R U_2^R V_0\, \d x  = 
- \frac{\sqrt{3} \,R X_i \pi}{16}  +0.203422 X_i+\lim_{R\ra\infty} \int_{-R}^R U V_0 \, \d x\label{U2R}
\eeq
since
\[
-\sqrt{3}\, \frac{X_i \pi}{16}+\frac{\sqrt{3}\,\pi\log 2}{32}+0.425644 \approx 0.203422.
\]
Using (\ref{Weqn}), (\ref{V1xV1p}), 
\beqa
\lim_{R\ra\infty}\int_{-R}^R W_3^R V_0 \, \d x 
& = &  \frac{\sqrt{3}\,\la_1}{2}\left(\frac{\ee^{X_{i}-X_{i+1}}}{G^2}+
 \frac{\ee^{X_{i-1}-X_i}}{G^2} \right) \left(\frac{1}{2}-\frac{\pi^2}{48} +R^2+R\right) \non\\&& \mbox{ }
+  0.173106\la_1
+  \frac{\sqrt{3}\,\pi^3 \la_1}{1024} X_i^2
+ \lim_{R\ra\infty}\int_{-R}^R W V_0\,\d x.\label{W3R}
\eeqa
Using (\ref{Uint}), (\ref{Wint}), (\ref{binfintegrals}), (\ref{extraterm}), 
\beqa
\int_{-R}^R (a_i^{(1)}U + b^{(0)}W)V_0\, \d x
& = & 
-   b^{(0)} 
\la_1(4-\la_1^2)\frac{\sqrt{3}\, \pi^3}{512}
\non\\
&&\mbox{ }+(R-1) \left(a_i^{(1)}\frac{(1-\la_1^2)}{16} + b^{(0)} \frac{3\la_1 X_i}{32}\right) \sqrt{3}\, \pi
\non\\
&&\mbox{ }+4  \sqrt{3}  \left(a_i^{(1)}-a_{i-1}^{(1)}-
  \frac{\la_1 (X_{i}-X_{i-1})}{2} b^{(0)}\right)\frac{\ee^{
  - X_{i}+X_{i-1}}}{G^2} (R - 1)\non\\
& = & 
-   b^{(0)} 
\la_1(4-\la_1^2)\frac{\sqrt{3}\, \pi^3}{512}
\non\\
&&\mbox{ }+\sqrt{3}\,(R-1) \left(
\la_1  b^{(0)}(X_i-X_{i+1})\frac{\ee^{X_i-X_{i+1}}}{G^2}
- \la_1b^{(0)} (X_{i}-X_{i-1}) \frac{\ee^{ - X_{i}+X_{i-1}}}{G^2}
\right)\non \\
&&\mbox{ } +\sqrt{3}\,(R-1) \left(
 2(a_{i+1}^{(1)}- a_i^{(1)})\frac{\ee^{X_i-X_{i+1}}}{G^2}
- 2(a_{i-1}^{(1)}- a_i^{(1)}) \frac{\ee^{ - X_{i}+X_{i-1}}}{G^2}
\right) \label{aUbW}
\eeqa
using (\ref{solnew2a}).

\subsection{Putting it all together}
Using (\ref{I1}), (\ref{I2}), (\ref{V2V0}), (\ref{U2R}), (\ref{W3R}) and (\ref{aUbW}) in (\ref{finalint}) gives, after some simplification,
\begin{multline*}
 \im \left[\fdd{f_4}{x}V_0 - f_4  \fdd{V_0}{x}
\right]^R_{-R}  =  
 0.170044\, a_i^{(1)} \la_1 X_i + \frac{\sqrt{3} \, a_i^{(1)}  R \la_1 X_i \pi}{16}  
+  b^{(0)} 
\la_1^2(4-\la_1^2)\frac{\sqrt{3}\, \pi^3}{512}
\\
-\sqrt{3}\,(R-1) \la_1^2\left(
  b^{(0)}(X_i-X_{i+1})\frac{\ee^{X_i-X_{i+1}}}{G^2}
- b^{(0)} (X_{i}-X_{i-1}) \frac{\ee^{ - X_{i}+X_{i-1}}}{G^2}
\right) \\
 -2\sqrt{3}\,(R-1)\la_1 \left(
 (a_{i+1}^{(1)}- a_i^{(1)})\frac{\ee^{X_i-X_{i+1}}}{G^2}
- (a_{i-1}^{(1)}- a_i^{(1)}) \frac{\ee^{ - X_{i}+X_{i-1}}}{G^2}
\right) 
 \\
- b^{(0)}\la_1^2 \frac{\sqrt{3}}{2}\left(\frac{\ee^{X_{i}-X_{i+1}}}{G^2}+
 \frac{\ee^{X_{i-1}-X_i}}{G^2} \right) \left(R^2+R\right) \\
+b^{(0)}\frac{\sqrt{3}}{96}\ee^{X_i - X_{i+1}}\left(-\pi^2 -
  16R^3-48X_iR^2 - 48 X_i^2 R - 16 X_i^3
  -48 X_{i+1}^2 + 16 X_{i+1}^3 + 48 \log 2\, X_{i+1} \right) \\
+ b^{(0)}\frac{\sqrt{3}}{96}\ee^{-X_i + X_{i-1}}\left(-\pi^2 -
  16R^3+48X_iR^2 - 48 X_i^2 R + 16 X_i^3
  -48 X_{i-1}^2 - 16 X_{i-1}^3 - 48 \log 2\, X_{i-1} \right)
\\
+b_i^{(2)}\left(\frac{4 \sqrt{3}}{G^2}\ee^{-X_{i+1}+X_i}+\frac{4
  \sqrt{3}}{G^2}\ee^{-X_{i}+X_{i-1}}
\right)
\end{multline*}
where we have set
\[ 0.173106 - 0.0113504 = \frac{5\sqrt{3}\,\pi^5}{16384}.\]
Thus the final solvability condition is
\begin{multline*}
\frac{4 \sqrt{3}}{G^2} \ee^{X_i-X_{i+1}} b_{i+1}^{(2)} -
\frac{\sqrt{3}}{G^2} \ee^{X_i-X_{i+1}}\la_1 a_{i+1}^{(1)} (1+2 R + 2 X_i - 2 X_{i+1}) \\ -
\frac{b^{(0)}}{32 \sqrt{3} G^2} \ee^{X_i-X_{i+1}}
\left(\pi^2 + 48 \la_1^2 R + 48 \la_1^2 R^2 + 16 R^3 + 48(\la_1^2+R)X_i^2
 + 16 X_i^3 
\right.\\ \left. - 48(\la_1^2+2\la_1^2 R + \log 2) X_{i+1} + 48(1+\la_1^2) X_{i+1}^2 
\right.\\ \left. - 16 X_{i+1}^3
 + 48 X_i(\la_1^2+2\la_1^2 R + R^2 - 2 \la_1^2 X_{i+1})\right)
\\ 
+ \frac{4 \sqrt{3}}{G^2} \ee^{X_{i-1}-X_i} b_{i-1}^{(2)}
- \frac{\sqrt{3} }{G^2} \ee^{X_{i-1}-X_i} \la_1 a_{i-1}^{(1)}(-1-2R-2X_{i-1}+2X_i) \\
- \frac{b^{(0)}}{32 \sqrt{3}G^2}  \ee^{X_{i-1}-X_i}\left(
\pi^2 + 48 \la_1^2 R + 48 \la_1^2 R^2 + 16 R^3 + 48(1+\la_1^2)X_{i-1}^2
\right.\\ \left. +
16 X_{i-1}^3 - 48(R^2 + \la_1^2(1+2R))X_i + 48(\la_1^2+R) X_i^2
\right.\\ \left. - 16 X_i^3 
- 48 X_{i-1}(\la_1^2(-1-2R) - \log 2 + 2\la_1^2 X_i)\right)\\
 =  
 0.170044\, a_i^{(1)} \la_1 X_i + \frac{\sqrt{3} \, a_i^{(1)}  R \la_1 X_i \pi}{16}  
+  b^{(0)} 
\la_1^2(4-\la_1^2)\frac{\sqrt{3}\, \pi^3}{512}
\\
\mbox{ }-\sqrt{3}\,(R-1) \la_1^2\left(
  b^{(0)}(X_i-X_{i+1})\frac{\ee^{X_i-X_{i+1}}}{G^2}
- b^{(0)} (X_{i}-X_{i-1}) \frac{\ee^{ - X_{i}+X_{i-1}}}{G^2}
\right) \\
\mbox{ } -2\sqrt{3}\,(R-1)\la_1 \left(
 (a_{i+1}^{(1)}- a_i^{(1)})\frac{\ee^{X_i-X_{i+1}}}{G^2}
- (a_{i-1}^{(1)}- a_i^{(1)}) \frac{\ee^{ - X_{i}+X_{i-1}}}{G^2}
\right) 
 \\
\mbox{ } 
- b^{(0)}\la_1^2 \frac{\sqrt{3}}{2}\left(\frac{\ee^{X_{i}-X_{i+1}}}{G^2}+
 \frac{\ee^{X_{i-1}-X_i}}{G^2} \right) \left(R^2+R\right) \\ \mbox{ }
+b^{(0)}\frac{\sqrt{3}}{96}\ee^{X_i - X_{i+1}}\left(-\pi^2 - 16R^3-48X_iR^2 - 48 X_i^2 R - 16 X_i^3\right. \\  \mbox{ } \hspace{4.5cm}\left.-48 X_{i+1}^2 + 16 X_{i+1}^3 + 48 \log 2\, X_{i+1} \right) \\
 \mbox{}+ b^{(0)}\frac{\sqrt{3}}{96}\ee^{-X_i + X_{i-1}}\left(-\pi^2 - 16R^3+48X_iR^2 - 48 X_i^2 R + 16 X_i^3\right. \\  \hspace{3.5cm} \left.-48 X_{i-1}^2 - 16 X_{i-1}^3 - 48 \log 2\, X_{i-1} \right)
\\
+b_i^{(2)}\left(\frac{4 \sqrt{3}}{G^2}\ee^{-X_{i+1}+X_i}+\frac{4
  \sqrt{3}}{G^2}\ee^{-X_{i}+X_{i-1}}
\right).
\end{multline*}
The terms proportional to $R^3$ are
\begin{multline*}
-\frac{b^{(0)}}{32 \sqrt{3}}\frac{\ee^{X_i-X_{i+1}}}{G^2}16 R^3
-\frac{b^{(0)}}{32 \sqrt{3}}\frac{\ee^{X_{i-1}-X_{i}}}{G^2}16 R^3 
= -\frac{b^{(0)}\sqrt{3}}{96}\frac{\ee^{X_i-X_{i+1}}}{G^2}16 R^3
-\frac{b^{(0)}\sqrt{3}}{96}\frac{\ee^{X_{i-1}-X_{i}}}{G^2}16 R^3
\end{multline*}
These cancel as they should.
The terms proportional to $R^2$ are
\begin{multline*}
-\frac{b^{(0)}}{32 \sqrt{3}}\frac{\ee^{X_i-X_{i+1}}}{G^2}(48 \la_1^2 R^2+48 X_i R^2)
-\frac{b^{(0)}}{32 \sqrt{3}}\frac{\ee^{X_{i-1}-X_{i}}}{G^2}(48 \la_1^2 R^2-48 X_i R^2) \\
= -b^{(0)} \frac{\sqrt{3}\la_1^2}{2}\left(\frac{\ee^{X_i-X_{i+1}}}{G^2} + \frac{\ee^{X_{i-1}-X_{i}}}{G^2}\right)R^2 \\
+ \frac{b^{(0)}\sqrt{3}}{96}\frac{\ee^{X_i-X_{i+1}}}{G^2}(-48 X_iR^2 )
+\frac{b^{(0)}\sqrt{3}}{96}\frac{\ee^{X_{i-1}-X_{i}}}{G^2}(48 X_i R^2)
\end{multline*}
These cancel as they should.
The terms proportional to $R$ are
\begin{multline*}
-\frac{\sqrt{3}}{G^2} \ee^{X_i-X_{i+1}}\la_1 a_{i+1}^{(1)} 2R
-\frac{b^{(0)}}{32 \sqrt{3}}\frac{\ee^{X_i-X_{i+1}}}{G^2}(48 \la_1^2 R + 48 X_i^2 R - 96 \la_1^2 X_{i+1}R+ 96 X_i \la_1^2 R)\\
- \frac{\sqrt{3}}{G^2} \ee^{X_{i-1}-X_i} \la_1 a_{i-1}^{(1)}(-2R)
- \frac{b^{(0)}}{32 \sqrt{3}}\frac{\ee^{X_{i-1}-X_{i}}}{G^2}(48 \la_1^2 R - 96 \la_1^2 X_i R + 48 X_i^2 R + 96 X_{i-1}\la_1^2 R)\\ = 
\la_1 a_i^{(1)} \frac{2 \sqrt{3}\, R \pi X_i}{32} - b^{(0)} \frac{\sqrt{3}\la_1^2}{2}\left(\frac{\ee^{X_i-X_{i+1}}}{G^2} + \frac{\ee^{X_{i-1}-X_{i}}}{G^2}\right)R\\
-\sqrt{3}\,R \la_1^2 b^{(0)}\left(
(X_i-X_{i+1})\frac{\ee^{X_i-X_{i+1}}}{G^2}
-  (X_{i}-X_{i-1}) \frac{\ee^{ - X_{i}+X_{i-1}}}{G^2}
\right) \\
 -2\sqrt{3}\,R\la_1 \left(
 (a_{i+1}^{(1)}- a_i^{(1)})\frac{\ee^{X_i-X_{i+1}}}{G^2}
- (a_{i-1}^{(1)}- a_i^{(1)}) \frac{\ee^{ - X_{i}+X_{i-1}}}{G^2}
\right) \\
+ \frac{b^{(0)}\sqrt{3}}{96}\frac{\ee^{X_i-X_{i+1}}}{G^2}(-48 X_i^2R )
+\frac{b^{(0)}\sqrt{3}}{96}\frac{\ee^{X_{i-1}-X_{i}}}{G^2}(- 48 X_i^2 R)
\end{multline*}
These cancel as they should.
Thus, finally,
\begin{multline*}
\frac{4 \sqrt{3}}{G^2} \ee^{X_i-X_{i+1}} b_{i+1}^{(2)} -
\frac{\sqrt{3}}{G^2} \ee^{X_i-X_{i+1}}\la_1 a_{i+1}^{(1)} (1 + 2 X_i -
2 X_{i+1}) \\
-
\frac{b^{(0)}}{32 \sqrt{3} G^2} \ee^{X_i-X_{i+1}}
\left(\pi^2 + 48\la_1^2X_i^2
 + 16 X_i^3 
  - 48(\la_1^2 + \log 2) X_{i+1} + 48(1+\la_1^2) X_{i+1}^2 
\right.\\ \left. - 16 X_{i+1}^3
 + 48 X_i(\la_1^2 - 2 \la_1^2 X_{i+1})\right)
\\ 
+ \frac{4 \sqrt{3}}{G^2} \ee^{X_{i-1}-X_i} b_{i-1}^{(2)}
- \frac{\sqrt{3} }{G^2} \ee^{X_{i-1}-X_i} \la_1 a_{i-1}^{(1)}(-1-2X_{i-1}+2X_i) \\
- \frac{b^{(0)}}{32 \sqrt{3}G^2}  \ee^{X_{i-1}-X_i}\left(
\pi^2  + 48(1+\la_1^2)X_{i-1}^2
  +
16 X_{i-1}^3 - 48\la_1^2X_i + 48\la_1^2 X_i^2
\right.\\ \left. - 16 X_i^3 
- 48 X_{i-1}(-\la_1^2 - \log 2 + 2\la_1^2 X_i)\right)\\
 =  
 0.170044\, a_i^{(1)} \la_1 X_i 
+  b^{(0)} 
\la_1^2(4-\la_1^2)\frac{\sqrt{3}\, \pi^3}{512}
\\
\mbox{ }+\sqrt{3}\, \la_1^2\left(
  b^{(0)}(X_i-X_{i+1})\frac{\ee^{X_i-X_{i+1}}}{G^2}
- b^{(0)} (X_{i}-X_{i-1}) \frac{\ee^{ - X_{i}+X_{i-1}}}{G^2}
\right) \\
\mbox{ } +2\sqrt{3}\,\la_1 \left(
 (a_{i+1}^{(1)}- a_i^{(1)})\frac{\ee^{X_i-X_{i+1}}}{G^2}
- (a_{i-1}^{(1)}- a_i^{(1)}) \frac{\ee^{ - X_{i}+X_{i-1}}}{G^2}
\right) 
 \\
 \mbox{ }
 +b^{(0)}\frac{\sqrt{3}}{96}\ee^{X_i - X_{i+1}}\left(-\pi^2  - 16 X_i^3
   -48 X_{i+1}^2 + 16 X_{i+1}^3 + 48 \log 2\, X_{i+1} \right) \\
 \mbox{}+ b^{(0)}\frac{\sqrt{3}}{96}\ee^{-X_i + X_{i-1}}\left(-\pi^2 
   + 16 X_i^3
   -48 X_{i-1}^2 - 16 X_{i-1}^3 - 48 \log 2\, X_{i-1} \right)\\
+b_i^{(2)}\left(\frac{4 \sqrt{3}}{G^2}\ee^{-X_{i+1}+X_i}+\frac{4
  \sqrt{3}}{G^2}\ee^{-X_{i}+X_{i-1}}
\right).
\end{multline*}
Simplifying gives
\begin{multline*}
 - 0.170044 a_i^{(1)} \la_1 X_i + 
 b^{(0)} 
\la_1^2(\la_1^2-4)\frac{\sqrt{3}\, \pi^3}{512}
\\
+4 \sqrt{3} (b_{i+1}^{(2)} - b_i^{(2)} ) \frac{\ee^{X_i - X_{i+1}}}{G^2}
+4 \sqrt{3} (b_{i-1}^{(2)} - b_i^{(2)} ) \frac{\ee^{X_{i-1} - X_{i}}}{G^2}
\\
+2 \sqrt{3} a_{i+1}^{(1)}(X_{i+1} -X_i) \la_1\frac{\ee^{X_i - X_{i+1}}}{G^2}
+2 \sqrt{3} a_{i-1}^{(1)}(X_{i-1} -X_i)\la_1\frac{\ee^{X_{i-1} - X_{i}}}{G^2}
\\
+ \sqrt{3} (2a_{i}^{(1)}-3a_{i+1}^{(1)}) \la_1\frac{\ee^{X_i - X_{i+1}}}{G^2}
+ \sqrt{3} (-2a_{i}^{(1)}+3a_{i-1}^{(1)})\la_1\frac{\ee^{X_{i-1} - X_{i}}}{G^2}
\\
+ \frac{\sqrt{3}}{2}b^{(0)}\frac{\ee^{X_i - X_{i+1}}}{G^2}\left(
- 3\la_1^2 X_i - \la_1^2 X_i^2  + 3 \la_1^2 X_{i+1} + 2 \la_1^2 X_i X_{i+1} 
 - \la_1^2 X_{i+1}^2 \right)\\
+ \frac{\sqrt{3}}{2}b^{(0)}\frac{\ee^{X_{i-1} - X_{i}}}{G^2}\left(
 3\la_1^2 X_i - \la_1^2 X_i^2 - 3 \la_1^2 X_{i-1} + 2 \la_1^2 X_i X_{i-1}  
 - \la_1^2 X_{i-1}^2 
\right)=0.
\end{multline*}
We can check our analysis and identify the constant 0.170044 using the exact eigenvalue $\la_1=2$, for which 
\[ b^{(0)} = 1, \qquad a_i^{(1)} = X_i, \qquad b_i^{(2)} = -\frac{X_{i}^2}{2}.\]
Substituting this in 
and simplifying using (\ref{Xeqns})
gives
\[
 - 0.170044 \times 2 X_i^2 +  \sqrt{3} X_i^2  \frac{\pi }{16}
=0,
\]
so that 
\[ 0.170044 = \frac{\sqrt{3}\pi}{32}.\]
Thus, using (\ref{Xeqns}) again to simplify,  the final solvability condition is
\begin{multline}
\frac{ b^{(0)} \la_1^2(\la_1^2-4) \pi^3}{512} + \frac{3\pi \la_1 X_i}{64} ( b^{(0)}\la_1  X_i-2a_{i}^{(1)}) 
\\
+4  (b_{i+1}^{(2)} - b_i^{(2)} ) \frac{\ee^{X_i - X_{i+1}}}{G^2}
+4  (b_{i-1}^{(2)} - b_i^{(2)} ) \frac{\ee^{X_{i-1} - X_{i}}}{G^2}
\\
+2  a_{i+1}^{(1)}(X_{i+1} -X_i) \la_1\frac{\ee^{X_i - X_{i+1}}}{G^2}
+2  a_{i-1}^{(1)}(X_{i-1} -X_i)\la_1\frac{\ee^{X_{i-1} - X_{i}}}{G^2}
\\
 -3 a_{i+1}^{(1)} \la_1\frac{\ee^{X_i - X_{i+1}}}{G^2}
+ 3 a_{i-1}^{(1)}\la_1\frac{\ee^{X_{i-1} - X_{i}}}{G^2}
\\
+ \frac{\la_1^2}{2}b^{(0)}\frac{\ee^{X_i - X_{i+1}}}{G^2}\left(
 -  X_i^2  + 3  X_{i+1} + 2  X_i X_{i+1} 
 -  X_{i+1}^2 \right)\\
+ \frac{\la_1^2}{2}b^{(0)}\frac{\ee^{X_{i-1} - X_{i}}}{G^2}\left(
  -  X_i^2 - 3  X_{i-1} + 2  X_i X_{i-1}  
 -  X_{i-1}^2 
\right)=0.\label{solvfinalapp}
\end{multline}
This is (\ref{finalsol}).

\end{document}